\documentclass[pre,reprint,amsmath,amssymb,preprintnumbers,aps,floatfix]{revtex4-2}
\usepackage[margin=0.75in]{geometry}
\usepackage[utf8]{inputenc}
\usepackage{xr}
\usepackage{placeins}
\usepackage{amsmath}
\usepackage{amssymb,float}
\usepackage{soul}
\usepackage[normalem]{ulem}
\usepackage{amsfonts}
\usepackage{amstext}
\usepackage{appendix}
\usepackage{mathtools}
\usepackage{amsthm}
\usepackage{subcaption}
\usepackage{graphicx}
\usepackage{color}
\usepackage{lipsum}
\usepackage{natbib}
\usepackage{xcolor}
\usepackage{float}
\usepackage{bm}
\usepackage[hidelinks]{hyperref}
\usepackage{comment}
\usepackage[normalem]{ulem}
\usepackage{ragged2e}
\usepackage[font={scriptsize}, skip=0pt,format=plain,singlelinecheck=false]{caption}

\usepackage{helvet}
\def\be{\begin{equation}}
\def\ee{\end{equation}}
\def\bea{\begin{eqnarray}}
\def\eea{\end{eqnarray}}

\def\be{\begin{equation}}
\def\ee{\end{equation}}
\def\bea{\begin{eqnarray}}
\def\eea{\end{eqnarray}}

\def\a{\alpha}

\begin{document}

\preprint{MIT-CTP/5955}
\title{Emergent self-inhibition governs the landscape of stable states in complex ecosystems}

\author{Nitesh Kumar Patro$^1$}  
\author{Washington Taylor$^2$}
\author{Akshit Goyal$^1$} 
\email{akshitg@icts.res.in}
\address{$^1$ International Centre for Theoretical Sciences, Tata Institute of Fundamental Research, Bengaluru 560089.}
\address{$^2$ Department of Physics, Massachusetts Institute of Technology, Cambridge, MA 02139.}

\begin{abstract}
\noindent Species-rich ecosystems often exhibit multiple stable states
with distinct species compositions. Yet, the factors determining the
likelihood of each state's occurrence remain poorly understood. Here,
we characterize and explain the landscape of stable states in the
random Generalized Lotka--Volterra (GLV) model, in which
multistability is widespread. 
We focus on the regime of the model where interspecies interactions are predominantly competitive and nearly symmetric. We find that the same pool of species with random initial abundances
can result in different stable states, whose likelihoods typically
differ by orders of magnitude. A state's likelihood
increases sharply with its total biomass, or inverse self-inhibition. We develop a simplified model to predict and explain this behavior, by coarse-graining ecological interactions so that each stable state behaves as a unit. In this setting, we can accurately predict the entire landscape of stable states using only two macroscopic properties: the biomass of each state and species diversity. Our analyses also provide insight into the biomass--likelihood relationship: High-biomass states have low self-inhibition and thus grow faster, outcompete others, and become much more likely. These results reveal emergent self-inhibition as a fundamental organizing principle for the attractor landscape of complex ecosystems—and
provide a path to predict ecosystem outcomes without knowing
microscopic interactions.   
\end{abstract}

\maketitle
\noindent
Species-rich ecosystems often exhibit multiple stable states with distinct species compositions and biomass levels \cite{Petraitis}. Examples range from lush versus barren forests
\cite{Staver-2011,Stritih_Seidl_Senf_2023_AlternativeStates,aleman2020floristic}
to healthy versus diseased gut microbiomes
\cite{LozuponeStombaughGordonJanssonKnight2012_diversity,VanDeGuchte_2020_AltStableIntestinal}. While
multistability is well documented
\cite{Schmitt2019AlternativeAttractors,Petraitis2009MultipleStates,Abreu2020AltStableStates,LopesAmorGore2024_multistability,Estrela2022FunctionalAttractors,TkaczCheemaChandraGrantPoole2015_rhizosphere_stability,KearneyEtAl2021_HeterotrophDiversityPicocyanobacteria},
the factors that govern which states are more \emph{likely} across a variety of initial conditions remain poorly understood—limiting our ability to predict or control ecological transitions
\cite{Scheffer2007Regime,zou2024positive,SANCHEZPINILLOS2024110409,dubinkina2019multistability}.

Understanding these likelihood patterns requires characterizing the sizes of attractor basins of different stable states. This becomes particularly challenging in diverse ecosystems, which are high-dimensional due to a large number of interacting species. 
In diverse ecosystems, attractor landscapes become complex and unintuitive \cite{Aguade-Gorgorio_2024}, and classical low-dimensional analysis offers limited insight. Recent work using statistical physics tools
has revealed rich phase behavior in disordered ecological models
\cite{Bunin2017_EcoComLv,GiralMartinez2024_stabilization,TikhonovMonasson2017PRL,AltieriRoyCammarotaBiroli2021_GlassyEquilibria,GarciaLorenzana2022_PRE_stronglyCompetitiveRLV,CuiMarslandMehta2021_PRE_TypicalRandomCRM,CuiMarslandMehta2024_LesHouchesCommunityEcology,MarslandMehta2020_PRL_SpeciesPacking,AltieriDeMonte2025_EPL_StatPhysCommunities,Ke2024Taxonomy,Advani2017Environmental,Marsland2019EnergyFluxes,allesina2015predicting,allesina2012stability}
, including stability-to-chaos transitions
\cite{ArnoulxDePirey2024_ManySpeciesEcologicalFluctuations,pearce2020stabilization,BlumenthalRocksMehta2024_PhaseTransitionChaos,MallminTraulsenDeMonte2024_ChaoticTurnoverRareAbundant,liu2025complex,mahadevan2023spatiotemporal}. Yet,
the multistable regime---where multiple attractors coexist---has
received comparatively little attention.  While previous studies have
characterized the number of stable states
\cite{Ros2023_TypicalNumberEquilibria,dubinkina2019multistability,goyal2018multiple,fried2017alternative,kessler2025interaction},
we lack understanding of which factors predict the sizes of different
attractor basins. Recent work on a certain class of
``niche-structured'' models \cite{taylor2024structure} found that larger
attractor basins correlate with increased steady-state biomass,
suggesting this relationship may hold more broadly. However, there
has been little
work that describes the landscape of attractor basins for more
general ecological models (though see \cite{gilpin1976multiple}).




In this Letter, we analyze the multistability landscape of the random Generalized Lotka–Volterra (GLV) model \textcolor{black}{with predominantly competitive, and nearly symmetric interspecies interactions}. We show that in this model,
a state's likelihood
increases sharply with its biomass in a strikingly consistent fashion, where biomass is computed as the sum of abundances of surviving species. 
We explain this relationship via a coarse-graining
approach, in terms of an effective model in which individual stable
states compete with each other. In this setting, each state behaves
like a single 
unit governed by an emergent self-inhibition, which
we find is inversely proportional to its
steady-state biomass. This self-inhibition
simultaneously slows growth and limits final abundance in opposing
ways, creating a tradeoff that favors high-biomass states. We
analytically derive the biomass–likelihood relationship for 
related monodominant and block-dominant situations, where only a single species or block of species survives in each
distinct stable state, and
use this to accurately predict the
likelihood of 100s of stable states in a complex disordered ecosystem. Notably, this prediction can be
made without precise details of the interaction matrix, which remains
extremely hard to infer for highly-diverse ecosystems
\cite{Srinivasan2024_ModelingMicrobialCommunity,Ona2025_disentangling,MarrecBravo-RuisecoZhouGodwin2025_interactions,MEROZ2024102511}. Instead,
it only requires macroscopic quantities---each state's biomass (or more generally, net growth rate)
and species
diversity---which are easy to measure experimentally
\cite{Mira2022,goldford2018emergent,dal2021resource}. Thus our results
provide a framework for estimating the relative likelihoods of
alternative ecological states, without requiring exhaustive
observational sampling or high-throughput experiments.

\begin{figure*}[ht!]    
   \centering\includegraphics[width=\textwidth]{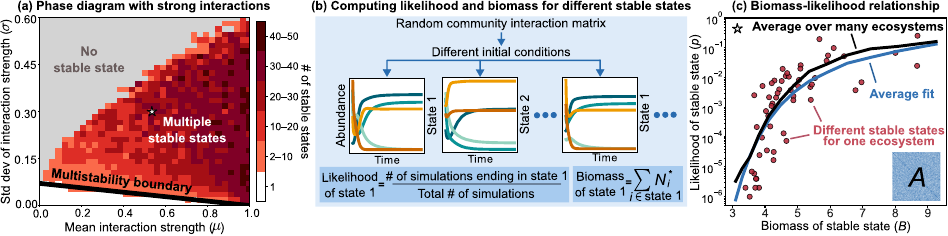}
    \vspace{0.05cm}\caption{\justifying\textsf{\textbf{Multistability and biomass--likelihood relationship in the random GLV model with strong interactions.}
(a) Phase diagram showing the typical number of stable states for each interaction matrix $A_{ij}$ with $S=100$ species given its mean $\mu$ and standard deviation $\sigma$. Multistability is widespread (red region); black line shows the analytically-derived multistability boundary (Appendix~\ref{app:multistabilityboundary}) separating the multistable and unique stable state (white) phase.
(b) For a fixed $A$, varying initial conditions reveals different
        stable states. Shown are example dynamics (each species is a
        different color). Each state's likelihood is estimated as the
        fraction of simulations that end in the state, while biomass
        is computed as the total abundance of surviving species.
(c) \textcolor{black}{The likelihood $p$ of a stable state increases sharply with biomass $B$. Data in red show results from $10^6$ simulations with random initial conditions for a single matrix; black curve shows biomass-likelihood relationship averaged over 110 matrices with $\mu = 0.5$, $\sigma = 0.3$.  Blue curve shows fit to a hyperbolic relation: $\log(p) \propto (B_c - B)^{-1}$ averaged over 110 matrices.}}}
\label{fig:fig1}
\end{figure*}

\textit{Setup.} Our starting point is the Generalized Lotka–Volterra (GLV) model, which describes the dynamics of $S$ interacting species:
\begin{equation}
\color{black}
\frac{dN_i}{dt} = N_i \left(r_i \left( 1- \frac{N_i}{K_i}\right) - \sum_{\substack{j=1,\ j\ne i}}^S A_{ij} N_j\right) , \quad i = 1, 2, \ldots, S.
\label{eq:GLVEq}
\end{equation}
Here, $N_i$ denotes the abundance (in units of biomass) of species $i$, \textcolor{black}{$r_i$ is its growth rate,} $K_i$ is its
carrying capacity, and $A_{ij}$ represents the effect of species $j$
on species $i$. The matrix $A$ thus encodes all interspecies
interactions and determines the stability landscape of the
ecosystem. Following the statistical physics tradition of replacing
complexity with randomness, we assume the parameters $A_{ij}$ are
drawn as quenched random variables
\cite{may1972will,GiralMartinez2024_stabilization,Bunin2017_EcoComLv,GarciaLorenzana2022_PRE_stronglyCompetitiveRLV,CuiMarslandMehta2024_LesHouchesCommunityEcology,ArnoulxDePirey2024_ManySpeciesEcologicalFluctuations,MallminTraulsenDeMonte2024_ChaoticTurnoverRareAbundant}. Specifically,
we take $A_{ij}$ to be distributed with mean $\mu$ and standard
deviation $\sigma$. We consider the \emph{strong interaction} regime:
both $\mu$ and $\sigma$ are finite constants that do not scale with
$S$. This contrasts with some recent statistical physics treatments
which focus on \textit{weak interactions} where $\mu \sim
\mathcal{O}(1/S)$ and $\sigma^2 \sim \mathcal{O}(1/S)$
\cite{Bunin2017_EcoComLv,ArnoulxDePirey2024_ManySpeciesEcologicalFluctuations,GiralMartinez2024_stabilization,CuiMarslandMehta2024_LesHouchesCommunityEcology,BlumenthalRocksMehta2024_PhaseTransitionChaos},
though there are exceptions
\cite{Hu2022_Science_EmergentPhases,GarciaLorenzana2022_PRE_stronglyCompetitiveRLV,MallminTraulsenDeMonte2024_ChaoticTurnoverRareAbundant}. Weak
interactions are assumed only to ensure a proper thermodynamic limit
as $S\to\infty$. Since natural ecosystems contain a finite number of
species, and intuitively the interaction between a given pair of species does
not change when a broader pool of species is available,
we instead stick to strong interactions. We further assume
that interactions are reciprocal, so $A$ is symmetric, and we fix \textcolor{black}{$r_i = K_i
= 1$ for random matrices. Our results are qualitatively robust to relaxing these simplifying assumptions (Appendix~\ref{app:robustness}). Note that when $r_i$'s differ by species, the growth-rate weighted biomass $\sum r_i N_i^*$ can be a slightly better predictor of likelihoods (see Appendix~\ref{app:robustness} for explanation).
}

By varying the two parameters $\mu$ and $\sigma$, we map out the phase diagram of the model (Fig.~\ref{fig:fig1}a), in agreement with past work \cite{Aguade-Gorgorio_2024,GiralMartinez2024_stabilization}. This diagram is dominated by a multistable phase (red), in which each community settles into one of multiple stable states depending on initial conditions. This region is separated from a unique stable state regime (white) by an analytically derived multistability boundary (black) using the cavity method (\textcolor{black}{by appropriate rescaling of the boundary obtained for weak interactions~\cite{Bunin2017_EcoComLv};} Appendix~\ref{app:multistabilityboundary}):

\begin{equation}
    \sigma = \frac{-\mu + 1}{\sqrt{2S}}.
    \label{eq:multistabBoundary}
\end{equation}

This shows that the multistable regime widens with increasing
$\mu$. Specifically, the critical $\sigma$ at which multistability
sets in decreases with $\mu$, unlike for models with weak interactions,
where it is constant \cite{Bunin2017_EcoComLv}. 
The third phase (grey) represents a dynamically infeasible phase
\cite{Bunin2017_EcoComLv,MallminTraulsenDeMonte2024_ChaoticTurnoverRareAbundant}
with unbounded growth
and no stable states
(Fig.~\ref{fig:fig1}a).
Due to the prevalence
of multistability, this model provides a natural setting to
investigate the attractor landscape, including the relative
likelihoods of stable states.

\textit{Biomass--likelihood relationship.} To probe the multistability
landscape, we choose a  random interaction matrix $A$ with $S=100$
species and simulate dynamics from $10^6$ random initial
conditions. We sample each species' initial abundance uniformly
between 0 and 1, uncorrelated across species. Our central results are
robust
to other sampling schemes (Appendix~\ref{app:robustness})
and to the specific choice of matrix $A$.
Each simulation converges to one of several possible stable states. We
define the likelihood of a stable state as the fraction of initial
conditions that converge to it (Fig.~\ref{fig:fig1}b,
Appendix~\ref{app:methods}). This allows us to numerically estimate ecological attractor basin sizes.

We find that the distribution of stable state
likelihoods is highly skewed: the most likely state is over six orders
of magnitude more probable than the least likely one \textcolor{black}{observed (in many cases, there are further unobserved states that we might not detect; see Appendix~\ref{app:methods})}
. These dramatic
differences in likelihood are strongly correlated with the total
biomass of each state. Specifically, the likelihood of a stable state $p$
increases sharply with its biomass $B$, with the sharpness progressively
increasing at higher biomass (note the log scale in
Fig.~\ref{fig:fig1}c). \textcolor{black}{Averaging results from many random matrices, we obtain an approximation to the biomass-likelihood relationship (Fig.~\ref{fig:fig1}c; Appendix~\ref{app:disorderAveraging}), which roughly looks like a hyperbola in $(-\log p, B)$ space: $\log\left(p\right) \propto (B_c -
B)^{-1}$ 
(Fig.~\ref{fig:fig1}c,
see Appendix~\ref{app:hyperboliclog} for a detailed discussion).}
%
The strong increase of the likelihood of a state with its biomass is
robust to changes in the number of species, interaction parameters, carrying
capacity variation, and degree of symmetry of $A$
(Appendices~\ref{app:robustness}, \ref{app:hyperboliclog}). \textcolor{black}{These
findings generalize earlier results on niche-structured ecosystems
\cite{taylor2024structure} (Appendix~\ref{app:hyperboliclog} for comparison), showing that biomass remains a strong
predictor of the likelihood of different stable states in a broad
range of complex ecosystems.
}

\begin{figure}
\centering\includegraphics[width=\linewidth]{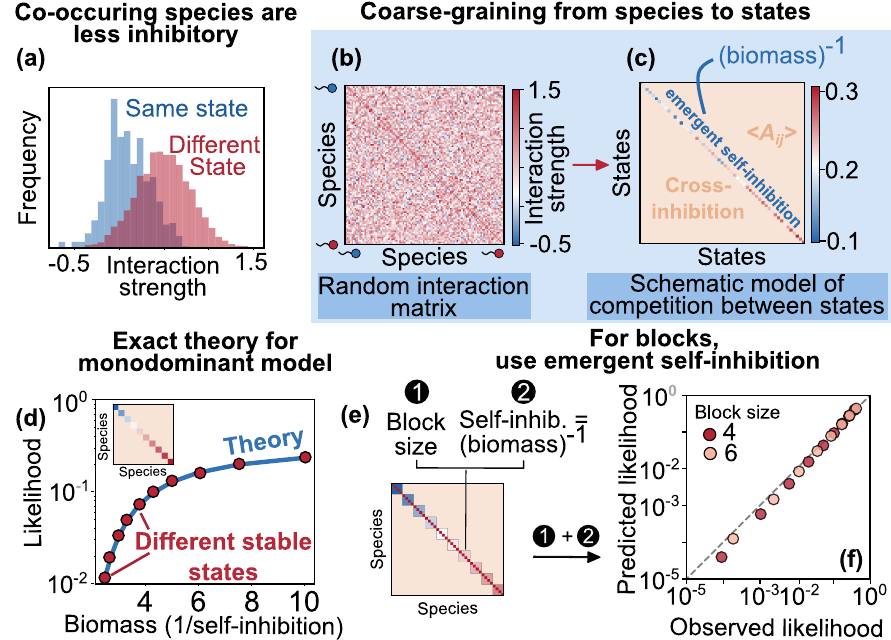}
    \vspace{3pt}\caption{\justifying \textsf{\textbf{Coarse-graining states reveals emergent self-inhibition as a predictor of biomass and likelihood.}
(a) Species that interact weakly with each other form uninvadable stable states. The interactions among coexisting species are separated from the original interaction distribution, with a reduced mean. (b) Species-level interaction matrix. (c) We can conceptualize this as an effective matrix where different states compete with each other and one eventually emerges as the winner.
        (d) In a monodominant matrix where different stable states contain only a single species, both biomass and likelihood can be computed analytically from the self-inhibition (Appendix~\ref{app:monodominantCalc}). 
        (e) In a block matrix where each different state corresponds to a block of $L$ species, we can treat each block as an effective species with self-inhibition inversely proportional to its biomass. (f) Predictions combining this self-inhibition and block size accurately predict simulated likelihoods. Shown are block sizes $L=\{4,6\}$.
}
}
\vspace{-10pt}
\label{fig:fig2}
\end{figure}

\textit{Coarse-graining species to states.} To understand the origin of this biomass--likelihood relationship, we focus on the interactions between species that coexist within the same stable state. Species that coexist in the same state inhibit each other much more weakly than expected by chance (Fig.~\ref{fig:fig2}a). Indeed, species in the same state can even be mutualistic and promote each other's growth ($A_{ij}<0$). This can be visualized as an emergent block structure in part of the interspecies interaction matrix (Figs.~\ref{fig:block_like}--\ref{fig:block}, Appendix~\ref{app:blockGrouping}). Together, these observations suggest that species within each state grow as a cohesive unit, motivating an effective model where different stable states compete with each other, and one state eventually emerges as the winner depending on initial conditions (Fig.~\ref{fig:fig2}b--c).

We develop this picture in stages, starting with the simplest case and
building toward random disordered matrices. We begin with a classic
\emph{monodominant} model \cite{lotka1956elements}
(also studied in this context in \cite{taylor2024structure}),
where individual species compete so
strongly that only one species survives in each stable state. We then
extend to block ecosystems with disjoint groups of coexisting
species. Finally, we apply these results to random matrices, where we again think of states as blocks of species. \textcolor{black}{Here, different states often overlap and contain species in common, which we can account for using a simple mean-field approximation under which the block model makes reasonable predictions even for random matrices
(Appendix~\ref{app:overlappingStates}).}

The monodominant model is exactly solvable. To construct it from
Eq.~\eqref{eq:GLVEq}, we consider a common interspecies interaction
strength $A_{ij} =D$ for all \textcolor{black}{$i \neq j$ (cross-inhibition) and ensure
that all self-inhibitions $A_{ii} = r_i/K_i$} are weaker than this: $A_{ii} < D$
for all $i$. \textcolor{black}{As in the random matrix case, here also we assume $r_i = 1$ (see Appendix~\ref{app:robustness} for generalization to variable growth rates $r_i$).} The per capita
growth rate of each species then becomes
\begin{equation}
\frac{1}{N_i} \frac{dN_i}{dt} = \left(1 - D \sum_{j=1}^S N_j\right) + (D - A_{ii}) N_i.
\label{monodomGrowthEq}
\end{equation}

The first term in parentheses is common to all species, while the
second is species-dependent. Specifically, the
relative growth rate of each
species depends on its abundance $N_i$ dressed with a prefactor
$\chi_i = D - A_{ii}$. If two species start at the same abundance, the
one with larger $\chi_i$ will grow faster and win. In particular, we
show that if all species start at abundances $N_i(0)$, the species
with the largest value of the dressed initial condition $\chi_i
N_i(0)$ will win (Fig.~\ref{fig:mono}). For constant $D$, this condition implies that
under uniform random initial conditions, species with lower
self-inhibition $A_{ii}$ will have a greater likelihood. Calculating
this likelihood in general is rather challenging. With uniform initial
conditions, the likelihood of each species winning is the fraction of
the volume of initial conditions that maps to its attractor basin. For
two species, calculating likelihoods is easy: a separatrix line
partitions the two-dimensional phase space into two
attractors. However, this method becomes computationally expensive
when the number of species $S$ is large. In this case, each
pair of species defines a separatrix hyperplane, and the basin volumes correspond to $S!$ high-dimensional solid
cones subtended by the arrangement of these
planes.

We circumvent this complexity by reformulating the problem as one of extreme value statistics. Rather than computing basin volumes directly, we consider the distributions of the dressed initial conditions $\chi_i N_i(0)$ and compute the probability that a given species has the maximum value. This simplification enables us to analytically calculate the likelihood $p_i$ of species $i$ winning in terms of $\chi_i$ (Appendix~\ref{app:monodominantCalc}) as:

\begin{equation}
    p_i =\sum_{k=1}^{i}\frac{\chi_k^{\,S-k+1}-\chi_{k-1}^{\,S-k+1}}{(S-k+1)\displaystyle\prod_{j=k}^{S}\chi_j}=p_{i-1}+\frac{\chi_i^{\,S-i+1}-\chi_{i-1}^{\,S-i+1}}{(S-i+1)\displaystyle\prod_{j=i}^{S}\chi_j}
    \label{piMonodomEq}
\end{equation}
Here, species are indexed by $i$ in increasing order of their $\chi_i$
(i.e., decreasing order of self-inhibition) and $\chi_0 = 0$. The
recursive relation between $p_i$ and $p_{i-1}$ shows that decreasing
self-inhibition (increasing $\chi$) increases likelihood. Further, the
dependence of $p_i$ purely on $\chi$ highlights that self-inhibition
$A_{ii}$ is the key determinant of attractor likelihood in
monodominant ecosystems. Notably, self-inhibition also directly sets
biomass as $B_i = 1/A_{ii}$. Thus, Eq.\eqref{piMonodomEq} can
equivalently be expressed in terms of biomass $B_i$ to reveal the
biomass--likelihood relationship explicitly
(Appendix~\ref{app:monodominantCalc}). Eq.~\eqref{piMonodomEq} shows
excellent agreement with simulations (Fig.~\ref{fig:fig2}d).  

\begin{figure}[ht!]
\centering\includegraphics[width=1.\linewidth]{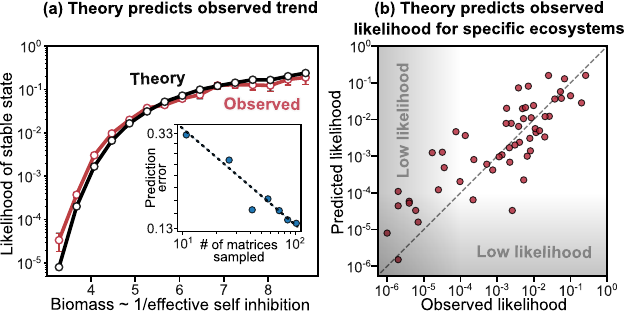}
    \vspace{3pt}\caption{\justifying \textsf{\textbf{Biomass predicts state likelihoods in ecosystems with disordered interactions.}
    \textcolor{black}{(a) Using the biomass of each state as a measure of its emergent
(inverse)
 self-inhibition, our theory (black) can predict the overall biomass–likelihood relationship (red). Inset shows how prediction error decreases with number of sampled matrices (roughly power-law with exponent $-0.4$; Appendix~\ref{app:disorderAveraging}).
For states with dramatically low likelihoods, our theory slightly deviates in likelihood predictions (gray region) (Appendix~\ref{app:likelihoodCalcRandMatrix}).
      (b) For a specific ecosystem, the predicted likelihoods of each state, without any detailed knowledge of interactions, match the observed likelihoods for most states. 
    }
}}
\label{fig:fig3}
\end{figure}

\textit{Generalizing monodominant results to block ecosystems.} The
monodominant model predicts the likelihood of stable states containing
only one species. In our random disordered ecosystems however, each
state typically consists of several species. We thus extend our
approach for computing state likelihoods to ecosystems with
block-structured interaction matrices. We consider block ecosystems
where each state $\alpha$ contains $L_\alpha$ species which inhibit each other with strength $A_{ij}^{(\alpha)}$. Just like in
the monodominant model, in a stable state one block outcompetes all
others, but which block that will be might differ based on initial
conditions. Species within each block $\alpha$ co-occur since their
inhibition $A_{ij}^{(\alpha)}$
is weaker than inhibition from others
outside the block $D$ (Fig.~\ref{fig:fig2}e). 

Similar to the monodominant model, to compute block likelihoods, we
treat each block as an effective unit. In this context, we can again show that with a proper definition of
effective self-inhibition
(Appendix~\ref{app:blockCalc}), the block with the largest value of the dressed initial condition will win. The initial abundance for a block now involves summing the
initial abundances of all $L_\alpha$ species within that block
$\sum_{i \in \alpha} N_i(0)$. By the central limit theorem, we expect
this sum to be Gaussian-distributed for large $L_\alpha$. Further,
each block's initial abundance is dressed by a prefactor $\chi_\alpha$
that depends on that block's effective self-inhibition $A_{\a\a}^{\text{eff}}$.
Similar to the monodominant model, this quantity is the inverse of the block's biomass
$B_\alpha$.
Combined, the dressed initial condition for block $\alpha$ becomes $\chi_\alpha \sum_{i \in \alpha}
N_i(0)$, where $\chi_\alpha = D - 1/B_\alpha$. This is well-approximated by a Gaussian with mean
$m_\alpha = L_\alpha \chi_\alpha/2$ and variance $s_\alpha^2 =
L_\alpha\chi_\alpha/12$ (Appendix~\ref{app:blockCalc}). Following a similar calculation as the
monodominant case, the likelihood
$p_\alpha$
of block state $\alpha$ winning
 becomes:

\begin{equation}
   p_\alpha = \int_{0}^\infty \frac{1}{s_\alpha}f\left( \frac{z_\alpha-m_\alpha}{s_\alpha}\right) \prod_{\substack{\beta=1\\\beta\neq \alpha}}^{N_B} \Phi \left( \frac{z_\alpha-m_\beta}{s_\beta}\right) dz_\alpha,
\label{eq:block_likelihood}
\end{equation}
where $f$ is the standard normal PDF, $\Phi$ is the standard normal
CDF, and $N_B$ is the total number of blocks. This integral has a
rather simple interpretation---it computes the probability that block
$\alpha$ has the largest dressed initial condition
$\chi_\alpha\sum_\alpha N_i(0)$ among all blocks for a random uniform
initial condition. To see this, notice that $f_\alpha$ gives the
density of block $\alpha$'s dressed initial condition being an
arbitrary $z_\alpha$. The remaining product goes over all other blocks
$\beta\neq\alpha$ and demands that their dressed initial conditions be
less than $z_\alpha$. We then simply integrate over all possible
values of $z_\alpha$. Thus, Eq.~\eqref{eq:block_likelihood} is really
a generalization of the monodominant Eq.~\eqref{piMonodomEq} to block
states containing multiple species. This expression 
predicts the likelihoods observed in simulations extremely well for a
range of block sizes $L_\alpha$ (Fig.~\ref{fig:fig2}e). Thus, for
block-structured ecosystems, each state's likelihood can be predicted
using only the inter-state inhibition $D$ and each
state’s steady-state (equilibrium) biomass
$B_\alpha$ and diversity $L_\alpha$.

\textit{Predicting likelihoods for disordered ecosystems.} Motivated
by these results, we return to the case of random unstructured
matrices from Fig.~\ref{fig:fig1}. We ask whether we can use the
framework developed so far to predict state likelihoods in this
case. Can we treat each observed state as an effective block which
competes with other states? \textcolor{black}{To map disordered matrices
  to the block model, we need to account for two major differences
  between these settings: (1) in disordered matrices, species often
  overlap between different states, unlike the block model with no
  overlaps; and (2) the inter-state inhibitions are now random
  variables. To this end, we generalized our block model to include
  random overlaps between blocks
  (Appendix~\ref{app:blockOverlaps}). We find
  that when overlaps are
  sufficiently widespread and randomly distributed, larger blocks tend
  to experience greater overall self-inhibition while smaller blocks
  tend to experience lower self-inhibition. This lowers the likelihood
  of larger blocks while increasing the likelihood of smaller blocks, thus reducing the effect of diversity variation between blocks
  (Fig.~\ref{fig:overlap_crossover}).
%
This motivates the following mean-field approximation:
  we replace the size $L_\alpha$ of each block with the mean diversity $\langle L_\alpha\rangle=L$ across all $N_B$ observed stable states, and the inter-block inhibition $D$ with the mean interspecies interaction strength
$\mu=\langle A_{ij}\rangle$. For each block's biomass $B_\alpha$, we use the biomass of each observed state.}
%

Using these assumptions, our model accurately predicts the
biomass--likelihood relationship in disordered interaction
matrices (Fig.~\ref{fig:fig3}a). The predictions match simulations
across a broad range of the multistable phase
(Appendix~\ref{app:likelihoodCalcRandMatrix},
Fig.~\ref{fig:general}). \textcolor{black}{Indeed, prediction error decreases roughly as a power-law as we sample more matrices, supporting the accuracy of our theory (Fig.~\ref{fig:fig3}a, inset; Appendix~\ref{app:disorderAveraging}
.} Furthermore, \textcolor{black}{for a particular ecosystem,} we can predict the likelihood
of each observed state using only its total biomass and typical species richness (Appendix~\ref{app:likelihoodCalcRandMatrix}).
These predictions align with simulated likelihoods for the majority of
states (Fig.~\ref{fig:fig3}b), with some deviations at low-likelihoods. Together, these results demonstrate that the
attractor landscape of complex
random ecosystems is largely predictable
without detailed knowledge of species-level interactions.

\textit{Discussion.} We have found a strong relationship between biomass and basin of
attraction size for multiple stable states in a broad class of GLV
models including monodominant, and block monodominant
models, as well as random matrix models \textcolor{black}{with predominantly competitive and nearly symmetric interactions}, generalizing the observations of Ref. \cite{taylor2024structure} for
niche models to a much broader class of systems.
Our findings furthermore 
give a quantitative underpinning to the idea that the likelihood of
observing a set of species in a community is governed
by
how they collectively inhibit each other during
growth. This collective self-inhibition of a community is a complex, emergent
quantity that can be estimated as the inverse of the community's final
biomass. This leads to a robust positive correlation between the
biomass and likelihood of a state. 
\textcolor{black}{Note that here we  focus on the relative probabilities of distinct stable states, giving the fundamental probability distribution on states.  
Since there are more states with intermediate biomass, computing properties of a ``typical'' state under this distribution would involve a balance of entropic factors favoring intermediate biomass and domain size, favoring larger biomass.}
Our results motivate
experimental validation in natural microbial communities. In
particular, communities derived from parallel enrichments—e.g., of
soils, human microbiomes or plant rhizospheres—often give rise to multiple
stable states under identical conditions
\cite{Estrela2022FunctionalAttractors,KearneyEtAl2021_HeterotrophDiversityPicocyanobacteria,TkaczCheemaChandraGrantPoole2015_rhizosphere_stability,LopesAmorGore2024_multistability,goyal2022interactions}. Quantifying the frequency of states across replicates enables one to estimate their likelihoods. Our
results predict that these frequencies should be predictable only
using each state’s total biomass (e.g., via cell counts or optical
density) and species diversity (e.g., from 16S amplicon
sequencing). Note that when individual species have different growth
rates, slightly different measurements such as total community growth
rate would be needed in our prediction method
(Appendix~\ref{app:robustness}). 
\textcolor{black}{We also note that our analysis applies directly to
  deterministic ecosystem dynamics and neglects stochastic
  fluctuations due to
  random births and deaths. While our results are robust to
  introducing such noise (Appendix~\ref{app:robustness}), it would be interesting to
  fully explore its consequences, especially how noise affects the
  interplay between the widths and depths of attractor basins.} 
Another possible direction is predicting which species subsets form stable states directly from interactions rather than simulations, although this may be computationally difficult (Appendix~\ref{app:hyperboliclog}).

Beyond empirical validation, our findings open several directions for future work. Can we build on our framework to find ways to control ecosystems to be in states with low likelihoods, which might otherwise be healthy or functionally desirable? How can we predict the likelihood of being near a steady-state in chaotic ecological dynamics, which can occur due to nonreciprocal interactions \cite{ArnoulxDePirey2024_ManySpeciesEcologicalFluctuations,BlumenthalRocksMehta2024_PhaseTransitionChaos}? Does community evolution significantly deform the landscape of ecological stable states? Addressing these questions will be key to understanding how interspecies interactions collectively determine the state of a complex ecosystem.

\emph{Acknowledgements.} We thank J.P. O'Dwyer, A. Altieri, J. Gore, and S.Y. Li for valuable discussions. A.G. acknowledges support from the DAE under project no. RTI4001, as well as a Ramanujan Fellowship.
The work of W.T. was supported in part by a grant from the
Schmidt Futures Foundation, and W.T. would like to thank the Santa Fe
Institute 
and the University of Auckland for hospitality while some of this work was done.

\bibliographystyle{apsrev4-2}
\bibliography{ref_1}

@article{good1953population,
  title={The population frequencies of species and the estimation of population parameters},
  author={Good, Irving J},
  journal={Biometrika},
  volume={40},
  number={3-4},
  pages={237--264},
  year={1953},
  publisher={Oxford University Press}
}

@article{chao1984nonparametric,
  title={Nonparametric estimation of the number of classes in a population},
  author={Chao, Anne},
  journal={Scandinavian Journal of statistics},
  pages={265--270},
  year={1984},
  publisher={JSTOR}
}

@article{chao2016species,
  title={Species richness: estimation and comparison},
  author={Chao, Anne and Chiu, Chun-Huo and others},
  journal={Wiley StatsRef: statistics reference online},
  volume={1},
  number={26},
  pages={10--1002},
  year={2016}
}

@article{chao2009sufficient,
  title={Sufficient sampling for asymptotic minimum species richness estimators},
  author={Chao, Anne and Colwell, Robert K and Lin, Chih-Wei and Gotelli, Nicholas J},
  journal={Ecology},
  volume={90},
  number={4},
  pages={1125--1133},
  year={2009},
  publisher={Wiley Online Library}
}

@article{chao2012coverage,
  title={Coverage-based rarefaction and extrapolation: standardizing samples by completeness rather than size},
  author={Chao, Anne and Jost, Lou},
  journal={Ecology},
  volume={93},
  number={12},
  pages={2533--2547},
  year={2012},
  publisher={Wiley Online Library}
}

@inproceedings{bollobas1976cliques,
  title={Cliques in random graphs},
  author={Bollob{\'a}s, B{\'e}la and Erd{\"o}s, Paul},
  booktitle={Mathematical Proceedings of the Cambridge Philosophical Society},
  volume={80},
  number={3},
  pages={419--427},
  year={1976},
  organization={Cambridge University Press}
}

@book{hofbauer1998evolutionary,
  title={Evolutionary Games and Population Dynamics},
  author={Hofbauer, Josef and Sigmund, Karl},
  year={1998},
  publisher={Cambridge University Press},
  address={Cambridge},
  doi={10.1017/CBO9781139173179},
  isbn={9780521625708}
}

@article{Aguade-Gorgorio_2024,
 title={Alternative cliques of coexisting species in complex ecosystems},
  author={Aguad{\'e}-Gorgori{\'o}, Guim and K{\'e}fi, Sonia},
  journal={Journal of Physics: Complexity},
  volume={5},
  number={2},
  pages={025022},
  year={2024},
  publisher={IOP Publishing}
}

@article{Mira2022,
  title={Estimating microbial population data from optical density},
  author={Mira, Portia and Yeh, Pamela and Hall, Barry G},
  journal={Plos one},
  volume={17},
  number={10},
  pages={e0276040},
  year={2022},
  doi = {10.1371/journal.pone.0276040},
  url = {https://doi.org/10.1371/journal.pone.0276040},
  publisher={Public Library of Science San Francisco, CA USA}
}

@article{dal2021resource,
  title={Resource--diversity relationships in bacterial communities reflect the network structure of microbial metabolism},
  author={Dal Bello, Martina and Lee, Hyunseok and Goyal, Akshit and Gore, Jeff},
  journal={Nature ecology \& evolution},
  volume={5},
  number={10},
  pages={1424--1434},
  year={2021},
  publisher={Nature Publishing Group UK London}
}

@article{goldford2018emergent,
  title={Emergent simplicity in microbial community assembly},
  author={Goldford, Joshua E and Lu, Nanxi and Baji{\'c}, Djordje and Estrela, Sylvie and Tikhonov, Mikhail and Sanchez-Gorostiaga, Alicia and Segr{\`e}, Daniel and Mehta, Pankaj and Sanchez, Alvaro},
  journal={Science},
  volume={361},
  number={6401},
  pages={469--474},
  year={2018},
  publisher={American Association for the Advancement of Science}
}

@article{Hu2022_Science_EmergentPhases,
 title={Emergent phases of ecological diversity and dynamics mapped in microcosms},
  author={Hu, Jiliang and Amor, Daniel R and Barbier, Matthieu and Bunin, Guy and Gore, Jeff},
  journal={Science},
  volume={378},
  number={6615},
  pages={85--89},
  year={2022},
  publisher={American Association for the Advancement of Science}
}

@book{Petraitis,
  title={Multiple Stable States in Natural Ecosystems},
  author={Petraitis, Peter},
  year={2013},
  publisher={Oxford University Press}
}

@article{may1972will,
  title={Will a large complex system be stable?},
  author={May, Robert M},
  journal={Nature},
  volume={238},
  number={5364},
  pages={413--414},
  year={1972},
  publisher={Springer}
}

@article{Staver-2011,
author = {A. Carla Staver  and Sally Archibald  and Simon A. Levin },
title = {The Global Extent and Determinants of Savanna and Forest as Alternative Biome States},
journal = {Science},
volume = {334},
number = {6053},
pages = {230-232},
year = {2011},
doi = {10.1126/science.1210465},
URL = {https://www.science.org/doi/abs/10.1126/science.1210465}}

@book{lotka1956elements,
  title={Elements of mathematical biology},
  author={Lotka, Alfred J},
  year={1956},
  publisher={Dover Publications}
}

@article{gilpin1976multiple,
  title={Multiple domains of attraction in competition communities},
  author={Gilpin, Michael E and Case, Ted J},
  journal={Nature},
  volume={261},
  number={5555},
  pages={40--42},
  year={1976},
  publisher={Nature Publishing Group UK London}
}

@article{macarthur1970species,
  title={Species packing and competitive equilibrium for many species},
  author={MacArthur, Robert},
  journal={Theoretical population biology},
  volume={1},
  number={1},
  pages={1--11},
  year={1970},
  publisher={Elsevier}
}

@article{gatto1990general,
  title={A general minimum principle for competing populations: Some ecological and evolutionary consequences},
  author={Gatto, Marino},
  journal={Theoretical Population Biology},
  volume={37},
  number={3},
  pages={369--388},
  year={1990},
  publisher={Elsevier}
}

@article{Stritih_Seidl_Senf_2023_AlternativeStates,
 title={Alternative states in the structure of mountain forests across the Alps and the role of disturbance and recovery},
  author={Stritih, Ana and Seidl, Rupert and Senf, Cornelius},
  journal={Landscape Ecology},
  volume={38},
  number={4},
  pages={933--947},
  year={2023},
  publisher={Springer}
}

@article{aleman2020floristic,
  title={Floristic evidence for alternative biome states in tropical Africa},
  author={Aleman, Julie C and Fayolle, Adeline and Favier, Charly and Staver, A Carla and Dexter, Kyle G and Ryan, Casey M and Azihou, Akomian Fortun{\'e} and Bauman, David and te Beest, Mariska and Chidumayo, Emmanuel N and others},
  journal={Proceedings of the National Academy of Sciences},
  volume={117},
  number={45},
  pages={28183--28190},
  year={2020},
  publisher={National Academy of Sciences}
}

@article{VanDeGuchte_2020_AltStableIntestinal,
  title={Alternative stable states in the intestinal ecosystem: proof of concept in a rat model and a perspective of therapeutic implications},
  author={Van de Guchte, Maarten and Burz, Sebastian D and Cadiou, Julie and Wu, Jiangbo and Mondot, Stanislas and Blotti{\`e}re, Herv{\'e} M and Dor{\'e}, Jo{\"e}l},
  journal={Microbiome},
  volume={8},
  number={1},
  pages={153},
  year={2020},
  publisher={Springer}
}

@article{LozuponeStombaughGordonJanssonKnight2012_diversity,
  title={Diversity, stability and resilience of the human gut microbiota},
  author={Lozupone, Catherine A and Stombaugh, Jesse I and Gordon, Jeffrey I and Jansson, Janet K and Knight, Rob},
  journal={Nature},
  volume={489},
  number={7415},
  pages={220--230},
  year={2012},
  publisher={Nature Publishing Group UK London}
}

@article{Srinivasan2024_ModelingMicrobialCommunity,
  title={Modeling microbial community networks: methods and tools for studying microbial interactions},
  author={Srinivasan, Shanchana and Jnana, Apoorva and Murali, Thokur Sreepathy},
  journal={Microbial ecology},
  volume={87},
  number={1},
  pages={56},
  year={2024},
  publisher={Springer}
}

@article{Ona2025_disentangling,
  title={Disentangling microbial interaction networks},
  author={O{\~n}a, Leonardo and Shreekar, Shryli K and Kost, Christian},
  journal={Trends in Microbiology},
  year={2025},
  publisher={Elsevier}
}

@article{MarrecBravo-RuisecoZhouGodwin2025_interactions,
  title={Exploring interactions in microbial communities},
  author={Marrec, Lo{\"\i}c and Bravo-Ruiseco, Gabriela and Zhou, Xingjian and Daodu, Adedamola G and Faust, Karoline},
  journal={Current Opinion in Biotechnology},
  volume={96},
  pages={103352},
  year={2025},
  publisher={Elsevier}
}

@article{MEROZ2024102511,
title={Quantifying microbial interactions: concepts, caveats, and applications},
  author={Meroz, Nittay and Livny, Tal and Friedman, Jonathan},
  journal={Current Opinion in Microbiology},
  volume={80},
  pages={102511},
  year={2024},
  publisher={Elsevier}
}

@article{Schmitt2019AlternativeAttractors,
title={Experimental support for alternative attractors on coral reefs},
  author={Schmitt, Russell J and Holbrook, Sally J and Davis, Samantha L and Brooks, Andrew J and Adam, Thomas C},
  journal={Proceedings of the National Academy of Sciences},
  volume={116},
  number={10},
  pages={4372--4381},
  year={2019},
  publisher={National Academy of Sciences}
}

@article{Petraitis2009MultipleStates,
  title={Experimental confirmation of multiple community states in a marine ecosystem},
  author={Petraitis, Peter S and Methratta, Elizabeth T and Rhile, Erika C and Vidargas, Nicholas A and Dudgeon, Steve R},
  journal={Oecologia},
  volume={161},
  number={1},
  pages={139--148},
  year={2009},
  publisher={Springer}
}

@article{Abreu2020AltStableStates,
title={Microbial communities display alternative stable states in a fluctuating environment},
  author={Abreu, Clare I and Andersen Woltz, Vilhelm L and Friedman, Jonathan and Gore, Jeff},
  journal={PLoS computational biology},
  volume={16},
  number={5},
  pages={e1007934},
  year={2020},
  publisher={Public Library of Science San Francisco, CA USA}
}

@article{LopesAmorGore2024_multistability,
  title={Cooperative growth in microbial communities is a driver of multistability},
  author={Lopes, William and Amor, Daniel R and Gore, Jeff},
  journal={Nature Communications},
  volume={15},
  number={1},
  pages={4709},
  year={2024},
  publisher={Nature Publishing Group UK London}
}

@article{Estrela2022FunctionalAttractors,
title={Functional attractors in microbial community assembly},
  author={Estrela, Sylvie and Vila, Jean CC and Lu, Nanxi and Baji{\'c}, Djordje and Rebolleda-G{\'o}mez, Maria and Chang, Chang-Yu and Goldford, Joshua E and Sanchez-Gorostiaga, Alicia and S{\'a}nchez, {\'A}lvaro},
  journal={Cell Systems},
  volume={13},
  number={1},
  pages={29--42},
  year={2022},
  publisher={Elsevier}
}

@article{KearneyEtAl2021_HeterotrophDiversityPicocyanobacteria,
  title={Microbial diversity of co-occurring heterotrophs in cultures of marine picocyanobacteria},
  author={Kearney, Sean M and Thomas, Elaina and Coe, Allison and Chisholm, Sallie W},
  journal={Environmental Microbiome},
  volume={16},
  number={1},
  pages={1},
  year={2021},
  publisher={Springer}
}

@article{TkaczCheemaChandraGrantPoole2015_rhizosphere_stability,
  title={Stability and succession of the rhizosphere microbiota depends upon plant type and soil composition},
  author={Tkacz, Andrzej and Cheema, Jitender and Chandra, Govind and Grant, Alastair and Poole, Philip S},
  journal={The ISME journal},
  volume={9},
  number={11},
  pages={2349--2359},
  year={2015},
  publisher={Oxford University Press}
}

@article{taylor2024structure,
  title={On the structure of multiple stable equilibria in competitive ecological systems},
  author={Taylor, Washington and O’Dwyer, James},
  journal={Theoretical Ecology},
  volume={18},
  number={1},
  pages={1--26},
  year={2025},
  publisher={Springer}
}

@article{goyal2022interactions,
  title={Interactions between strains govern the eco-evolutionary dynamics of microbial communities},
  author={Goyal, Akshit and Bittleston, Leonora S and Leventhal, Gabriel E and Lu, Lu and Cordero, Otto X},
  journal={Elife},
  volume={11},
  pages={e74987},
  year={2022},
  publisher={eLife Sciences Publications Limited}
}

@article{Ros2023_TypicalNumberEquilibria,
  title={Generalized lotka-volterra equations with random, nonreciprocal interactions: The typical number of equilibria},
  author={Ros, Valentina and Roy, Felix and Biroli, Giulio and Bunin, Guy and Turner, Ari M},
  journal={Physical Review Letters},
  volume={130},
  number={25},
  pages={257401},
  year={2023},
  publisher={APS}
}

@article{dubinkina2019multistability,
  title={Multistability and regime shifts in microbial communities explained by competition for essential nutrients},
  author={Dubinkina, Veronika and Fridman, Yulia and Pandey, Parth Pratim and Maslov, Sergei},
  journal={Elife},
  volume={8},
  pages={e49720},
  year={2019},
  publisher={eLife Sciences Publications, Ltd}
}

@article{fried2017alternative,
  title={Alternative steady states in ecological networks},
  author={Fried, Yael and Shnerb, Nadav M and Kessler, David A},
  journal={Physical Review E},
  volume={96},
  number={1},
  pages={012412},
  year={2017},
  publisher={APS}
}

@article{kessler2025interaction,
  title={Interaction network structures in competitive ecosystems},
  author={Kessler, David A and Shnerb, Nadav M},
  journal={Physical Review E},
  volume={111},
  number={3},
  pages={034408},
  year={2025},
  publisher={APS}
}

@article{goyal2018multiple,
  title={Multiple stable states in microbial communities explained by the stable marriage problem},
  author={Goyal, Akshit and Dubinkina, Veronika and Maslov, Sergei},
  journal={The ISME journal},
  volume={12},
  number={12},
  pages={2823--2834},
  year={2018},
  publisher={Oxford University Press}
}

@article{ArnoulxDePirey2024_ManySpeciesEcologicalFluctuations,
  title={Many-species ecological fluctuations as a jump process from the brink of extinction},
  author={Arnoulx de Pirey, Thibaut and Bunin, Guy},
  journal={Physical Review X},
  volume={14},
  number={1},
  pages={011037},
  year={2024},
  publisher={APS}
}

@article{BlumenthalRocksMehta2024_PhaseTransitionChaos,
  author       = {Emmy Blumenthal and Jason W. Rocks and Pankaj Mehta},
  title        = {Phase Transition to Chaos in Complex Ecosystems with Nonreciprocal Species-Resource Interactions},
  journal      = {Physical Review Letters},
  year         = {2024},
  volume       = {132},
  number       = {12},
  pages        = {127401},
  doi          = {10.1103/PhysRevLett.132.127401},
  url          = {https://doi.org/10.1103/PhysRevLett.132.127401}
}

@article{MallminTraulsenDeMonte2024_ChaoticTurnoverRareAbundant,
  title={Chaotic turnover of rare and abundant species in a strongly interacting model community},
  author={Mallmin, Emil and Traulsen, Arne and De Monte, Silvia},
  journal={Proceedings of the National Academy of Sciences},
  volume={121},
  number={11},
  pages={e2312822121},
  year={2024},
  publisher={National Academy of Sciences}
}

@article{liu2025complex,
  title={Complex ecosystems lose stability when resource consumption is out of niche},
  author={Liu, Yizhou and Hu, Jiliang and Lee, Hyunseok and Gore, Jeff},
  journal={Physical Review X},
  volume={15},
  number={1},
  pages={011003},
  year={2025},
  publisher={APS}
}

@article{pearce2020stabilization,
  title={Stabilization of extensive fine-scale diversity by ecologically driven spatiotemporal chaos},
  author={Pearce, Michael T and Agarwala, Atish and Fisher, Daniel S},
  journal={Proceedings of the National Academy of Sciences},
  volume={117},
  number={25},
  pages={14572--14583},
  year={2020},
  publisher={National Academy of Sciences}
}

@article{mahadevan2023spatiotemporal,
  title={Spatiotemporal ecological chaos enables gradual evolutionary diversification without niches or tradeoffs},
  author={Mahadevan, Aditya and Pearce, Michael T and Fisher, Daniel S},
  journal={Elife},
  volume={12},
  pages={e82734},
  year={2023},
  publisher={eLife Sciences Publications Limited}
}

@article{Bunin2017_EcoComLv,
  author       = {Guy Bunin},
  title        = {Ecological communities with Lotka--Volterra dynamics},
  journal      = {Physical Review E},
  year         = {2017},
  volume       = {95},
  number       = {4},
  pages        = {042414},
  doi          = {10.1103/PhysRevE.95.042414},
  url          = {https://doi.org/10.1103/PhysRevE.95.042414}
}

@article{GiralMartinez2024_stabilization,
  title={Stabilization of macroscopic dynamics by fine-grained disorder in many-species ecosystems},
  author={Mart{\'\i}nez, Juan Giral and De Monte, Silvia and Barbier, Matthieu},
  journal={arXiv preprint arXiv:2411.14969},
  year={2024}
}

@article{AltieriRoyCammarotaBiroli2021_GlassyEquilibria,
  author       = {Ada Altieri and Felix Roy and Chiara Cammarota and Giulio Biroli},
  title        = {Properties of Equilibria and Glassy Phases of the Random Lotka-Volterra Model with Demographic Noise},
  journal      = {Physical Review Letters},
  year         = {2021},
  volume       = {126},
  number       = {25},
  pages        = {258301},
  doi          = {10.1103/PhysRevLett.126.258301},
  url          = {https://doi.org/10.1103/PhysRevLett.126.258301}
}

@article{TikhonovMonasson2017PRL,
  title={Collective phase in resource competition in a highly diverse ecosystem},
  author={Tikhonov, Mikhail and Monasson, Remi},
  journal={Physical review letters},
  volume={118},
  number={4},
  pages={048103},
  year={2017},
  publisher={APS}
}

@article{GarciaLorenzana2022_PRE_stronglyCompetitiveRLV,
  title={Well-mixed Lotka-Volterra model with random strongly competitive interactions},
  author={Garcia Lorenzana, Giulia and Altieri, Ada},
  journal={Physical Review E},
  volume={105},
  number={2},
  pages={024307},
  year={2022},
  publisher={APS}
}

@article{CuiMarslandMehta2021_PRE_TypicalRandomCRM,
title = {Diverse communities behave like typical random ecosystems},
author = {Cui, Wenping and Marsland, Robert and Mehta, Pankaj},
journal = {Physical Review E},
volume = {104},
number = {3},
pages = {034416},
year = {2021},
doi = {10.1103/PhysRevE.104.034416},
url = {https://doi.org/10.1103/PhysRevE.104.034416},
publisher = {American Physical Society}
}

@article{CuiMarslandMehta2024_LesHouchesCommunityEcology,
  title={Les houches lectures on community ecology: From niche theory to statistical mechanics},
  author={Cui, Wenping and Marsland III, Robert and Mehta, Pankaj},
  journal={ArXiv},
  pages={arXiv--2403},
  year={2024}
}

@article{MarslandMehta2020_PRL_SpeciesPacking,
  title={Effect of resource dynamics on species packing in diverse ecosystems},
  author={Cui, Wenping and Marsland III, Robert and Mehta, Pankaj},
  journal={Physical review letters},
  volume={125},
  number={4},
  pages={048101},
  year={2020},
  publisher={APS}
}

@article{AltieriDeMonte2025_EPL_StatPhysCommunities,
author = {Altieri, Ada and De Monte, Silvia},
title = {Statistical physics approaches to ecological communities},
journal = {EPL (Europhysics Letters)},
year = {2025},
volume = {150},
number = {5},
pages = {51002},
doi = {10.1209/0295-5075/addae2},
url = {https://iopscience.iop.org/article/10.1209/0295-5075/addae2}
}

@article{Ke2024Taxonomy,
  title={A taxonomy of multiple stable states in complex ecological communities},
  author={Aguad{\'e}-Gorgori{\'o}, Guim and Arnoldi, Jean-fran{\c{c}}ois and Barbier, Matthieu and K{\'e}fi, Sonia},
  journal={Ecology Letters},
  volume={27},
  number={4},
  pages={e14413},
  year={2024},
  publisher={Wiley Online Library}
}

@article{Marsland2019EnergyFluxes,
  title={Available energy fluxes drive a transition in the diversity, stability, and functional structure of microbial communities},
  author={Marsland III, Robert and Cui, Wenping and Goldford, Joshua and Sanchez, Alvaro and Korolev, Kirill and Mehta, Pankaj},
  journal={PLoS computational biology},
  volume={15},
  number={2},
  pages={e1006793},
  year={2019},
  publisher={Public Library of Science, San Francisco, CA USA}
}

@article{Advani2017Environmental,
  author = {Madhu Advani and Guy Bunin and Pankaj Mehta},
  title = {Environmental engineering is an emergent feature of diverse ecosystems and drives community structure},
  journal = {arXiv},
  volume = {1707.03957},
  year = {2017},
  url = {https://arxiv.org/abs/1707.03957}
}

@article{allesina2015predicting,
  title={Predicting the stability of large structured food webs},
  author={Allesina, Stefano and Grilli, Jacopo and Barab{\'a}s, Gy{\"o}rgy and Tang, Si and Aljadeff, Johnatan and Maritan, Amos},
  journal={Nature communications},
  volume={6},
  number={1},
  pages={7842},
  year={2015},
  publisher={Nature Publishing Group UK London}
}

@article{allesina2012stability,
  title={Stability criteria for complex ecosystems},
  author={Allesina, Stefano and Tang, Si},
  journal={Nature},
  volume={483},
  number={7388},
  pages={205--208},
  year={2012},
  publisher={Nature Publishing Group UK London}
}

@article{Scheffer2007Regime,
  author = {Marten Scheffer and Erik Jeppesen},
  title = {Regime Shifts in Shallow Lakes},
  journal = {Ecosystems},
  volume = {10},
  number = {1},
  pages = {1--3},
  year = {2007},
  doi = {10.1007/s10021-006-9002-y},
  url = {https://doi.org/10.1007/s10021-006-9002-y}
}

@article{zou2024positive,
  title={Positive feedbacks and alternative stable states in forest leaf types},
  author={Zou, Yibiao and Zohner, Constantin M and Averill, Colin and Ma, Haozhi and Merder, Julian and Berdugo, Miguel and Bialic-Murphy, Lalasia and Mo, Lidong and Brun, Philipp and Zimmermann, Niklaus E and others},
  journal={Nature Communications},
  volume={15},
  number={1},
  pages={4658},
  year={2024},
  publisher={Nature Publishing Group},
  doi={10.1038/s41467-024-48676-5},
  url={https://www.nature.com/articles/s41467-024-48676-5}
}

@article{SANCHEZPINILLOS2024110409,
title = {Ecological dynamic regimes: A key concept for assessing ecological resilience},
journal = {Biological Conservation},
volume = {289},
pages = {110409},
year = {2024},
issn = {0006-3207},
doi = {https://doi.org/10.1016/j.biocon.2023.110409},
url = {https://www.sciencedirect.com/science/article/pii/S0006320723005104},
author = {Martina Sánchez-Pinillos and Vasilis Dakos and Sonia Kéfi}
}

\appendix
\captionsetup{font=normalsize}

\onecolumngrid

\newpage

\section*{Supplementary Information}
\renewcommand{\thefigure}{S\arabic{figure}}
\setcounter{figure}{0} 

\section{Computing the likelihoods and biomass of stable states}
\label{app:methods}
{\color{black}
In this appendix we describe our algorithm to compute likelihoods and biomasses for different uninvasible stable states corresponding to a chosen interaction matrix $A$. We first generate a random interaction matrix $A$ with mean $\mu$ and
standard deviation $\sigma$. Note that in most of the results, we use a symmetric interaction matrix, $A_{ij} = A_{ji}$, but relax this assumption in  Fig.~\ref{fig:assymetry}.  
 Symmetric interaction matrices yield a Lyapunov function for the dynamics in the GLV model~\cite{macarthur1970species,gatto1990general}, given by

\begin{equation}
   H[\mathbf N] = -\sum_i r_i N_i + \frac{1}{2}\sum_i \frac{r_i}{K_i}N_i^2 + \frac{1}{2}\sum_{i\neq j} A_{ij}N_iN_j.
\end{equation}

Here $N_i$ is the abundance of species $i$,
$r_i$ is its intrinsic growth rate,
$K_i$ is its carrying capacity and $A_{ij}$ represents the strength of
the interaction between species $i$ and $j$. The existence of this
Lyapunov function guarantees that at long time, the dynamics always
flow to a stable state, rather than having limit cycles or
chaos~\cite{hofbauer1998evolutionary}; this makes it very tractable to
study multiple stable states. In this setting, each stable state is a
local minima of this function, and this allows us to use the Lyapunov
function to envision a landscape of stable
states~\cite{AltieriRoyCammarotaBiroli2021_GlassyEquilibria}.  We also
note that in what follows, we will not attempt to find every stable
state corresponding to a given matrix $A$.
This is because there are a
substantially large number of states with exceedingly small basins of
attraction, and
while it is  in many cases
possible to  identify these states (see Appendix~\ref{app:overlappingStates}), statistically determining their likelihoods
requires a computationally infeasible
number of simulations when $A$ is a large matrix (in our case,
typically $100\times100$). We typically use $10^6$ initial conditions,
while we estimate that  
 computing likelihoods for all states might require roughly
$10^{10}$ initial conditions (see
Appendix~\ref{app:overlappingStates}). Thus we  
 analyze probabilities for a large but not
 complete subset of all the stable states; for example, we 
 encounter
 55
 states in our simulations for the matrix in Fig.~\ref{fig:fig1}c while
  the total number of states is around 80 (Fig.~\ref{fig:block-number-matching}).}

To find stable states, we start by initializing all species with abundances
sampled from a uniform distribution between $0$ and $1$. We then solve
the GLV equations using the \texttt{Radau} method, which has adaptive
time steps. After running the simulation for sufficient time until
$t=300$, we check whether we have reached a stable, uninvasible steady
state in the following ways:

\textit{Identifying surviving species.}---To determine steady state convergence, we first identify the surviving species.  In our numerical simulations, species abundances never reach exactly zero, and species that are destined to go extinct may nevertheless take a long time to do so. To ascertain which species are on their way to extinction, we scan over extinction threshold abundances for surviving species over a wide range between $10^{-5}$ and $10^{-1}$. \textcolor{black}{For most initial conditions, a threshold of $10^{-5}$ successfully identifies the surviving species.} We begin by checking for feasibility. We start with a threshold of $10^{-5}$. We then classify all species with abundances above this threshold as surviving species. From the surviving species, we construct a reduced interaction matrix $A_{\text{reduced}}$ that contains only the surviving species \textcolor{black}{; note that the diagonal elements of this matrix are $A_{ii} = r_i/K_i$}. We then compute the steady-state abundances using the fixed point conditions of the GLV model in Eq.~\eqref{eq:GLVEq} as:
\textcolor{black}{
\begin{equation}
    \vec{N}^* = A_{\text{reduced}}^{-1} \vec{r}_m,
    \label{eq:nStar}
\end{equation}
where $m$ represents the number of surviving species and $\vec{r}_m$ is the $m$-dimensional vector of growth rates of surviving species. For a given set of surviving species, the GLV equations have a unique fixed point, as given by Eq.~\eqref{eq:nStar}.}

\textit{Feasibility verification.}---If computing $\vec{N}^*$  yields
negative abundances for any species, then the set of surviving species
is not feasible. If the set is feasible, then we ascertain if the
abundances of survivors from simulations and Eq.~\eqref{eq:nStar} are
close enough (within numerical error). If we find that our guess of
the surviving species is either infeasible or outside numerical error,
then we assume that our extinction threshold was incorrect. In this
case, we increase our threshold by a factor of 10 and repeat our
procedure to identify surviving species. We do so until we reach a
threshold abundance of $10^{-1}$. If we still fail to find a feasible
steady state whose predicted abundances match our simulations, then we
assume that we have not yet reached steady-state. In this case, we
continue the simulation for additional time $t=1000$ from this point,
and repeat the process of identifying survivors after this additional
time using the same procedure above. Once we find a feasible set of
survivors with their steady-state abundances, we check their
stability. We discard the small minority of simulations (approximately~0.1\%) that fail to converge to a uninvasible steady state by $t = 50300$.

\begin{figure}[ht!]
    \centering
    \includegraphics[width=0.7\textwidth]{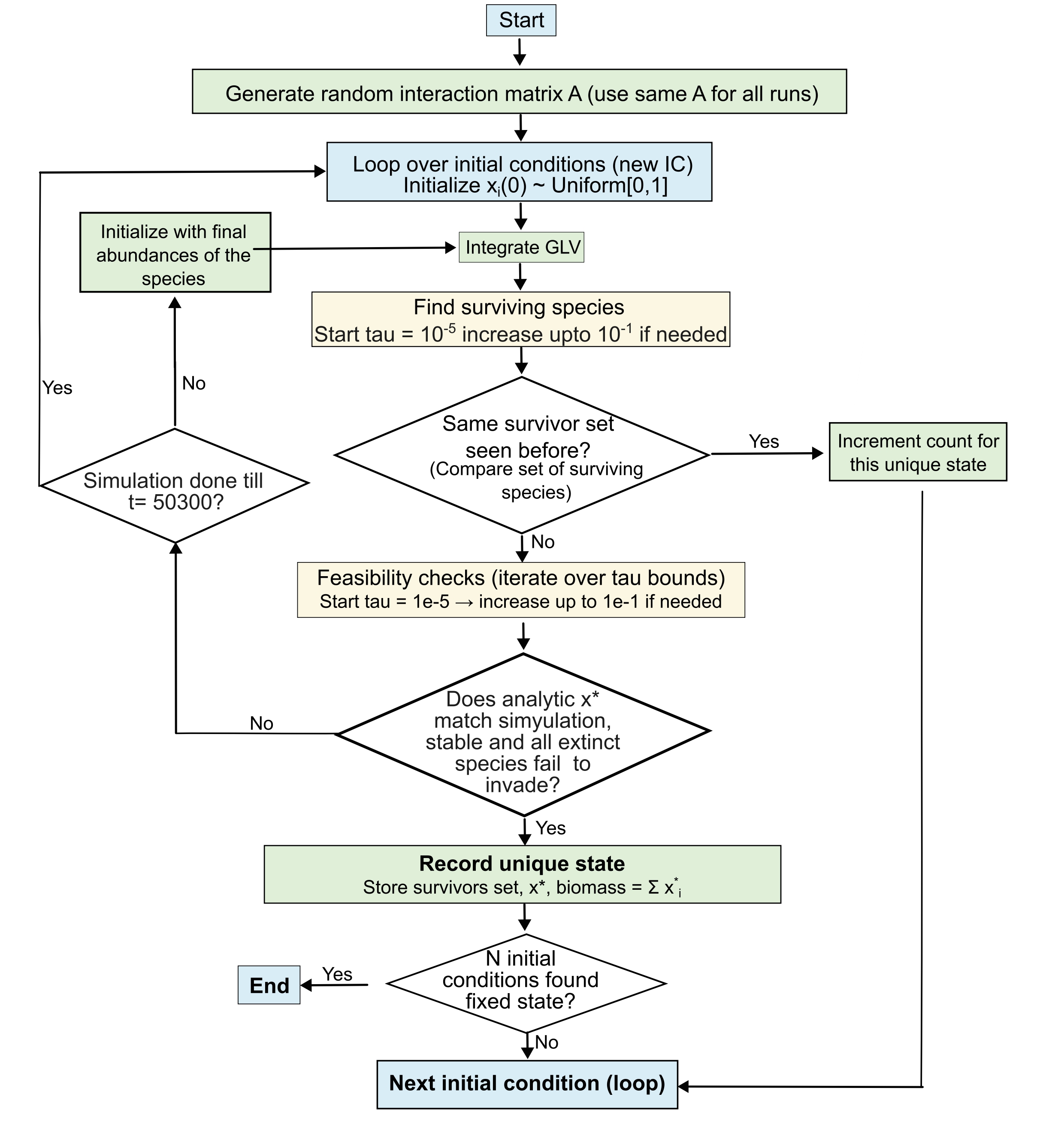} 
    \caption{\justifying Flowchart of our algorithm to identify feasible, stable and uninvadable states for a given GLV interaction matrix in our model. Briefly, we choose a random initial condition, numerically integrate the dynamics, and detect surviving species using an self-consistently determined extinction threshold. We then finally ensure the resulting state is feasible and stable both to perturbations in surviving species abundances and to invasions of extinct species.}
    \label{fig:flow}
\end{figure}

\textit{Stability verification.}---To check stability, we calculate the Jacobian matrix at the steady state:
\textcolor{black}{
\begin{equation}
    J  = -D(\vec{N}^*) A_{\text{reduced}}.
\end{equation}
}
If the largest real part of all eigenvalues of $J$ is negative, then the steady state is linearly stable to perturbations in abundances. We ensure that all steady states we obtain pass this test.

\textit{Uninvasibility verification.}---We next check whether the stable steady state is uninvasible. Uninvasibility refers to the idea that extinct species should not be able to reinvade the community when introduced in small numbers. If $m$ is the number of surviving species so that $S-m$ species are extinct, then the state is uninvasible if all extinct species have negative invasion fitness, given by their per capita growth rate at low abundance:
\textcolor{black}{
\begin{equation}
    r_i - \sum_{j\in \text{survivors}} A_{ij} N_j^* < 0 \ \text{for all extinct species with index $i$}
\end{equation}
}
Once a steady-state passes even this criterion, it is certain to be a feasible, stable and uninvasible steady-state for the model with the given interaction matrix $A$. We add this state to our list of stable states.

\textit{Cataloging unique states.}---We then start from a new randomly chosen initial condition, and repeat this entire process. After ascertaining which state this initial condition arrives at, we check whether the state is a new, unique state that we had not identified before, or if it matches an existing state. We track the number of random initial conditions that converge to each of these states. After repeating these simulations for $10^6$ random initial conditions, we collect all the unique stable states and the fraction of initial conditions that converge to them. Note that for convenience, we use $10^6$ initial conditions in the main text, but for several SI figures, we use $10^5$ initial conditions instead.

\textit{Computing likelihoods and biomass.}---We define the likelihood of each unique stable state as:
\begin{equation}
    \frac{\text{number of simulations ending in that state }}{\text{total number of simulations}}.
\end{equation}
We calculate the biomass of each state as the sum of all steady-state species abundances for that state, i.e., $\sum_{i=1}^S N_i^*$.

\textcolor{black}{Using the Lyapunov function,  
the likelihood of a stable state is related to the  
size
of its basin of attraction, while the biomass at steady-state is related to its depth~\cite{hofbauer1998evolutionary,taylor2024structure}. However, this Lyapunov function does not guarantee any clear relation between the  
 size in abundance space
 and the depth of attractor basins. In this work, we find a robust positive biomass--likelihood relationship, showing that the  
 size and depth of basins may indeed be related
 in the GLV model.}

\FloatBarrier

{\color{black}
\section{Disorder averaging of the biomass--likelihood relationship}
\label{app:disorderAveraging}

In the main text, we show that for a single random interaction matrix $A$, the likelihoods $p$ of different stable states increase with biomass $B$ and are well-described by our block-model. However, for any single realization of $A$, we observe error due to disorder in the interaction matrices, which are drawn randomly. To isolate the average biomass-likelihood relationship, we perform a quenched disorder average over many independent matrices drawn with the same $(\mu, \sigma)$. Here we use $N_I=110$ independent matrices.

\textit{Averaging protocol.}---For each disordered matrix $A^{(k)}$ ($k = 1, \ldots, N_I$), we bin the stable states by biomass $B$ and compute the mean likelihood $\langle p \mid B \in b, A^{(k)} \rangle$ within each bin $b$ (see Appendix~\ref{app:simulationdetails} for details of binning procedure). The disorder-averaged conditional likelihood is then
\begin{equation}
\overline{\langle p \mid B \in b \rangle}
= \frac{1}{N_I} \sum_{k=1}^{N_I} \langle p \mid B \in b, A^{(k)} \rangle,
\end{equation}
where
\begin{equation}
\langle p \mid B \in b, A^{(k)} \rangle
= \frac{1}{|\mathcal{J}_{k,b}|} \sum_{j:\, B_{kj} \in b} p_j^{(k)}.
\end{equation}
Here, $|\mathcal{J}_{k,b}|$ is the observed number of states for matrix $A^{(k)}$ whose biomass falls in bin $b$, and $j$ indexes the specific stable states with biomass $B_{kj}$ observed for matrix $k$. An alternative is to pool all data points across interaction matrix realizations and compute a single conditional average:
\begin{equation}
\langle p \mid B \in b \rangle_{\mathrm{pooled}}
= \frac{1}{\sum_{k=1}^{N_I} |\mathcal{J}_{k,b}|}
\sum_{k=1}^{N_I} \sum_{j:\, B_{kj} \in b} p_j^{(k)}.
\end{equation}
We find that both methods produce very similar results when the number of states per bin does not vary strongly across matrix interaction matrix realizations (Fig.~\ref{fig:averaging_methods}), so we use the first method throughout.

\begin{figure}[H]
    \centering
    \begin{subfigure}[b]{0.4\textwidth}
        \centering
        \includegraphics[width=\textwidth]{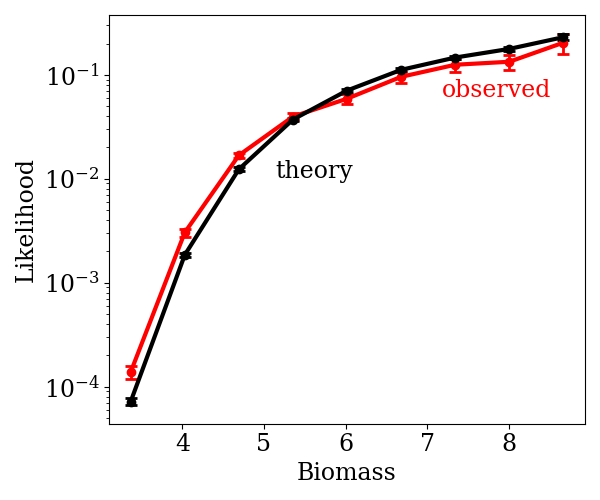}
        \caption{}
        \label{fig:averaging_individual}
    \end{subfigure}
    \hfill
    \begin{subfigure}[b]{0.4\textwidth}
        \centering
        \includegraphics[width=\textwidth]{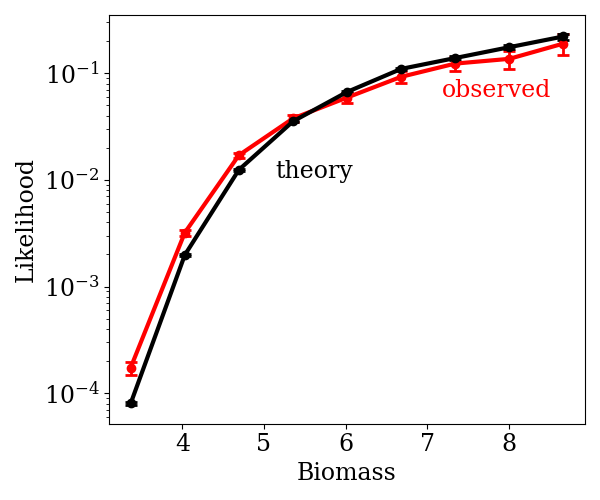}
        \caption{}
        \label{fig:averaging_pooled}
    \end{subfigure}
    \caption{\justifying\textcolor{black}{Comparison of averaging methods: (a) individual averaging, where the mean likelihood is first computed within each interaction matrix realization $A^{(k)}$ for a given biomass bin and then averaged over matrices, and (b) pooled averaging, where all data points in each biomass bin are combined across interaction matrix realizations before averaging. Both methods produce the same mean trend and agree with the theoretical prediction, confirming that the result does not depend on the order of averaging. The error bars represent the standard error of the mean (SEM).}}
    \label{fig:averaging_methods}
\end{figure}

\textit{Arithmetic vs.\ geometric mean.}---A subtlety arises in comparing disorder-averaged data with theory. Our block-model prediction yields the likelihood $p$ directly, not $\log p$. Since the scatter around the predicted value is approximately symmetric in $p$ (not in $\log p$), the appropriate comparison is between the theory and $\log \langle p \rangle$, the logarithm of the arithmetic mean. Comparing instead with $\langle \log p \rangle$ (the geometric mean) would overweight rare, low-likelihood states. In general, $\log \langle p \rangle \neq \langle \log p \rangle$; we use the arithmetic mean throughout this appendix.

\textit{Comparison with theory.}---Fig.~\ref{fig:disorder_theory}(a) shows the disorder-averaged likelihood $\langle p \rangle$ in each biomass bin, computed from $N_I = 110$ independent matrices with $\mu = 0.5$, $\sigma = 0.3$, compared with the block-model prediction from Eq.~\eqref{eq:block_likelihood} in the main text (using the mean block size as discussed in Appendix~\ref{app:likelihoodCalcRandMatrix}). Error bars indicate the standard error of the mean, $\mathrm{SEM}_\beta = \sigma_\beta / \sqrt{n_\beta}$, where $\sigma_\beta$ is the standard deviation of likelihoods within biomass bin $\beta$ and $n_\beta$ is the number of contributing data points across all interaction matrix realizations. The averaged data closely follow the theoretical curve, with only minor deviations at the lowest biomass bins where states are rare and sampling is limited.

To quantify prediction error, we compute the root-mean-square-log error ($\mathrm{RMSE}_{\log}$) between our theoretical prediction and observations from simulations across biomass bins:

\begin{equation}
\mathrm{RMSE}_{\log} = \sqrt{\frac{1}{N_b} \sum_{\beta=1}^{N_b} \left( \log_{10} p_\beta^{\mathrm{obs}} - \log_{10} p_\beta^{\mathrm{theory}} \right)^2},
\label{eq:rmse_log}
\end{equation}
where $N_b$ is the number of biomass bins and the index $\beta$  labels bins.Here, $p_\beta^{\mathrm{obs}}$ is the average likelihood obtained from simulations, while $p_\beta^{\mathrm{theory}}$ is the mean calculated likelihood across all states within that bin. This is sensitive to errors at low likelihoods. Since state likelihoods vary over orders of magnitude, one might prefer this as a metric of prediction error. We find that this error decrease monotonically as we sample more interaction matrices, i.e., as $N_I$ increases (Fig.~\ref{fig:disorder_theory}(c)). This confirms that our theory captures the average biomass-likelihood  relationship rather accurately, and any scatter in likelihood estimates for a specific realization of an interaction matrix is likely a reflection of quenched disorder fluctuations. 


\begin{figure}[ht!]
\centering
\includegraphics[width=1\linewidth]{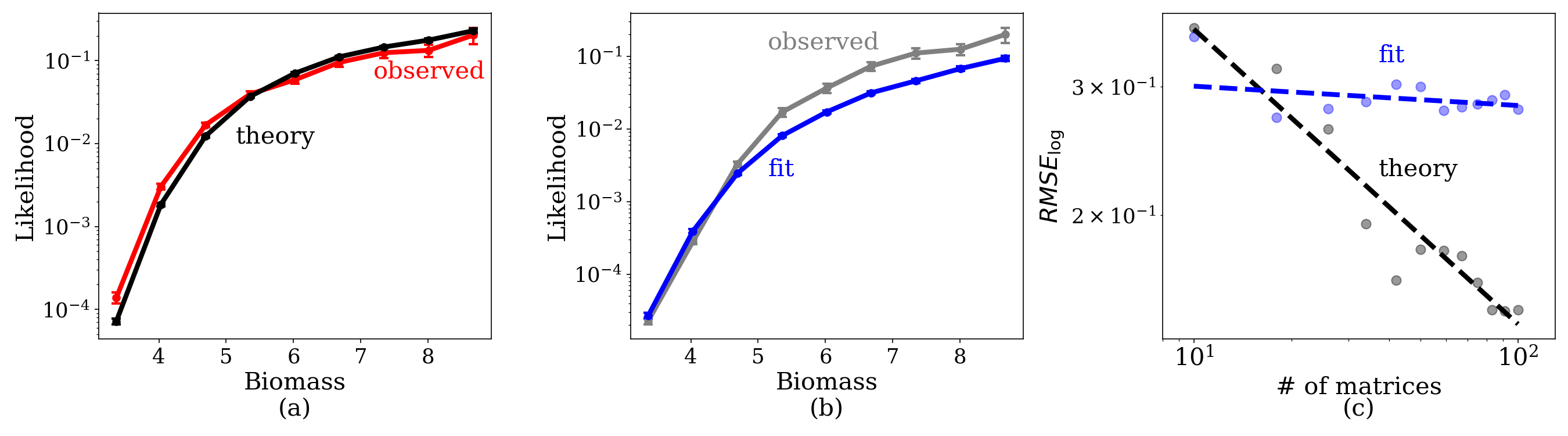}

\caption{\justifying\textcolor{black}{Our theory captures the biomass--likelihood relationship better than the log-hyperbolic fit. (a) The predicted block-model prediction (black) matches the disorder-averaged likelihood $\langle p \rangle$ (red)
    across biomass bins, with error bars showing the standard error of the mean. The theoretical 
    curve captures the observed trend with only minor deviations at low-biomass.
    (b) The disorder-averaged log-hyperbolic fitted curve (blue) broadly follows the disorder-averaged 
    log-likelihood $\log \langle p \rangle$ (gray) across biomass bins, with error bars showing 
    the standard error of the mean. The theoretical curve reproduces the overall trend, though 
    some deviations are observed at high-biomass. (c) The $\mathrm{RMSE_{\log}}$ as a function of number of matrices shows that block-model prediction (black dots) captures the trend more compared to the log-hyperbolic fit (blue dots), with error decreasing systematically faster as the number of matrices increases. The dashed lines are power-law fits to the data points, with exponent $-0.4$ (blue) and $-0.03$ (red).  The disordered 
    average is for the $110$ matrices used in Fig.~\ref{fig:fig1}c.
}}
\label{fig:disorder_theory}
\end{figure}

\textit{Comparison with log-hyperbolic fit.}---We can perform a similar disorder average for the log-hyperbolic fit $\log p \propto (B_c - B)^{-1}$. Since this fit is performed in log-likelihood space, the appropriate comparison is between the averaged fit and $\langle \log p \rangle$. We find that the fit decently approximates the data at low likelihoods but deviates systematically for states with high likelihoods (Fig.~\ref{fig:disorder_theory}(b)). Furthermore, in stark contrast with the prediction from our theory, the RMSE for the log-hyperbolic fit decreases very slowly compared to the theory, showing
that
our theory continues to predict better with increase in number of matrix realizations (Fig.~\ref{fig:disorder_theory}(c)). This shows that the log-hyperbolic fit, while a simplified phenomenological approximation, does not capture the true biomass-likelihood relationship nearly as accurately as our theory. 

\textit{Visualizing the trend across many matrices.}---When all data points from $N_I = 110$ matrices are plotted together, the large spread of biomass values across interaction matrix realizations makes it difficult to discern the underlying trend by eye (Fig.~\ref{fig:disorder_raw}a). To make the trend visible, we plot a kernel density estimate (KDE) of the likelihood within each biomass bin, normalizing the distribution separately in each bin. This visualization clearly reveals the predicted biomass--likelihood relationship even in the raw multi-matrix data (Fig.~\ref{fig:disorder_raw}b).

\begin{figure}[h]
    \centering
    \begin{subfigure}[b]{0.45\textwidth}
        \centering
        \includegraphics[width=\textwidth]{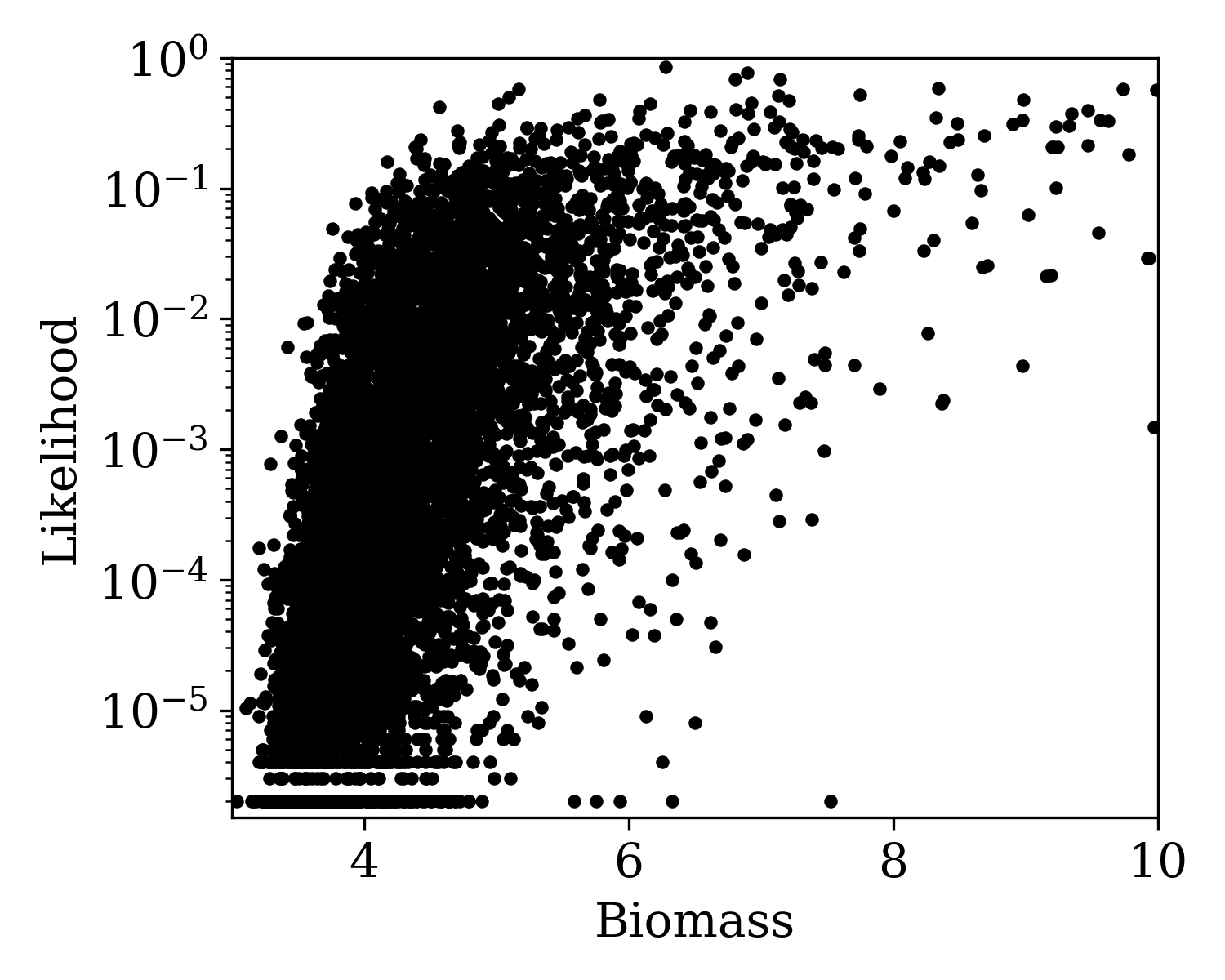}
        \caption{}
        \label{fig:disorder_raw_a}
    \end{subfigure}
    \hfill
    \begin{subfigure}[b]{0.45\textwidth}
        \centering
        \includegraphics[width=\textwidth]{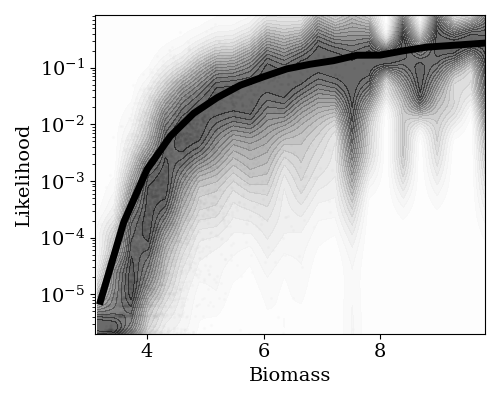}
        \caption{}
        \label{fig:disorder_raw_b}
    \end{subfigure}
    \caption{\justifying\textcolor{black}{(a) Likelihood versus biomass for all stable states from $N_I = 110$ independent matrices at $\mu = 0.5$, $\sigma = 0.3$. The large spread of biomass values across realizations obscures the underlying trend. (b) KDE of the likelihood in each biomass bin, with each bin normalized separately. The predicted biomass--likelihood trend is clearly visible in this representation. The black curve shows our theoretical prediction, also shown in \ref{fig:fig3}(a)}}
    \label{fig:disorder_raw}
\end{figure}

} 

\FloatBarrier

\section{Robustness of biomass--likelihood relationship}
\label{app:robustness}
In the main text, all our results have been shown for symmetric
interaction matrices, \textcolor{black}{fixed growth rates $r_i = 1$ and carrying capacities $K_i=1$, number of species in the pool
$S=100$} and uniform, random initial conditions between 0 and 1. \textcolor{black}{ Further, we also assume deterministic dynamics with no demographic noise.} Here we
systematically relax these assumption and show that our central
result---namely the biomass-likelihood relationsip---is \textcolor{black}{qualitatively
robust to including interaction asymmetry, allowing hetereogeneity in growth rates and
carrying capacities, modifying the number of species in the pool, sampling initial species
abundances from a truncated Gaussian, and including demographic noise in the dynamics. Towards the end of this Appendix, we also discuss the reliability and uncertainty in our likelihood estimates for different stable states.}

\begin{figure}[ht!]
    \centering
    \begin{subfigure}[b]{0.45\textwidth}
        \centering
        \includegraphics[width=\textwidth]{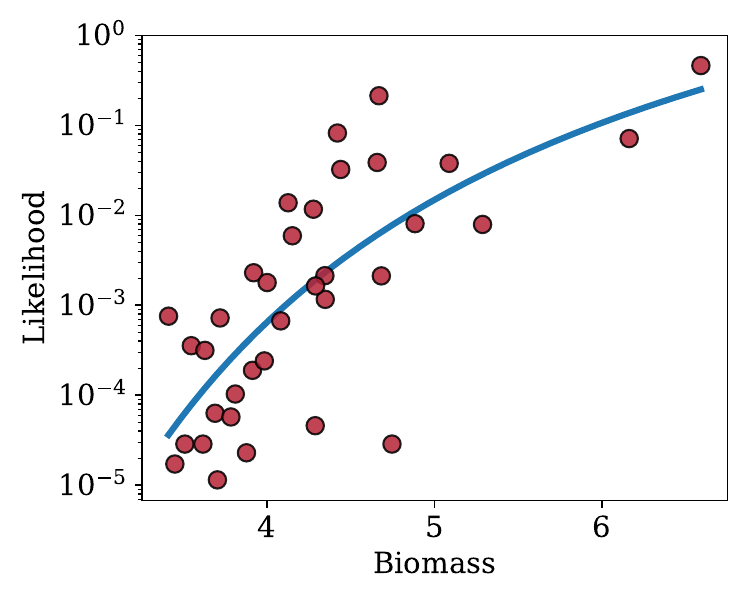}
        \caption{Correlation $\rho = 0.9$}
        \label{fig:rho09}
    \end{subfigure}
    \hfill
    \begin{subfigure}[b]{0.45\textwidth}
        \centering
        \includegraphics[width=\textwidth]{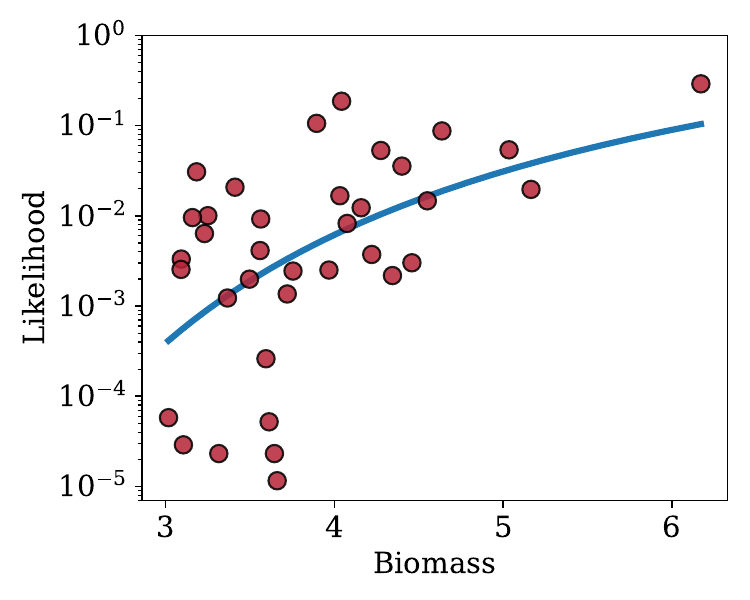}
        \caption{Correlation $\rho = 0.8$}
        \label{fig:rho08}
    \end{subfigure}
    \caption{\justifying Biomass--likelihood distributions for asymmetric interaction matrices. (a) Correlation $\rho=0.9$, (b) correlation $\rho=0.8$. In both cases a positive correlation between likelihood and biomass is observed, although the strength decreases as $\rho$ becomes smaller. Curves indicate fits to the log-hyperbolic relation $\log\left(p\right) \propto (B_c - B)^{-1}$.}
    \label{fig:assymetry}
\end{figure}

First, we discuss interaction asymmetry. We repeat our analysis in
Fig.~\ref{fig:fig1}c with a slightly asymmetric matrix, i.e., with
across-diagonal correlation
$\text{corr}(A_{ij},A_{ji})=\{0.9,0.8\}$. Just as in symmetric
matrices, in these cases we also observe that states with higher
biomass are much more likely (Fig.~\ref{fig:assymetry}a--b).

In general, we expect that as long as asymmetries are not so big that
the qualitative nature of the dynamics changes, such as by introducing
limit cycles or chaos, asymmetries will be either neutral or enhance
the tendency to favor states  with lower self-inhibition.  For
example, in the block models studied in Appendix~\ref{app:blockCalc},
introducing an asymmetry in a matrix entry between blocks will favor
one block over the other, but will not affect the
biomass/self-inhibition of either block, which is determined solely by
the in-block matrix elements, so such a change is neutral to the
probability/biomass correlation.  On the other hand, introducing a
slight asymmetry of order $\epsilon$ in a given block has the effect of slightly
increasing (by order $\epsilon^2$) the self-inhibition of the block
and correspondingly decreasing the biomass.
An analysis similar to that of the next Appendix shows that this
correlates with a slight deviation of the separatrix from other blocks
that disfavors the block with the added asymmetry; thus, in such a
case the asymmetry  causes same-sign changes in biomass and
probability, maintaining the correlation between these quantities.

\begin{figure}[h]
    \centering
    \begin{subfigure}[b]{0.4\textwidth}
        \centering
        \includegraphics[width=\textwidth]{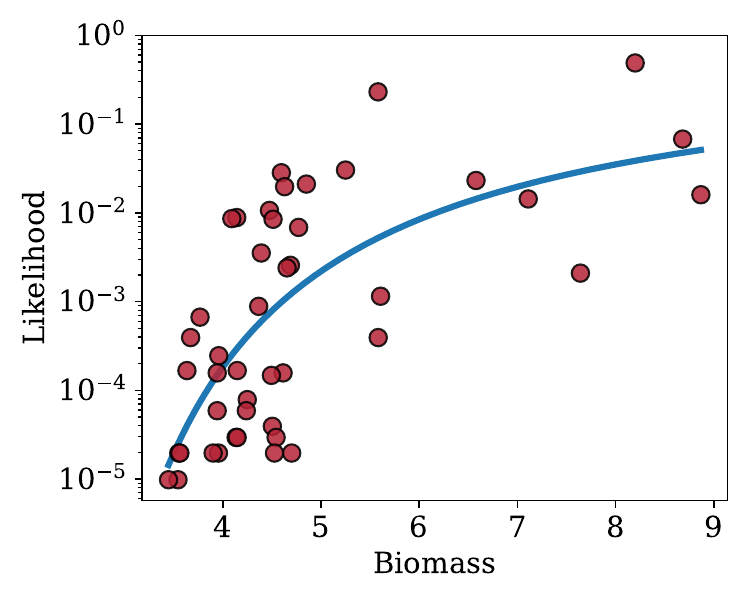}
        \caption{$\epsilon_k = 0.01$.}
        \label{subfig:ki-0.01}
    \end{subfigure}
    \hfill
    \begin{subfigure}[b]{0.4\textwidth}
        \centering
        \includegraphics[width=\textwidth]{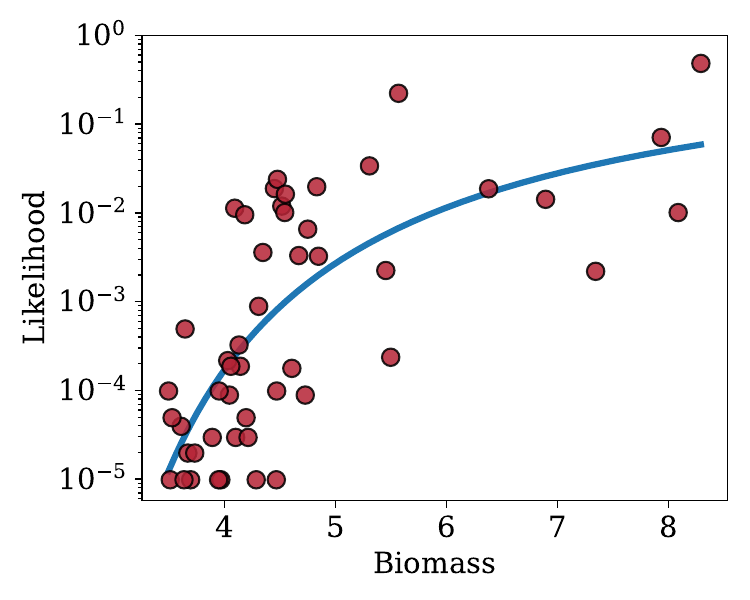}
        \caption{$\epsilon_k = 0.05$. }
        \label{subfig:ki-0.05}
    \end{subfigure}
    \caption{\justifying \textcolor{black}{Biomass--likelihood relation when carrying capacity $K_i$ varies around 1 with small variation $\epsilon_k$. 
    In both cases, the positive correlation of likelihood biomass relation remains. Curves indicate fits to the log-hyperbolic relation $\log\left(p\right) \propto (B_c - B)^{-1}$.}}
    \label{fig:variation_k}
\end{figure}

\textcolor{black}{Next, instead of fixed carrying capacity $K_i = 1$ for all species, we introduce species-specific heterogeneity in $K_i$  by sampling them from a uniform distribution on $[1-\epsilon_k,1+\epsilon_k]$. Note that this change affects the self-inhibition $A_{ii} = r_i/K_i$ of each species. Thus, now both the self-inhibition $A_{ii}$ and interspecies inhibition $A_{ij}$ are disordered in the interaction matrix. Here $\epsilon_k$ is the measure of variation in the carrying capacities. In the main text, we study the case $\epsilon_k=0$. We observe that for $\epsilon_k=\{0.01, 0.05\}$, the biomass--likelihood relationship still shows a positive correlation and is qualitatively robust to species-specific $K_i$ (Fig.~\ref{fig:variation_k}).}

\begin{figure}[h]
    \centering
    \begin{subfigure}[b]{0.4\textwidth}
        \centering
        \includegraphics[width=\textwidth]{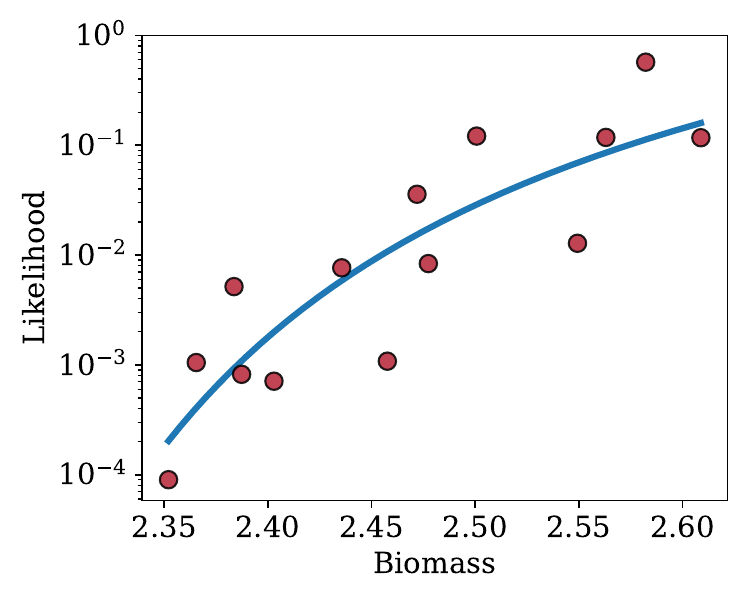}
        \caption{$S = 50$.}
        \label{subfig:S-50}
    \end{subfigure}
    \hfill
    \begin{subfigure}[b]{0.4\textwidth}
        \centering
        \includegraphics[width=\textwidth]{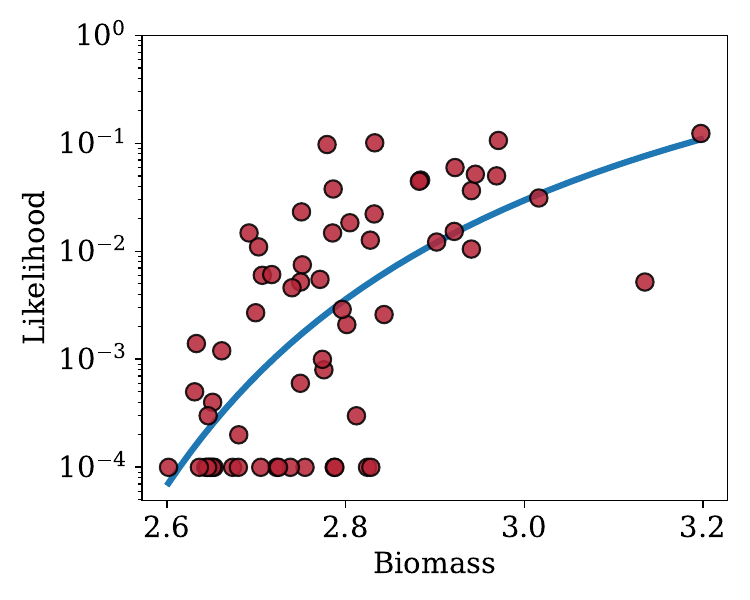}
        \caption{$S = 150$. }
        \label{subfig:S-150}
    \end{subfigure}
    \caption{\justifying Biomass--likelihood relation when number of species ($S$) varies.
    In both cases, we continue to observe a positive correlation between likelihood and biomass. Curves show fits to the log-hyperbolic relation $\log\left(p\right) \propto (B_c - B)^{-1}$.}
    \label{fig:variation_S}
\end{figure}

Next, instead of fixed $S = 100$ , we take different system size
We observe that for $S=\{50, 150\}$, the biomass--likelihood relationship still shows a positive correlation. As we decrease the system size the number of stable states decreases.(Fig.~\ref{fig:variation_S}).

\begin{figure}[h]
    \centering
    \begin{subfigure}[b]{0.4\textwidth}
        \centering
        \includegraphics[width=\textwidth]{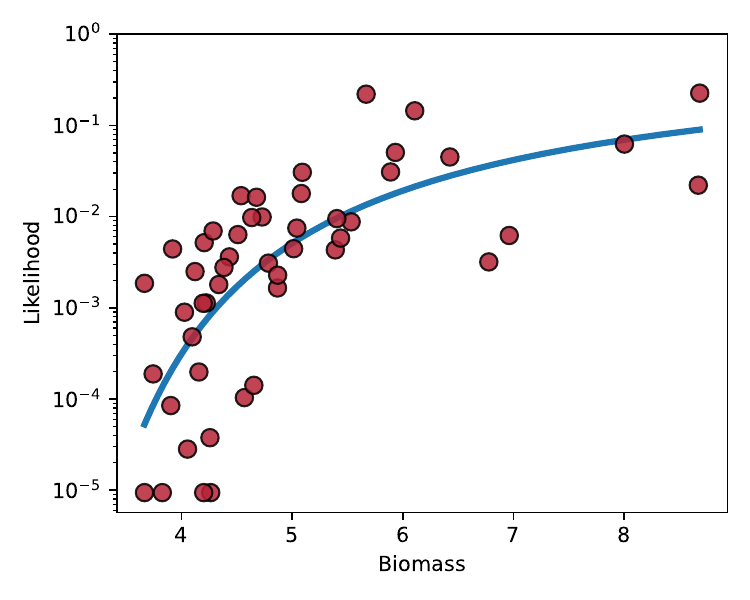}
        \caption{Uniform distribution}
        \label{fig:ini_uni}
    \end{subfigure}
    \hfill
    \begin{subfigure}[b]{0.4\textwidth}
        \centering
        \includegraphics[width=\textwidth]{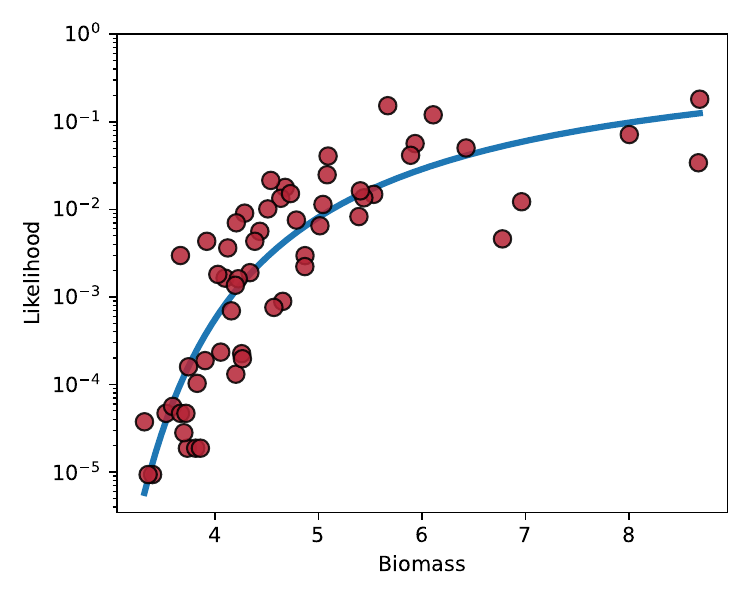}
        \caption{Truncated normal distribution}
        \label{fig:uni_nor}
    \end{subfigure}
    \caption{\justifying \textcolor{black}{Our results are robust to different initial condition sampling schemes: shown are both (a) initial conditions sampled from a uniform distribution in $[0,1]$ and (b) a standard normal distribution truncated below 0. Both initial condition sampling schemes show a similar likelihood-–biomass relationship. Curves show fits to the log-hyperbolic relation $\log\left(p\right) \propto (B_c - B)^{-1}$.}}
    \label{fig:initial}
\end{figure}

Next, in the main text, we sample each species' initial abundance
from a uniform distribution between 0 and 1. Here we show that the
biomass--likelihood relationship is robust to another reasonable
sampling scheme: that of sampling the initial conditions from a
Gaussian centred at 0 with standard deviation $0.3$, truncated 
to the positive orthant to ensure nonnegative abundances. We still find that the biomass--likelihood is qualitatively similar (\ref{fig:initial}). Even
though sampling initial conditions differently does change the
likelihoods of some states \textcolor{black}{and we gain few extra states of low likelihood}, the overall relationship is robust.

{\color{black}

We also test robustness to heterogeneous growth rates $r_i$ across
species. In the main text, we set $r_i = K_i = 1$; here we relax
this by drawing $r_i$ from a uniform distribution on $[1-\epsilon_r,
1+\epsilon_r]$. As noted in \cite{taylor2024structure}, when the $r_i$ values are
different, the value of the Lyapunov function in a local extremum
becomes $L = - \sum_i r_i N_i^*$ (weighted biomass), so it is actually this quantity
rather than biomass that we expect to be more closely
correlated with basin of
attraction size.  This quantity, which represents an alternative
weighting over abundances, can also be interpreted as the net growth
rate of the stable subset community in the absence of inhibition. We observe that for $\epsilon_r=\{0.01, 0.05\}$, the weighted biomass--likelihood relationship shows a positive correlation (Fig.~\ref{fig:variation_r}). \textcolor{black}{With increasing $\epsilon_r$, the correlation only slightly decreases and eventually saturates at roughly 0.5, still significantly positive (Fig.~\ref{fig:variation_r}(c)). Further, this correlation persists both when considering the biomass of each state as well as the weighted biomass, with the latter showing a slightly stronger correlation (Fig.~\ref{fig:variation_r}(c)). These results show that the biomass--likelihood relationship is robust to the inclusion of growth-rate variation between species.} 

\begin{figure}[h]
    \centering
    \begin{subfigure}[b]{0.4\textwidth}
        \centering
        \includegraphics[width=\textwidth]{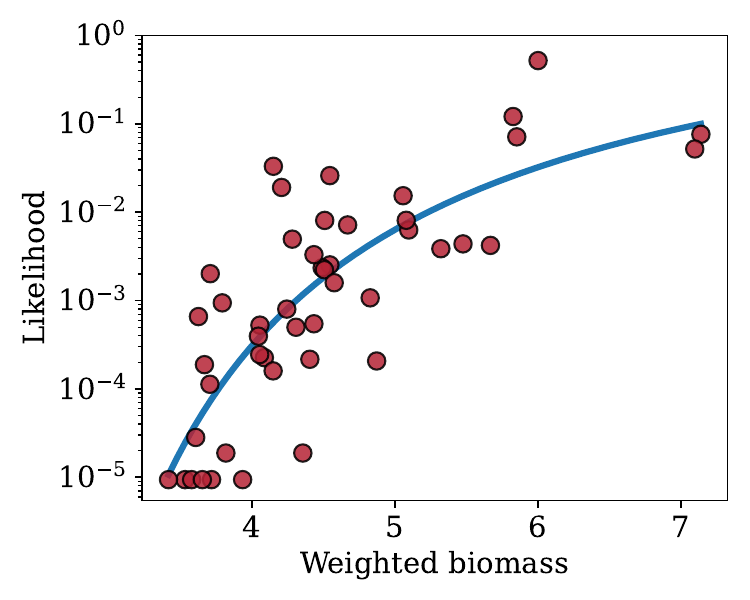}
        \caption{$\epsilon_r = 0.01$.}
        \label{subfig:r-0.01}
    \end{subfigure}
    \hfill
    \begin{subfigure}[b]{0.4\textwidth}
        \centering
        \includegraphics[width=\textwidth]{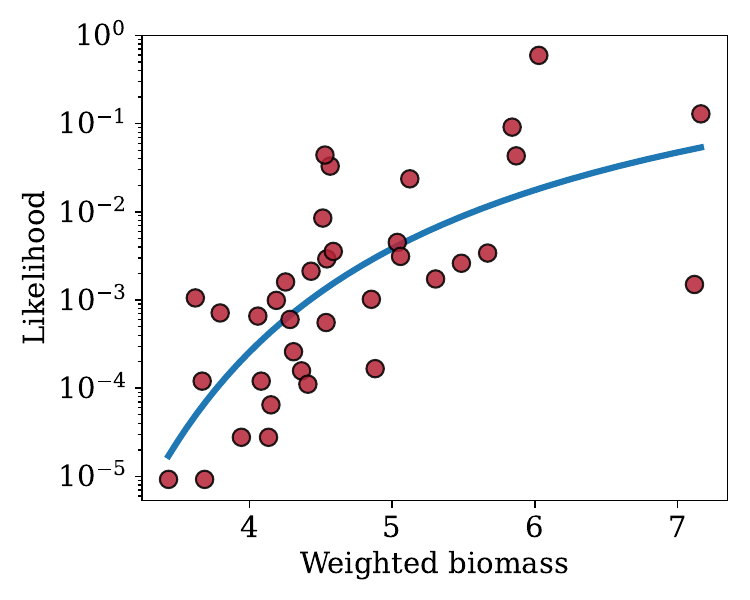}
        \caption{$\epsilon_r = 0.05$.}
        \label{subfig:r-0.05}
    \end{subfigure}

    \vspace{1em}

    \begin{subfigure}[b]{0.47\textwidth}
        \centering
        \includegraphics[width=\textwidth]{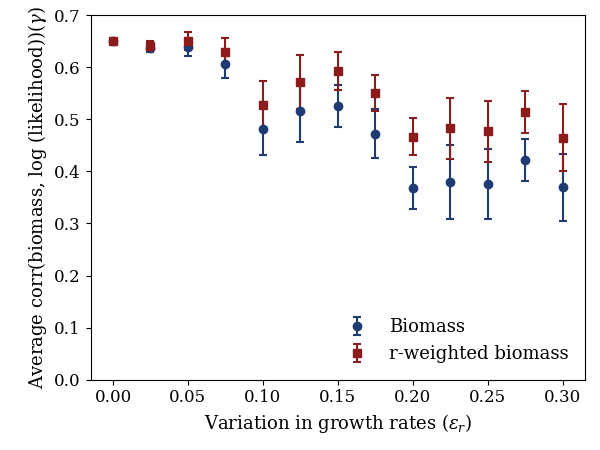}
        \caption{}
        \label{subfig:r-corr}
    \end{subfigure}
    \caption{\justifying Biomass--likelihood relation when growth rate $r_i$ varies around 1 with small variation $\epsilon_r$.\textcolor{black}{(a,b)} In both the case likelihood and growth-rate weighted biomass $\sum r_i N_i^*$ shows positive correlation. Curves indicate fits to the log-hyperbolic relation $\log\left(p\right) \propto (B_c - B)^{-1}$. \textcolor{black}{(c) Increasing variation in growth rates $r_i$ maintains a positive Pearson correlation $\gamma$ between biomass and log-likelihood (blue) for random interaction matrices, while slightly decreasing its magnitude. The correlation eventually saturates to roughly 0.4. Using growth rate weighted biomass $\sum_i r_i N_i^*$ results in a slightly stronger correlation (roughly 0.5, red). Results are shown for 5 independent realizations of sampling $r_i$ for each $\epsilon_r$, while keeping the interaction matrix $A$ random but fixed for all realizations. Error bars indicate SEM across realizations.}}
    \label{fig:variation_r}
\end{figure}

To confirm that the biomass--likelihood relationship is not
specific to the particular $(\mu, \sigma)$ used in the main text, we
sweep over a grid of values spanning the multistable phase and compute $\gamma$, the Pearson correlation between biomass and log-likelihood, for each. We find that $\gamma$ remains positive throughout
the multistable regime (Fig.~\ref{fig:phase_correlation}), confirming
that the positive relationship between biomass and likelihood is a
generic feature of this phase.

To understand how reliable our likelihood estimates are, we measured the uncertainty in them when repeating our simulations for the same interaction matrix. We repeated the simulation procedure described in Appendix~\ref{app:methods} 5 times for the same interaction matrix $A$ used in the main text Figs.~\ref{fig:fig1}c and ~\ref{fig:fig3}, each time sampling a different set of $10^6$ random initial conditions. We quantified uncertainty using the standard error in the mean (SEM) in the predicted likelihoods for each state. We observed a negligible uncertainty ($<1\%$) for the majority of states. Only for states with the lowest likelihoods ($\leq10^{-5}$) did we observe noticeable uncertainty (Fig.~\ref{fig:uncertainity}).

\begin{figure}[ht!]
    \centering
    \includegraphics[width=0.5\textwidth]{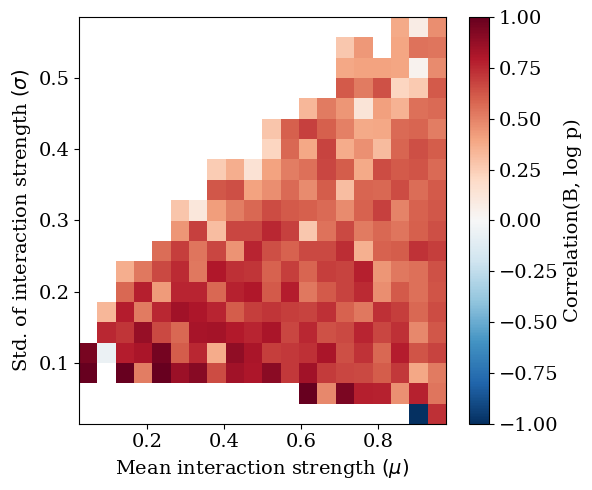}
    \caption{\justifying\textcolor{black}{The Pearson     correlation $\gamma$ between
        biomass and log-likelihood across the multistable
        phase space is positive. Each point corresponds to a different $(\mu,
        \sigma)$ value; $\gamma$ remains positive throughout,
        confirming that the biomass--likelihood relationship is a
        robust feature of the multistable regime independent of the
        specific parameter choice.}}
    \label{fig:phase_correlation}
\end{figure}

\begin{figure}[ht!]
\centering
\includegraphics[width=0.5\linewidth]{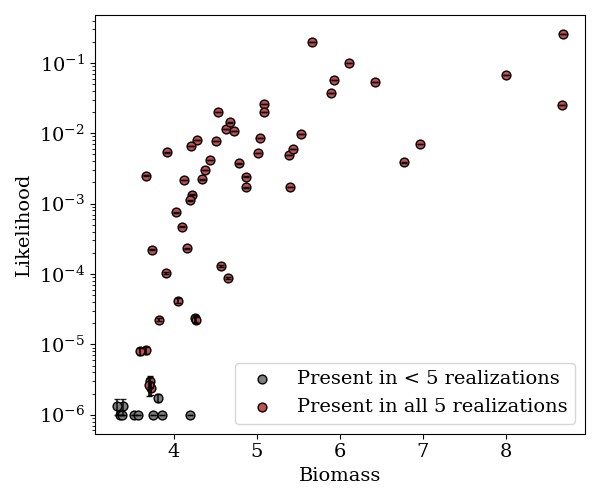}
\caption{\justifying\textcolor{black}{Uncertainty estimates for state likelihoods. Each point shows the likelihood of a state across multiple realizations of initial conditions for the same matrix. High-likelihood states appear in all five realizations (red) and show very small variation. In contrast, low-likelihood states show larger fluctuations, and some of them (gray) do not appear in all realizations. Results are shown for five different realizations. Error bars show the SEM across the realizations in which the state is observed.}}
\label{fig:uncertainity}
\end{figure}

Finally, we added demographic noise to the GLV dynamics to study the effect of fluctuations on the likelihood--biomass relationship. We incorporated noise into the dynamics through
\begin{equation}
    \frac{dN_i}{dt} = N_i \left(r_i \left(1 - \frac{N_i}{K_i}\right)
    - \sum_{\substack{j=1,\\ j\ne i}}^{S} A_{ij} N_j\right)
    + \beta_i \sqrt{N_i} \, \xi_i(t),
    \label{eq:LangevinGLV}
\end{equation}
where $\beta_i \geq 0$ is the noise amplitude of species $i$,
$\sqrt{N_i}$ reflects the Poisson scaling of birth--death fluctuations
(vanishing at $N_i = 0$, preserving the extinction boundary), and
$\xi_i(t)$ is Gaussian white noise with 
$\langle \xi_i(t) \rangle = 0$
and 
$\langle \xi_i(t)\,\xi_j(t') \rangle = \delta_{ij}\,\delta(t-t')$. We still observed a positive likelihood--biomass relationship after introducing noise, showing that high-biomass states generally remain more likely. The likelihood--biomass relationship remains qualitatively similar after adding noise, showing that low-biomass states do not become significantly less likely due to fluctuations (Fig.~\ref{fig:demographic_10}). 

We also directly compared the likelihoods of the same states with and without fluctuations, and found that low-biomass states can also be surprisingly resilient to fluctuations. In some cases, fluctuations actually increased the likelihood of the low-biomass states (Fig.~\ref{fig:demographic_compare}), making them even more resilient than high-biomass states. These observations may reflect the interplay between the widths and depths of basins of attraction, but understanding this in detail requires further study, which we leave for future work.

\begin{figure}[t]
\centering
\includegraphics[width=0.5\linewidth]{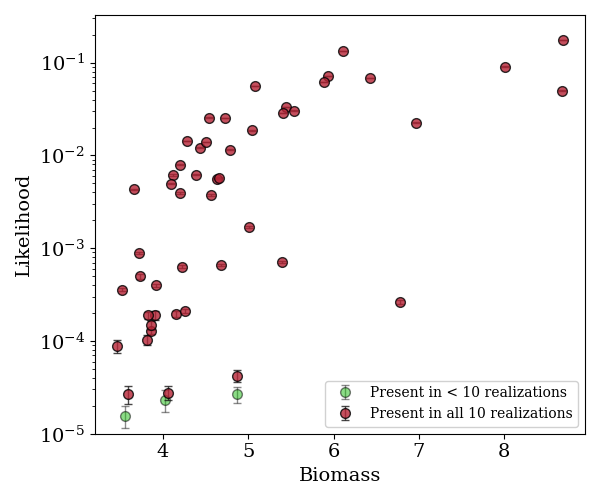}
\caption{\justifying \textcolor{black}{In the presence of demographic noise, the likelihood and biomass of different stable states are still positively correlated. We find that both low-biomass and high-biomass states are resilient to fluctuations. In all 10 realizations, we observe roughly the same set of low-biomass and high-biomass states. Each point shows the average likelihood of a state across 10 realizations of initial conditions for the same matrix and the same noise strength. Error bars show the SEM across the realizations in which the state is observed.}}
\label{fig:demographic_10}
\end{figure}

\begin{figure}[t]
\centering
\includegraphics[width=0.5\linewidth]{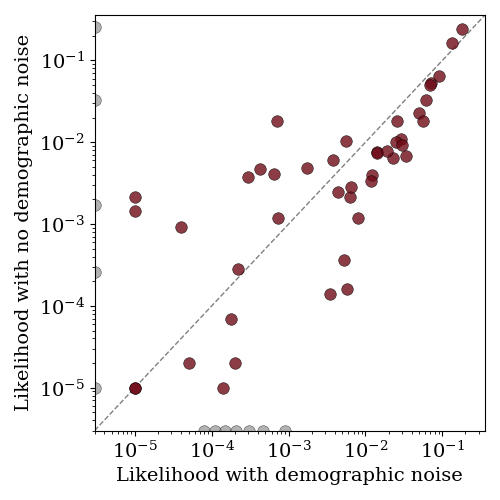}
\caption{\justifying \textcolor{black}{Demographic noise changes the likelihood of ecological states. States observed under both conditions (red) show that high-likelihood states remain highly likely even in the presence of noise, while low-likelihood states can shift, with some becoming more likely and others less likely. States that disappear under noise (gray, left edge) or appear only with noise (gray, bottom edge) suggest that demographic stochasticity can make rare states dominant and dominant states rare.}}
\label{fig:demographic_compare}
\end{figure}

} 

\FloatBarrier

\section{Exact solution of likelihoods and biomasses in the monodominant model}
\label{app:monodominantCalc}
In the monodominant model, the likelihood and biomass can be
calculated exactly, which is not possible for random matrices. Since
the monodominant model is the simplest one in which we can compute the
likelihood--biomass relationship exactly, we start there. Afterwards,
we will extend our calculation to block-structured interaction
matrices, where more than one species can survive. Although random
matrices do not obviously show clear block patterns, species that coexist in a
state derived from a random disordered interaction matrix still
contain block-like features, as 
we show explicitly in Appendix~\ref{app:blockGrouping}.
In the subsequent Appendices, we
generalize the monodominant analysis to block models, and
use the block model as a framework 
for
developing a technique to
predict likelihoods for such random matrices. As we show in the main
text, this allows us to predict likelihoods for random matrices
reasonably well using only the biomass and diversity of each state,
along with the mean interaction strength $\mu$ which is a coarse
parameter.

In the monodominant model all species $i$ interact with the same interspecific strength $D$, while they differ in self-inhibition $K_{i}$. \textcolor{black}{For simplicity, we take $r_i =1$ for all species (though we relax this in a later subsection in this Appendix). Thus, the self-inhibition of species $i$ is given by $A_{ii} = 1/K_i$, and species with higher carrying capacity have weaker self-inhibition.} To ensure that at most one species survives at steady state we consider only cases where $D>A_{ii}$ for all $i$.

The GLV equation in the monodominant case can be written as

\begin{equation}
    \frac{1}{N_i}\frac{dN_i}{dt} = 1 + (D - A_{ii})N_i - D\sum_k N_k .
    \label{eq:mono_modi}
\end{equation}

The last term $-D\sum_k N_k$ is identical for all species. Thus the species-specific contribution is $(D-A_{ii})N_i$. Define
\begin{equation}
    \chi_i := D - A_{ii}.
\end{equation}

From (\ref{eq:mono_modi}), we can see that the hyperplane defined by
$\chi_i N_i = \chi_j N_j$ is a fixed locus of the dynamics.
Assuming this relation holds at any time $t$, we have $\chi_i d
N_i/d t = \chi_j d N_j/d t$, since the right-hand side of
(\ref{eq:mono_modi}) is the same for both species.  This shows that
this hyperplane is a separatrix between the two species, and any
initial condition with $\chi_i N_i >\chi_j N_j$ must end in a
steady-state solution where $N_j = 0$.  Because this geometric
relation holds for every pair of species, for 
any initial condition the species that dominates must be the one with 
the largest value of $\chi_i N_i(0)$, which we call the dressed initial condition.

Geometrically, we can interpret this as a division of the positive
orthant by $S (S -1)/2$ hyperplanes into $S!$ cones, each associated
with an ordering of the $\chi_i N_i$ values.  Within each cone, all
initial conditions lead to the final steady state $i$ with the largest
value of $\chi_i N_i$.
To compute the probability of a given final state for a given
distribution of initial conditions, we need to determine the fraction
of initial conditions associated with the set of cones that lead to
the given final state.
(Note that while this general cone structure was described in
\cite{taylor2024structure}, the exactness of this geometric
decomposition for the monodominant systems was not realized there.)

In the subsequent analysis we consider this question from two
different perspectives: first, given the initial conditions on each
species taken from a uniform distribution from 0 to 1, giving a
hypercube of initial conditions within which we calculate the fraction
in each domain of attraction; second, scaling the initial conditions
by $\chi_i$, so the sides of the initial condition boxes vary, but
such that we simply determine the largest value of $\chi_i N_i (0)$
directly.

\subsection*{Two species}
\begin{figure}[ht!]
    \centering
    \begin{subfigure}[b]{0.35\textwidth}
        \centering
        \includegraphics[width=\textwidth]{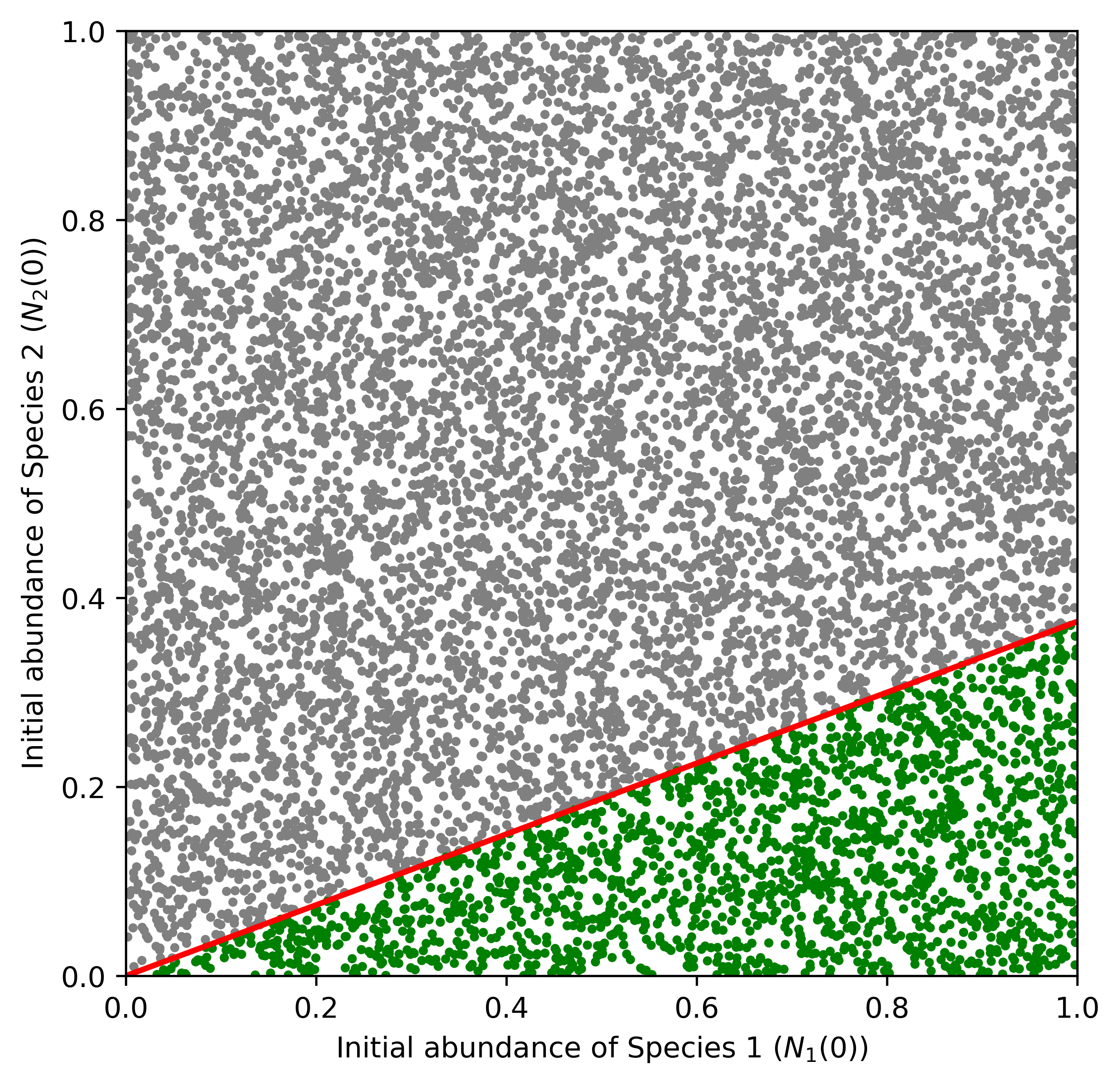}
        \caption{Raw initial abundances: the initial-condition square in $(N_1,N_2)$ and the separatrix (red).}
        \label{fig:mon}
    \end{subfigure}
    \hskip 40pt
    \begin{subfigure}[b]{0.35\textwidth}
        \centering
        \includegraphics[width=\textwidth]{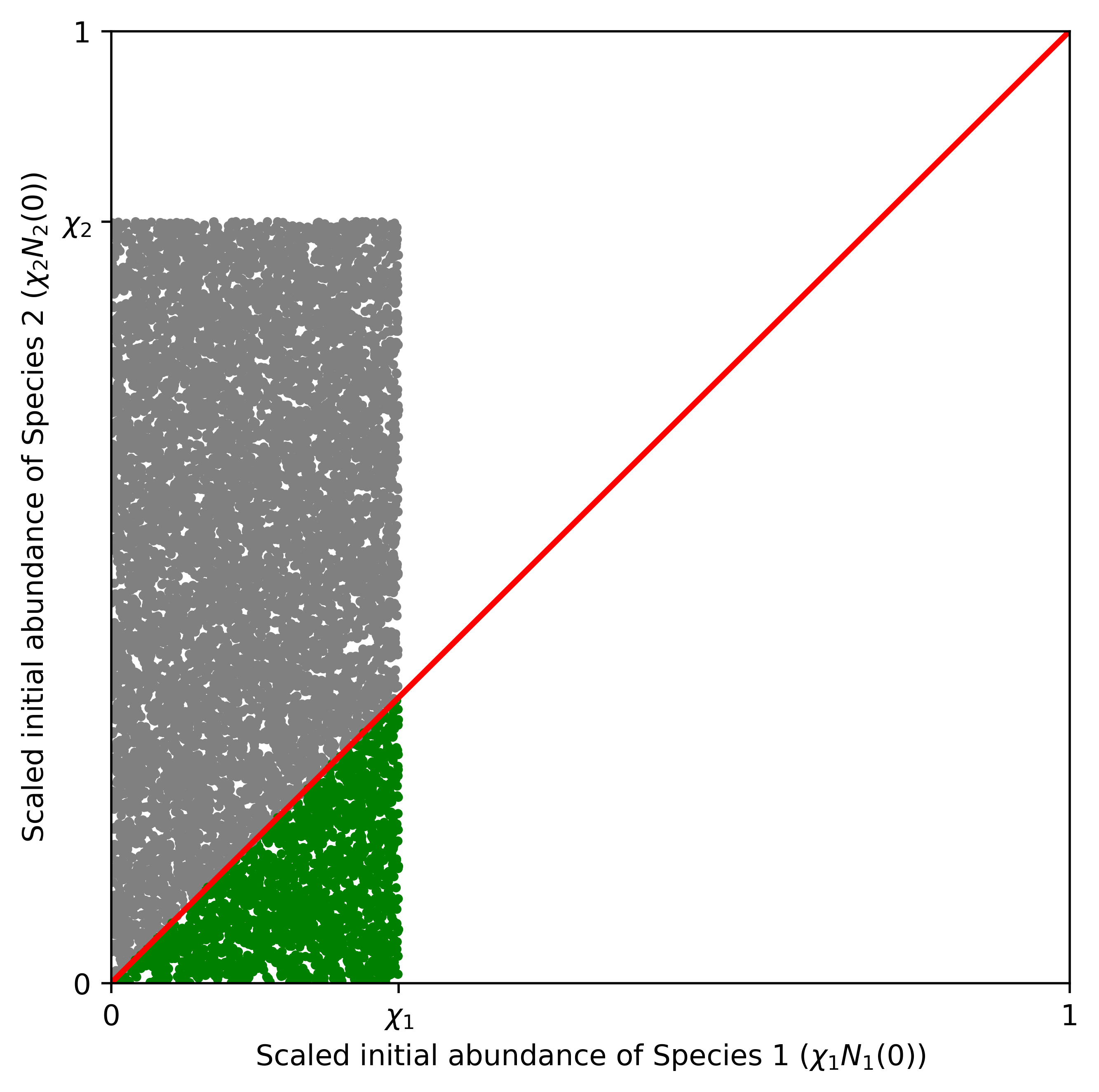}
        \caption{Dressed initial conditions: after scaling by $\chi_i$ the separatrix becomes the diagonal $X_1=X_2$.}
        \label{fig:mons}
    \end{subfigure}
    \caption{\justifying Green regions indicate outcomes where species 1 wins; gray regions indicate outcomes where species 2 wins. In panel (a) the separatrix has slope $\chi_1/\chi_2$ in the $(N_1,N_2)$ plane. In panel (b) the dressed variables $(X_1,X_2)$ make the separatrix the diagonal $X_1=X_2$.}
    \label{fig:mono}
\end{figure}

Consider two species with self-inhibition parameters satisfying $A_{11}>A_{22}$ (so $\chi_1<\chi_2$). In our simulations we draw initial abundances for both species from the same uniform distribution between 0 and 1, so the set of initial conditions is a square in the $(N_1,N_2)$ plane (see Fig.~\ref{fig:mon}). The separatrix can be found analytically; it passes through the point $((D-A_{22}),(D-A_{11}))$ in the coordinate system used in the simulation and partitions the square into two basins.

Viewed in the dressed-initial-condition picture, set $X_1\sim\mathrm{Uniform}(0,\chi_1)$ and $X_2\sim\mathrm{Uniform}(0,\chi_2)$ where $X_i=\chi_i N_i(0)$. The separatrix is then given by the line $\chi_1 N_1(0)=\chi_2 N_2(0)$, which in the dressed $(X_1,X_2)$ coordinates is simply $X_1=X_2$ (see Fig.~\ref{fig:mons}). The question ``what is the likelihood of species $2$ winning?'' reduces to computing

\begin{equation}
    P(X_2>X_1),\qquad X_1\sim\mathrm{Uniform}(0,\chi_1),\; X_2\sim\mathrm{Uniform}(0,\chi_2).
\end{equation}

\begin{figure}[ht!]
    \centering
    \includegraphics[width=0.7\textwidth]{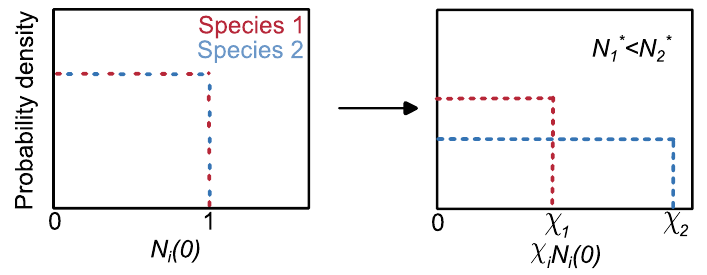} 
    \caption{\justifying Left: PDFs of the raw initial abundances for species 1 and 2 (both drawn from $\mathrm{Uniform}(0,1)$) are identical. Right: PDFs of the dressed initial conditions $X_i=\chi_i N_i(0)$ are uniform on $[0,\chi_i]$. Because $\chi_2>\chi_1$, species 2 has extra support on the interval $(\chi_1,\chi_2]$ that species 1 cannot access, giving species 2 an advantage in the maximum.}
    \label{fig:scale2}
\end{figure}

We compute $P(X_2>X_1)$ by splitting into cases. If $X_2$ lies between $\chi_1$ and $\chi_2$ then $X_2$ is always larger than $X_1$. If $X_2$ lies between $0$ and $\chi_1$ then $X_2$ is larger than $X_1$ with probability $1/2$ (by symmetry of two identical $\mathrm{Uniform}(0,\chi_1)$ draws). Therefore
\begin{equation}
P(X_2>X_1) = P(\chi_1 < X_2 < \chi_2)\,P(X_2>X_1 \mid \chi_1 < X_2 < \chi_2) + P(0 < X_2 < \chi_1)\,P(X_2>X_1 \mid 0 < X_2 < \chi_1).    
\end{equation}

We also have that

\begin{equation}
    P(\chi_1 < X_2 < \chi_2)=\dfrac{\chi_2-\chi_1}{\chi_2}, \ P(0<X_2<\chi_1)=\dfrac{\chi_1}{\chi_2},    
\end{equation}
and the conditional probabilities are $1$ and $1/2$ respectively. Hence

\begin{equation}
    P(X_2>X_1)= \frac{\chi_2-\chi_1}{\chi_2}\cdot 1 + \frac{\chi_1}{\chi_2}\cdot\frac{1}{2} = 1 - \frac{\chi_1}{2\chi_2}.
\end{equation}

By complementarity,

\begin{equation}
    P(X_1>X_2)=1-P(X_2>X_1)=\frac{\chi_1}{2\chi_2}.
\end{equation}

\subsection*{Many species}

\begin{figure}[ht!]
    \centering
    \includegraphics[width=0.4\textwidth]{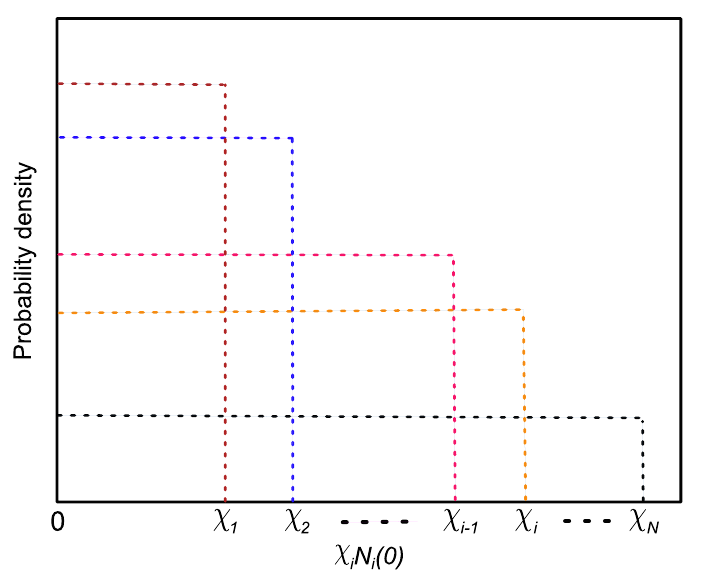} 
    \caption{\justifying PDF of dressed initial conditions for $S$ species. The self-inhibition values are ordered as $A_{11}>A_{22}>\dots>A_{SS}$, so the corresponding upper bounds satisfy $\chi_1<\chi_2<\dots<\chi_S$. A larger $\chi_i$ (smaller self-inhibition) gives species $i$ a wider range of dressed initial values and thus a larger chance to be the maximum.}
    \label{fig:many_n}
\end{figure}

Consider $S$ species with self-inhibition values ordered as $A_{11}>A_{22}>\dots>A_{SS}$. Define $\chi_i=D-A_{ii}$ so that
$$
\chi_1<\chi_2<\dots<\chi_S,
$$
i.e. species with smaller self-inhibition have larger $\chi_i$.

For each species $j$ let
\begin{equation}
    X_j \sim \mathrm{Uniform}(0,\chi_j),\qquad j=1,\dots,S.
\end{equation}

The density of $X_i$ is
\begin{equation}
    f_i(x)=
    \begin{cases}
    \dfrac{1}{\chi_i}, & 0\le x\le\chi_i,\\[6pt]
    0, & \text{otherwise,}
    \end{cases}
\end{equation}
and the cumulative distribution of $X_j$ is
\begin{equation}
    P(X_j\le x)=
    \begin{cases}
    \dfrac{x}{\chi_j}, & 0\le x\le\chi_j,\\[6pt]
    1, & x>\chi_j.
    \end{cases}
\end{equation}

The probability that $X_i$ is the largest among the $S$ variables is
\begin{equation}
    p_i \;=\; \int_0^{\chi_i} f_i(x)\prod_{j\ne i} P(X_j\le x)\,dx.
\end{equation}

Because $P(X_j\le x)$ changes form at the breakpoints $\chi_j$, partition the interval $[0,\chi_i]$ using
\begin{equation}
    I_k=(\chi_{k-1},\chi_k],\qquad k=1,\dots,i,
\end{equation}
with the convention $\chi_0=0$. On the subinterval $I_k$ exactly the first $k-1$ variables have reached their full support (so $P(X_j\le x)=1$ for those), while the remaining variables satisfy $P(X_j\le x)=x/\chi_j$. Hence on $I_k$ the product becomes
\begin{equation}
    \prod_{j\ne i} P(X_j\le x)
    \;=\; \frac{x^{\,S-k}}{\displaystyle\prod_{\substack{j=k\\ j\ne i}}^{S}\chi_j}.
\end{equation}

Using $f_i(x)=1/\chi_i$, the contribution of the interval $I_k$ to $p_i$ is
\begin{equation}
    I_k \;=\; \int_{\chi_{k-1}}^{\chi_k} \frac{1}{\chi_i}\cdot\frac{x^{\,S-k}}{\displaystyle\prod_{\substack{j=k\\ j\ne i}}^{S}\chi_j}\,dx
    \;=\;
    \frac{\chi_k^{\,S-k+1}-\chi_{k-1}^{\,S-k+1}}{(S-k+1)\displaystyle\prod_{j=k}^{S}\chi_j}.
\end{equation}
Note that this contribution can be understood in a simple geometric
fashion: $\chi_k^{S - k +1} - \chi_{k -1}^{S - k +1}$ is the volume of the region in which all
values of $\chi_i \leq \chi_k$, for $i \geq k$, removing the volume
where all $\chi_i <\chi_{k -1}$. The product over $\chi_j$'s is the
appropriate normalization factor for this volume.  Within this volume
there are $S - k +1$ species that are equally likely to have the
greatest value of $\chi_i N_i$.  So this volume gives an equal
contribution to each of those species for the probability to be greatest.

Therefore the exact probability that species $i$ is the largest is the sum over these contributions:
\begin{equation}
    p_i \;=\; \sum_{k=1}^{i} I_k
    \;=\; \sum_{k=1}^{i}\frac{\chi_k^{\,S-k+1}-\chi_{k-1}^{\,S-k+1}}{(S-k+1)\displaystyle\prod_{j=k}^{S}\chi_j},
\end{equation}
The difference in probability between species $i$ and $i-1$ is  
\begin{equation}
    p_i - p_{i-1} \;=\; I_i
    \;=\; \frac{\chi_i^{\,S-i+1}-\chi_{i-1}^{\,S-i+1}}{(S-i+1)\displaystyle\prod_{j=i}^{S}\chi_j}.
\end{equation}
The above expression shows that when species $i$ has lower self-inhibition than species $i-1$, i.e.\ $A_{(i-1)(i-1)} > A_{ii}$ (or equivalently $\chi_i > \chi_{i-1}$), its likelihood always increases.

{\color{black}

\subsection*{Extension of \textcolor{black}{the monodominant model} to heterogeneous growth rates}

In the main text and throughout the analysis above, we set $r_i = 1$
for all species \textcolor{black}{(except see Appendix C, Fig.~S9)}. Here we \textcolor{black}{extend} the monodominant model to allow species-specific growth rates $r_i$, which is important because in the GLV model (Eq.~\eqref{eq:GLVEq} in the main text) the growth rate and carrying capacity can play distinct roles. With heterogeneous $r_i$, the
monodominant GLV equation becomes
\begin{equation}
    \frac{dN_i}{dt} = N_i \left( \left(r_i - \frac{r_i}{K_{i}} N_i\right) - D \sum_{j \neq i} N_j \right),
    \label{eq:mono_ri}
\end{equation}
%

\textcolor{black}{To test if we can quantitatively predict the winner in this modified monodominant model, we attempt to use a similar approach as before: that of defining an appropriate dressed initial condition. Rewriting Eq.~\eqref{eq:mono_ri} in terms of the per capita growth rate of species $i$ we get}
\begin{equation}
    \frac{1}{N_i}\frac{dN_i}{dt} = r_i + \left(D - \frac{r_i}{K_{i}}\right)N_i - D\sum_j N_j.
    \label{eq:mono_ri_rewrite}
\end{equation}
The last term is again common to all species, but the
species-dependent part now involves $r_i$ explicitly. 
\textcolor{black}{We can now define a modified version of the dressed initial condition to account for growth rate differences as:} 
\begin{equation}
    C_i = r_i + \left(D - \frac{r_i}{K_i}\right) N_i\textcolor{black}{(0)},
\end{equation}
where $K_i$ is the steady-state abundance or biomass of species $i$. When $r_i = 1$, we get $C_i = 1 + \chi_i N_i\textcolor{black}{(0)}$, and like in the previous section, the species with the largest $\chi_i N_i(0)$ wins. However, when $r_i$ varies between species, the \textcolor{black}{modified dressed initial condition $C_i$} acquires a complicated dependence on both $r_i$ and $K_i$. The species 
{\color{black}with the largest $C_i$ has the largest per capita growth rate at $t = 0$.

To test whether the modified dressed initial condition $C_i$ can quantitatively predict the winner in this modified monodominant model, we perform simulations in which we systematically control the extent of growth rate variation. We introduce a parameter $\varepsilon$ and for each species $i$ we draw its growth rate $r_i$ from a uniform distribution centered at $1$ with width $2\varepsilon$, such that $r_i \in [1-\varepsilon, 1+\varepsilon]$. When $\varepsilon=0$, all species have the same growth rate $r_i = 1$ and we recover the original monodominant model, Eq.~\eqref{eq:mono_modi}. Increasing $\varepsilon$ increases the extent of growth rate variation. 

In the presence of growth rate variation between species, we find that the modified dressed initial condition $C_i$ only approximately predicts which species wins, as shown in Fig.~\ref{fig:top_ci}. Specifically, while $C_i$ perfectly predicts the winner in the absence of growth rate variation, in the presence of substantial growth rate variation, $C_i$ correctly predicts the winner only $\sim50\%$ of the time. This is still significantly better than the null expectation of $1/S = 5\%$ for $S=20$ species. Furthermore, the winner predicted using the largest $C_i$ is still typically among the top 2 or 4 species ($70\%$ and $85\%$ of the time, respectively), significantly better than the null expectation (dashed lines in Fig.~\ref{fig:top_ci}). This shows that the modified dressed initial condition $C_i$ remains a somewhat good approximation of the winner. This reduction in predictive power might arise to the fact that in the presence of growth rate variation, the separatrices between the different basins of attraction are no longer straight lines, but become curved. This may result in the transient nonlinear dynamics mattering a lot more in predicting the winner, which cannot be captured by the initial conditions alone. We leave a full exploration of this for future work.}

\begin{figure}
    \centering
    \includegraphics[width=0.55\textwidth]{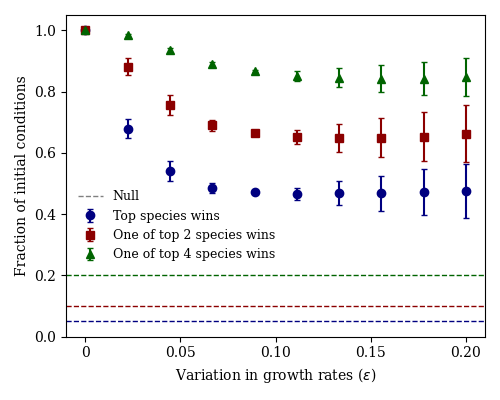}
    \caption{\justifying\textcolor{black}{The species with the largest $C_i(0)$ remains a good approximation of the winner in the monodominant model with heterogeneous growth rates. The plot shows the fraction of initial conditions in which the winning species is the one with the largest $C_i(0)$ (blue), one of the two largest (red), or one of the four largest (green), as a function of the variation in growth rates $\epsilon$ among a total of $S=20$ species. Here, $r_i$ is drawn from a uniform distribution on $[1-\epsilon, 1+\epsilon]$. Dashed lines of the same colors show the corresponding null expectation, $k/S$ for the top $k$ species, obtained if the winner were chosen uniformly at random among all $S$ species. Error bars indicate SEM for 30 monodominant interaction matrix realizations.}}
    \label{fig:top_ci}
\end{figure}

} 

\begin{figure}[ht!]
    \centering
    \includegraphics[width=0.5\textwidth]{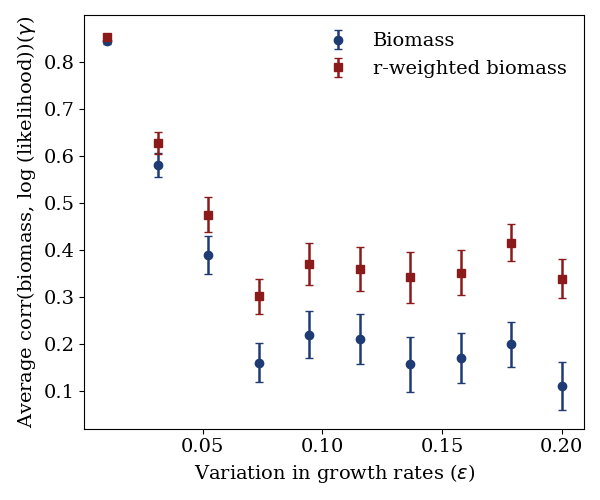}
    \caption{\justifying\textcolor{black}{Increasing variation in growth
        rates $r_i$ maintains a positive Pearson correlation $\gamma$ between biomass
        and log-likelihood in the monodominant model, but decreases its magnitude. Here, $r_i$ is drawn from a uniform distribution on $[1-\epsilon, 1+\epsilon]$ in the monodominant model. The correlation saturates eventually, but to a positive value and never becomes negative. Using growth rate weighted biomass $\sum_i r_i N_i^*$ produces systematically \textcolor{black}{higher correlations}, especially as the variation in growth rates increases. Error bars indicate SEM for 30 monodominant interaction matrix realizations.}}
    \label{fig:ri_variation}
\end{figure}

\textcolor{black}{Beyond predicting the winning species, we also examine how the correlation between biomass and likelihood depends on the growth rate variation. We find that in the monodominant model, with increasing growth rate variation,
the average correlation between biomass and likelihood $\gamma=\text{corr(biomass, log(likelihood))}$ decreases significantly from $\gamma\sim0.85$ and saturates to $\gamma\sim0.15$ around $\varepsilon=0.20$ (Fig.~\ref{fig:ri_variation}). Notably, while its magnitude decreases, the correlation still remains positive. Using the growth rate-weighted biomass $\sum r_i N_i^*$ instead of the true biomass $\sum_i N_i^*$ increases this correlation to $\gamma\sim0.35$ in the presence of substantial growth rate variation. These results show that the correlation between biomass and likelihood in the monodominant model always remains positive, but its magnitude is sensitive to growth rate variation.}

\textcolor{black}{Note that this decrease in the correlation is qualitatively similar to what we observed in the case of random disordered matrices (Fig.~\ref{fig:variation_r}(c)). However, the effect is much more pronounced in the monodominant model than the random matrix case, where the decrease in the correlation is relatively small. This might be because the monodominant model is much more sensitive to the growth rates of individual species, since each stable state contains only one species, while stable states in the random disordered matrices contain multiple species. We leave a full investigation of this effect for future work.}


\FloatBarrier    

\section{Solution of likelihoods and biomasses for block models}
\label{app:blockCalc}
In the monodominant model, the chance of a species winning depends on
its self-inhibition $A_{ii}$ (or equivalently its eventual biomass
$B_i = 1/A_{ii}$). If all species' initial conditions are the same
uniform distribution, then this difference in self-inhibitions causes
changes in the slope of the separatrices dividing their basins of
attraction. But, a random disordered matrix case does not map easily
to the monodominant model since in each state, the former has several
coexisting species. This motivates us move on to an enhanced variant of
the monodominant model that allows more than one species to coexist
in a state.

The simplest such model is to have the interaction matrix consist of
blocks, where species within a block $\alpha$ inhibit both each
other and themselves with a common strength $A_{\a\a}$, and species in
different blocks inhibit each other with a different, stronger
strength $D$. We assume no overlap between blocks, which means that any
species belongs exclusively to one block state. In this case only one
block can win given any set of initial conditions, and all species
within the winning block survive. We call this a ``neutral block''
model since each species in a block is identical to all others and
thus in ecology jargon, all species within a block are ``neutral''. We
can show that the neutral block model maps exactly to a monodominant
model. To see this, note that we can write down the GLV equations
governing the dynamics of the total abundance (biomass) $B_\a= \sum_{i\in \a} N_i$ of each
block $\alpha$ as
\begin{equation}
    \frac{1}{B_{\alpha}}\frac{dB_\alpha}{dt} = 1+(D-A_{\alpha\alpha})B_{\alpha} -D\sum_{\beta}B_{\beta}
    \label{eq:neutral_block}
\end{equation}

\begin{figure}[ht!]
    \centering
    \includegraphics[width=0.7\textwidth]{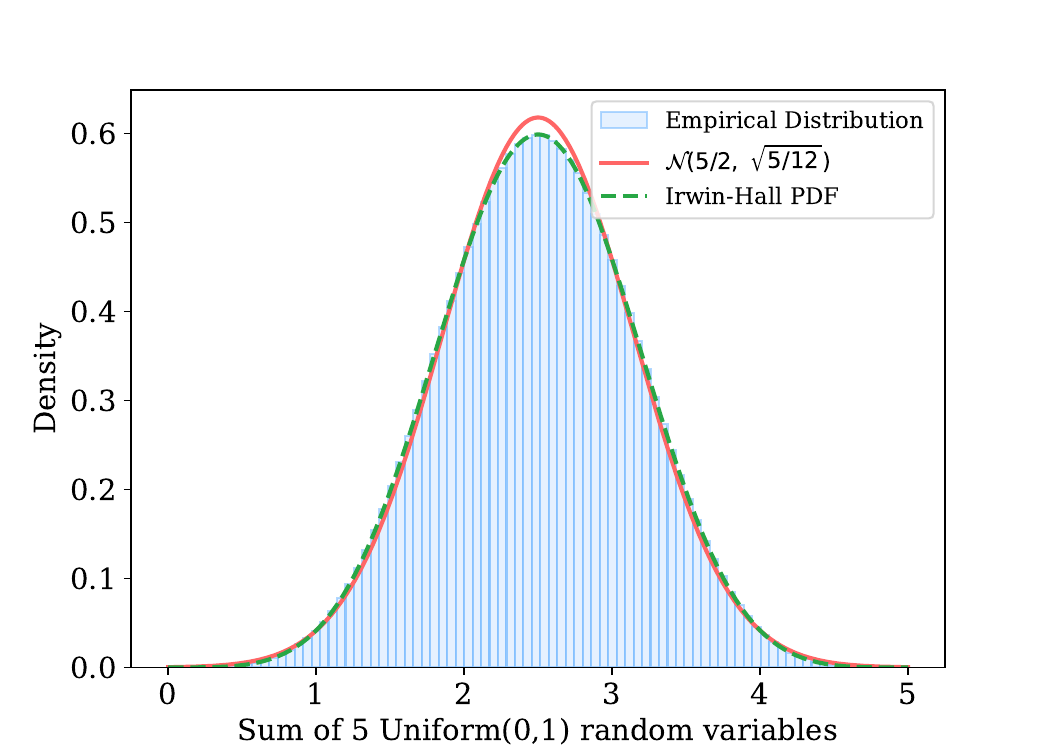} 
    \caption{\justifying The sum of 5 independent uniform random variables (blue) compared with a Gaussian distribution with mean $5/2$ and standard deviation $\sqrt{5/12}$ (red). The Irwin--Hall distribution matches the Gaussian approximation very well. The empirical distribution contains $10^4$ samples.}
    \label{fig:irwin}
\end{figure}

At steady state the final abundance of the winning block $\alpha$ will
be $B^*_\alpha=\frac{1}{A_{\alpha\alpha}}$.
Eq.~\eqref{eq:neutral_block} maps exactly to
Eq.~\eqref{eq:mono_modi}. The steady state abundance of each block
$\a$, now the inverse of $A_{\a\a}$, replaces the
steady-state abundance of each species, which was the inverse of
$A_{ii}$. Following the monodominant results, the winning block will
be the one with the largest value of the dresssed initial condition,
now $\chi_\alpha B_\alpha$, where $\chi_\alpha = D - A_{\a\a}$ and
$B_\a = \sum_{i\in \a} N_i$
being the sum of initial abundances of
species within block $\a$. If each block $\a$ contains $L_\a$ species,
then given uniform initial conditions for each species' abundances,
the block's initial abundance follows the Irwin--Hall distribution.
At sufficiently large $L_\a$ (here, 5), this distribution agrees
extremely well with a Gaussian with mean $\tfrac{L_\a}{2}$ and
standard deviation $\sqrt{\tfrac{L_\a}{12}}$
(Fig.~\ref{fig:irwin}). Thus, for blocks of sufficient size (as shown,
5 is enough), the key difference between this neutral block model and
the monodominant model is that the distribution of dressed initial
conditions for each block will be a Gaussian rather than a uniform
distribution. This will affect the likelihood of each block
quantitatively, but not change the central insight that the dressed
initial condition, which as we have shown depends crucially on the
self-inhibition, perfectly predicts which block wins.
Note that this choice of initial distribution will strongly favor larger blocks;
 with truly neutral interactions within each block, depending upon the
 context another distribution such as uniform across $B_\alpha (0)$
 might be natural, but in the context of the random matrix systems for
 which we are developing this theory, where each species is treated as
 a priori the same regarding initial conditions, the Irwin-Hall
 distribution is what we want to analyze.  Note also that if the
 distribution on each species is taken to be something else that is
 the same across species, such as a Gaussian truncated to the positive
 orthant, for blocks of sufficient size we will again have a Gaussian
 distribution on the dressed initial conditions.

The neutral block model assumes that all species' self-inhibition is
the same as their inhibition  of each other species in the same block, but in
our random disordered matrices all species' self-inhibitions are 1. To
account for this, we now consider a different block interaction
matrix, which only differs from the neutral block in that the
self-inhibitions of all species 
is 1. All other
interspecies inhibitions of species in the same block $\a$ are
$A_{\a\a}$ and across blocks are $D$. This is the block model we use in
the main text. We will show that likelihoods in this block model
are determined in a very similar fashion to that of the completely
neutral block model, with an effective self-inhibition that again
corresponds to the inverse of the equilibrium biomass for each block.
To see this, let us begin with the species-level dynamics. The
abundances of each species $N_i$ evolve according to:

\begin{equation}
    \frac{dN_i}{dt} = N_i\left( 1 - N_i 
    - A_{\alpha\alpha} \sum_{\substack{j \in \text{block}~\alpha \\ j \neq i}} N_j 
    - D \sum_{\substack{\beta \neq \alpha \\ k \in \text{block}~\beta}} N_k \right).
    \label{eq:block_main_equation}
\end{equation}

At steady state, 
all species in block $\alpha$ have the same abundance, i.e.\ $N_i^* = N_j^*$ for all $i,j \in \text{block}~\alpha$. The steady state condition becomes

\begin{equation}
    1 - N_i^* - (L_\alpha-1)A_{\alpha\alpha}N_i^* = 0,
\end{equation}

which gives

\begin{equation}
    N_i^* = \frac{1}{(L_\alpha-1)A_{\alpha\alpha}+1}.
\end{equation}

Hence the steady state biomass of block $\alpha$ is

\begin{equation}
    B_\alpha^* = \sum_{i\in \text{block}~\alpha}N_i^* 
    = \frac{L_\alpha}{(L_\alpha-1)A_{\alpha\alpha}+1}.
    \label{eq:block_biomass}
\end{equation}

Summing equation (\ref{eq:block_main_equation}) over all $i$ in block $\alpha$, we obtain the effective dynamics for the blocks in terms of their current abundances $B_\a$ as

\begin{equation}
    \frac{1}{B_{\alpha}}\frac{dB_\alpha}{dt} 
    = 1
    - D\sum_{\beta}{B}_{\beta} 
    + \left(D - A_{\alpha \alpha}\right)B_{\alpha} 
    + (A_{\alpha\alpha}-1)\frac{\sum_{i\in\text{block}~\alpha}N_i^2}{B_\alpha}.
    \label{eq:neutral_block_1}
\end{equation}

Note that the dynamics within each block in this model is no longer
neutral, but
tends to equilibrate species within each block when $1 >A_{\alpha \alpha}$, so that
for a given total block biomass $B_\alpha$ the species abundances are driven within each block to the local quasi-equilibrium
\begin{equation}
  N_\alpha \cong B_\alpha/L_\alpha \,.
\label{eq:}
\end{equation}
It is thus helpful to rewrite Eq.~(\ref{eq:neutral_block_1}) in the form
\begin{equation}
    \frac{1}{B_{\alpha}}\frac{dB_\alpha}{dt} 
    = 1
    - D\sum_{\beta}{B}_{\beta} 
    + \left(D - {A}^{{\rm eff}}_{\alpha \alpha}\right)B_{\alpha} 
    + (A_{\alpha\alpha}-1)\frac{\sum_{i\in\text{block}~\alpha}(N_i- N_\alpha)^2}{B_\alpha},
    \label{eq:neutral_block_2}
\end{equation}

where the effective self-inhibition
\begin{equation}
{A}^{\rm eff}_{\alpha \alpha} = A_{\alpha \alpha} - \frac{(A_{\alpha
    \alpha} -1)}{L_\alpha}
= \frac{(L_\a -1) A_{\a\a} +1}{L_\a}= \frac{1}{B_\alpha^*}
\label{eq:}
\end{equation}
is exactly the inverse of the equilibrium biomass for each block.  The
last term in Eq.~(\ref{eq:neutral_block_2}) corresponds to statistical fluctuations from the initial conditions that rapidly dissipate over
time as each block locally equilibrates, and add some noise but should
not significantly impact the relative probabilities of ending up in
different equilibria with random initial conditions, over most ranges
of the parameters.  Thus, we see that the effective dynamics even with
the unit self-inhibition is essentially identical to that of the block
monodominant model with neutral conditions within each block, with
the proper value of the effective self-inhibition matching the inverse biomass.

As an illustration, consider two blocks. Just as in the monodominant model, we might describe the dynamics in terms of a phase portrait where we follow the abundances of the two blocks $B_1$ and $B_2$. The basins of attraction for the two blocks will be separated by a separatrix in the $(B_1, B_2)$ plane given by the line:
\begin{equation}
    B_1 = \frac{D - A_{11}^{\text{eff}}}{D - A_{22}^{\text{eff}}}\,B_2.
\end{equation}

\begin{figure}[ht!]
    \centering
    \begin{subfigure}[b]{0.45\textwidth}
        \centering
        \includegraphics[width=\textwidth]{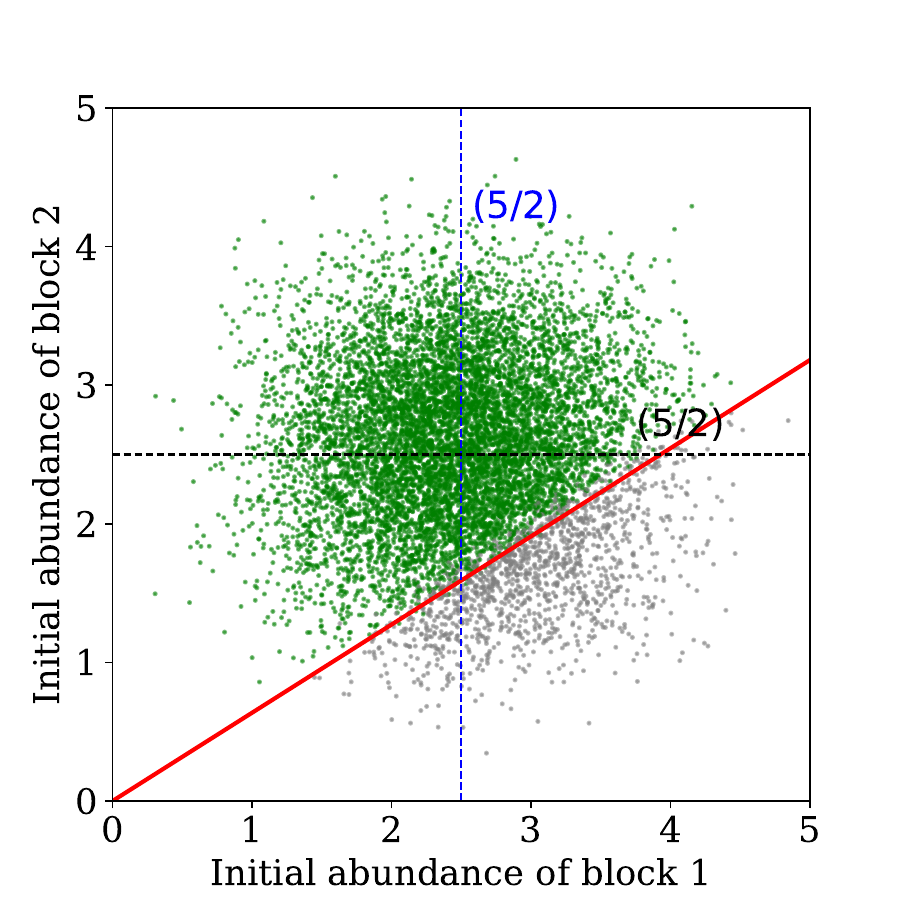}
        \caption{Equal block sizes ($L_1=L_2=5$) with different equilibrium biomass: the initial distribution is circular and centered at $B_1=B_2$, but the separatrix (red) is tilted.}
        \label{fig:block_1}
    \end{subfigure}
    \hfill
    \begin{subfigure}[b]{0.45\textwidth}
        \centering
        \includegraphics[width=\textwidth]{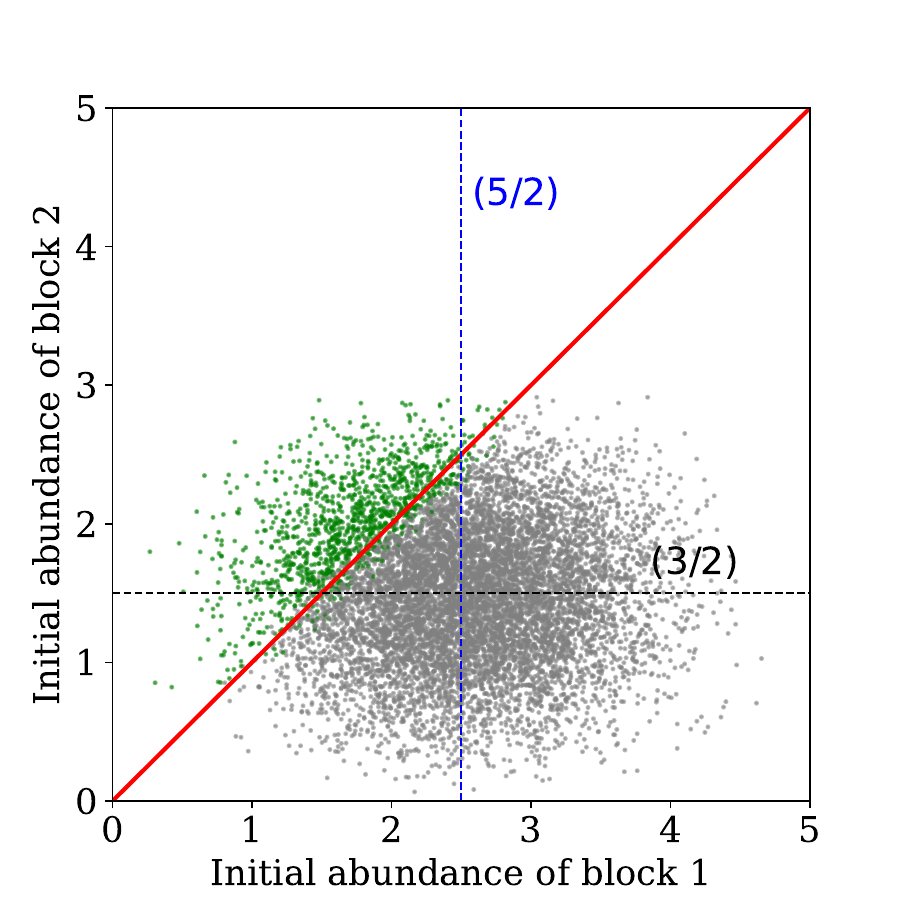}
        \caption{Different block sizes ($L_1 \neq L_2$) with equal
          equilibrium
          biomass: the separatrix is fixed at $B_1=B_2$, but the spread is stretched more for the larger block size, centered at $(L_1/2, L_2/2)$.}
        \label{fig:blocks_1}
    \end{subfigure}
    \vskip 10pt
    \caption{\justifying Effect of equilibrium biomass and block size on the likelihood. Green regions show outcomes where block 1 wins; gray regions where block 2 wins.}
\end{figure}

\begin{figure}[ht!]
    \centering
    \begin{subfigure}[b]{0.45\textwidth}
        \centering
        \includegraphics[width=\textwidth]{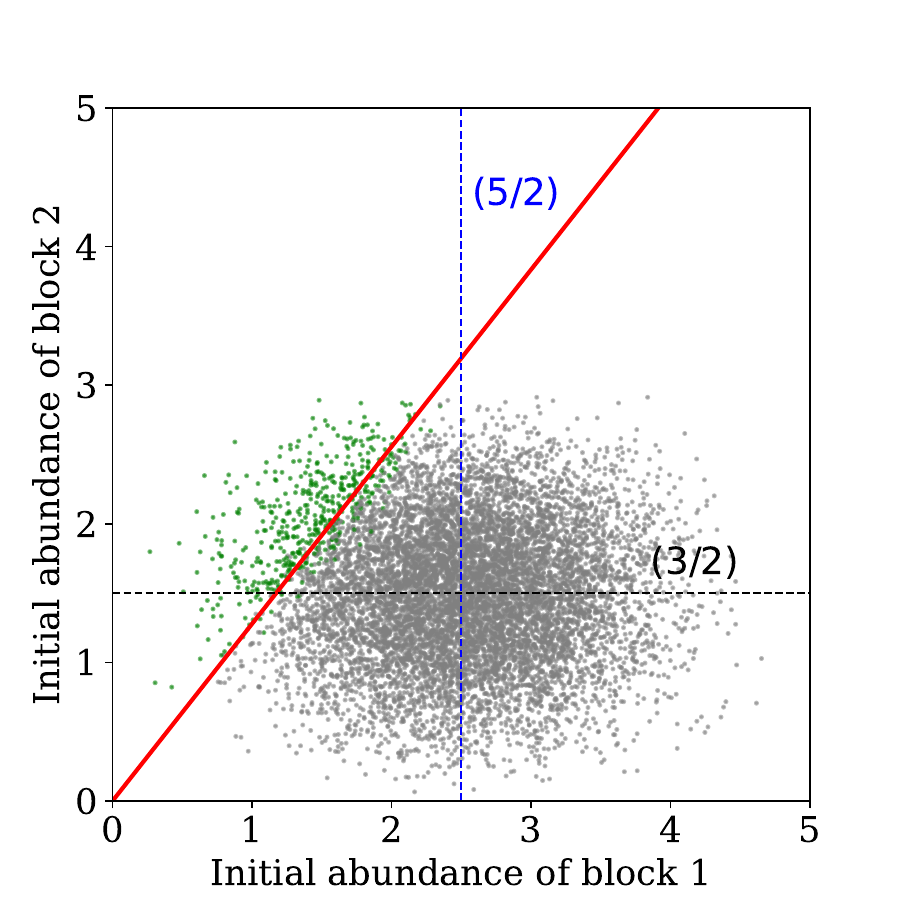}
        \caption{}
        \label{fig:block2}
    \end{subfigure}
    \hfill
    \begin{subfigure}[b]{0.45\textwidth}
        \centering
        \includegraphics[width=\textwidth]{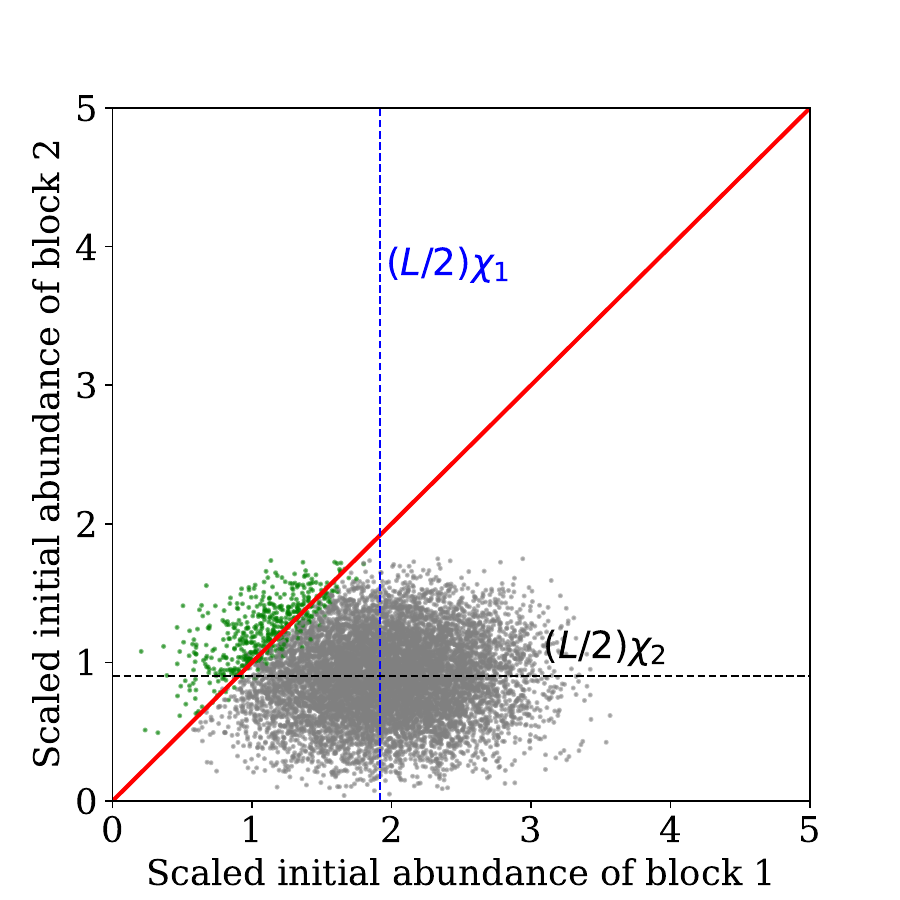}
        \caption{}
        \label{fig:block2s}
    \end{subfigure}
    \vskip 10pt
    \caption{\justifying
Two different ways of graphically depicting likelihoods
 of different blocks
  surviving in a case with 
different block sizes and different equilibrium biomasses.
  Scaling block abundances captures the effects of both
  equilibrium    biomass and block size
  in determining relative likelihoods.
Block with largest dressed initial biomass
      wins.
   Green regions: block 1 wins; gray: block
      2 wins. (a) Without scaling, the separatrix has slope
      $\chi_1/\chi_2$, with initial abundance centered at
      $(L_1/2,L_2/2)$, with $L_1 = 5$ and $L_2 = 3$. (b) After scaling, the separatrix becomes $X_1
      = X_2$,with dressed initial abundance centered at
      $(\chi_1L_1/2,\chi_2L_2/2)$.
    }
    \label{fig:blocks}
\end{figure}

Based on our argument above, we take the effective self-inhibition of
each block as the inverse of its equilibrium biomass. From
Fig.~\ref{fig:blocks_1} and Fig.~\ref{fig:block2} we see that this
works well.
If both blocks have the same size $L$
but different equilibrium biomass, the initial distribution is
Gaussian with a roughly circular spread with its center on the line
$B_1 = B_2$, while the separatrix is tilted towards the $x$-axis
compared to this line (Fig.~\ref{fig:block_1}). On the other hand, if
the blocks have equal equilibrium biomass but different sizes, the
separatrix coincides with $B_1 = B_2$, but the initial distribution
bivariate Gaussian has a spread which is both shifted and stretched
due to differences in block size (Fig.~\ref{fig:blocks_1}). This shows
that, in block models, both the equilibrium biomass and size of a
block affect its likelihood: equilibrium biomass tilts the separatrix,
while block size moves and streches a block's initial abundance
distribution.

Similar to our
treatment of the monodominant model, we can simplify
the analysis
by scaling the initial condition distribution in an
appropriate way while also avoiding the need to tilt the
separatrix. The key insight is that---when both the biomass and block
size vary (Fig.~\ref{fig:block2})---their combined effect can be
captured by scaling a block's total initial abundance by the factor
\begin{equation}
    \chi_\alpha = D - A_{\alpha \alpha}^{\text{eff}}.
\end{equation}

As shown earlier in this Appendix, for sufficiently large blocks, a
block's total  initial abundance can be well-approximated by a
Gaussian with a mean $\frac{L_\alpha}{2}$ and standard deviation $\sqrt{\frac{L_\alpha}{12}}$. The block with the largest dressed initial abundance is the winner (Fig.~\ref{fig:blocks}). Thus, the distribution of dressed initial condition $X_\a$ for block $\alpha$ also follows a Gaussian with
\begin{equation}
    X_\alpha \sim N\!\left(\frac{L_\alpha}{2}\chi_\alpha, \; \left(\sqrt{\tfrac{L_\alpha}{12}}\chi_\alpha \right)^2\right) \sim N\!\left(m_\a, \; s_\a^2\right).
\end{equation}

Now consider $N_B$ independent blocks with dressed initial conditions $X_1, X_2, \dots, X_{N_B}$, where
\begin{equation}
    X_\beta \sim \mathcal{N}(m_\beta, s_\beta^2), \qquad 
    m_\beta = \frac{L_\beta}{2}\chi_\beta, \quad 
    s_\beta = \sqrt{\frac{L_\beta}{12}}\,\chi_\beta.
\end{equation}

Following the same approach as in the monodominant many-species case, the probability that block $\alpha$ has the largest dressed initial condition $X_\a$ is
\begin{equation}
    p_\alpha = \mathbb{P}\!\big(X_\alpha > X_\beta \ \text{for all } \beta \neq \alpha\big).
\end{equation}

This can be expressed as the following integral
\begin{equation}
    p_\alpha = \int_{0}^{\infty} f_{X_\alpha}(z_\alpha)\,\prod_{\substack{\beta=1\\\beta\neq \alpha}}^{N_B} \mathbb{P}(X_\beta\le z_\alpha)\,dz_\alpha,
\end{equation}
where $f_{X_\alpha}$ is the pdf of $X_\alpha$. Explicitly, this reads as
\begin{equation}
    p_\alpha = \int_{0}^{\infty}
    \frac{1}{\sqrt{2\pi}\,s_\alpha}\exp\!\Big[-\frac{(z_\alpha-m_\alpha)^2}{2s_\alpha^2}\Big]\;
    \prod_{\substack{j=1\\j\neq \alpha}}^{N_B}
    \Bigg(
    \int_{0}^{z_\alpha}\frac{1}{\sqrt{2\pi}\,s_\beta}\exp\!\Big[-\frac{(u-m_\beta)^2}{2s_\beta^2}\Big]du
    \Bigg)\,dz_\alpha .
\end{equation}

\begin{figure}[t!]
    \centering
    \begin{subfigure}[b]{0.5\textwidth}
        \centering
        \includegraphics[width=\textwidth]{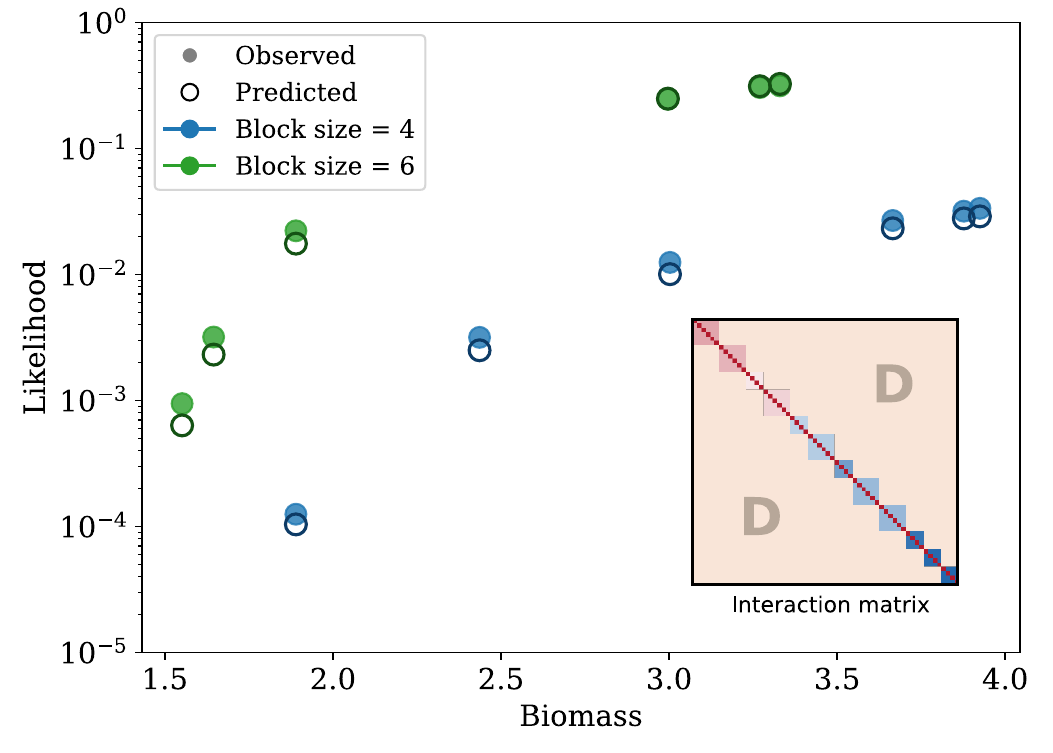}
        \caption{}
        \label{fig:many_dist}
    \end{subfigure}
    \hfill
    \begin{subfigure}[b]{0.4\textwidth}
        \centering
        \includegraphics[width=\textwidth]{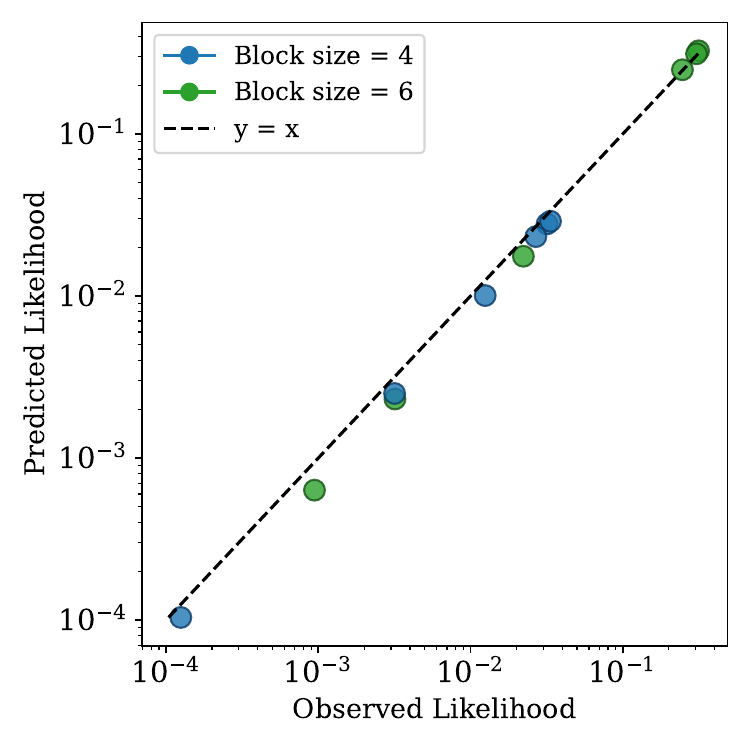}
        \caption{}
        \label{fig:many_comp}
    \end{subfigure}
    
    \caption{\justifying(a) Likelihood versus biomass for different block sizes, using an 
interaction matrix with 12 blocks (inset). Blocks were either of 
size 4 or 6, with randomly chosen self-inhibitions and inter-block 
inhibition $D = 1.1$. Some blocks had high likelihood despite lower 
biomass due to their larger size. Eq.~\eqref{eq:final_block} predicted 
likelihoods extremely well. (b) Comparison of observed likelihoods 
with predicted likelihoods for the same matrix as in (a).
}
    \label{fig:many}
\end{figure}

\begin{figure}[h!]
    \centering
    \includegraphics[width=0.4\textwidth]{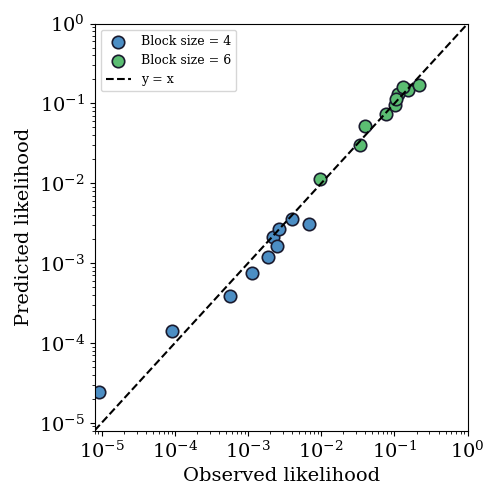}
    \caption{\justifying\textcolor{black}{Variability in inter-state inhibitions did not significantly affect the predictions of our block model. We constructed a matrix with $10$ blocks of size $4$ and $10$ blocks of size $6$, with $10$ biomass values linearly spaced from $3$ to $7$. We took the inter-state interaction from a gaussian distribution $\mathcal{N}(1.1, 0.1)$. We used $D = \langle A_{ij} \rangle =1.1$ in Eq.~\eqref{eq:final_block} to compute the predicted likelihoods. Our predictions agreed well with the observed likelihoods across both block sizes, with only minor deviations .}}
    \label{fig:var_block_interaction}
\end{figure}

Equivalently,
\begin{equation}\label{eq:main_block}
p_\alpha \;=\; \int_{0}^{\infty}\frac{1}{s_\alpha}\,f\!\left(\frac{z_\alpha-m_\alpha}{s_\alpha}\right)
\prod_{\substack{j=1\\j\neq \alpha}}^{N_B}
\Bigg(
\int_{0}^{z_\alpha}\frac{1}{\sqrt{2\pi}\,s_\beta}\exp\!\Big[-\frac{(u-m_\beta)^2}{2s_\beta^2}\Big]du
\Bigg)dz_\alpha
\end{equation}
where $f(y)=\tfrac{1}{\sqrt{2\pi}}e^{-y^2/2}$ is the standard normal
density. Note that the lower limit of the inner integrals in
Eq.~\eqref{eq:main_block} can be replaced by $-\infty$ (for blocks of size $L\sim 5$ or larger)
without significantly affecting the quantitative contribution from these
integrals. This is because virtually all of their mass is concentrated
in the positive space of abundances. Thus, for convenience we may
replace the inner integrals with the CDF of the Gaussian distribution,
using which we get

\begin{equation}\label{eq:final_block}
p_\alpha = \int_{0}^\infty \frac{1}{s_\alpha}f\left( \frac{z_\alpha-m_\alpha}{s_\alpha}\right) \prod_{\substack{\beta=1\\\beta\neq \alpha}}^{N_B} \Phi \left( \frac{z_\alpha-m_\beta}{s_\beta}\right) dz_\alpha,
\end{equation}
which is Eq.~\eqref{eq:block_likelihood} in the main text. Here 

\begin{equation}
    \Phi(y) = \int_{-\infty}^{y}\frac{1}{\sqrt{2\pi}}e^{-u^2/2}du
\end{equation}
is the Gaussian CDF. 
In Fig.~\ref{fig:many}b--c, we show that the prediction from Eq.~\eqref{eq:final_block} agrees well with the observed likelihoods for an example block matrix shown in Fig.~\ref{fig:many}a. This matrix has 12 blocks, each of size either 4 or 6 (chosen randomly). We set the inter-block inhibition $D=1.1$ and all diagonal entries (representing species' self-inhibitions) to 1. To assign within-block self-inhibitions $A_{\a\a}$ for each block $\a$, we randomly picked a biomass values uniformly randomly between 1.5 and 4. We then used Eq.~(\ref{eq:block_biomass}) to compute the corresponding within-block inhibition $A_{\alpha\alpha}$ to get that equilibrium biomass. We then computed the likelihoods for each block numerically using simulations (Fig.~\ref{fig:many}a, closed circles) and compared them with the predicted likelihoods from Eq.~\eqref{eq:final_block} (Fig.~\ref{fig:many}a, open circles). \textcolor{black}{ When the inter-state inhibition in a block-structured matrix was not constant, as in the random matrix case, we used  $D = \langle A_{ij} \rangle$ in Eq.~\eqref{eq:final_block} to predict the likelihood of the blocks (Fig.~\ref{fig:var_block_interaction}).} Overall, our formula captures the likelihoods well, with some minor deviations at low likelihoods, likely due to numerical errors.
Replacing our Gaussian approximation with the exact Irwin–Hall distribution for block initial conditions yields very similar results. Taken together, we have obtained an analytic formula for the likelihood of a block state, which is analogous to the many-species monodominant case. This formula in Eq.~\eqref{eq:final_block} accounts for blocks (states) with multiple species and Gaussian initial distributions. Each block's probability to win depends on both
its biomass and its block size, whereas in the monodominant model it
depends only on biomass.

\FloatBarrier

\section{Overlaps between states
\textcolor{black}{ and a simple statistical model of states}
 in ecosystems with random disordered interaction matrices}
\label{app:overlappingStates}

\begin{figure}[ht!]
    \centering
    \includegraphics[width=0.7\textwidth]{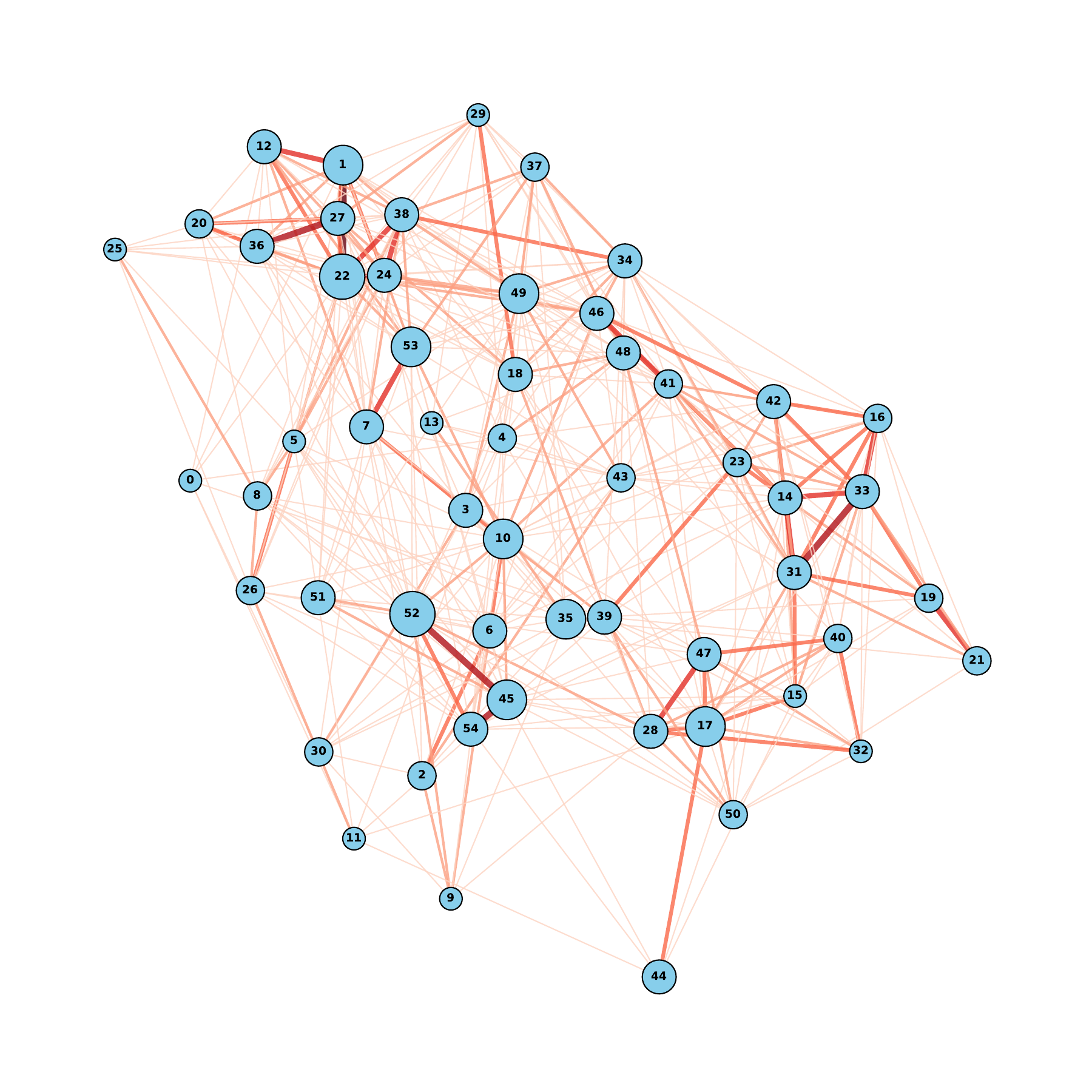} 
    \caption{\justifying Each node represents a state and each edge shows that two states share common species. Thickness of the edge shows the number of common species between the states. The overlapping goes from 1 to 6 species.  The node index follows the order of likelihood.The size of the nodes shows the diversity of the state,which goes from 4 to 8 species. We can see that states overlap with each other, so the random matrix does not form a perfect block structure.}
    \label{fig:network}
\end{figure}


\textcolor{black}{ In this Appendix we describe the overlaps between states in these random disordered ecosystems, and describe a simple statistical model that captures some basic aspects of the number and size of the individual block states in these models}

When we predict the likelihood of different states from a random matrix, we apply the theory of block-structured matrices developed in Appendix~\ref{app:blockCalc}. In a perfect block-structured matrix, each species belongs to only one state. As we show in Fig.~\ref{fig:fig2}a, the theoretical motivation for treating states as blocks comes from the observation that species coexisting within the same stable state interact much more weakly with each other than expected from the overall interaction distribution. This reduced intra-state inhibition suggests that species within each state should form cohesive, block-like units.

In  
 random matrices, we observe a richer structure where states
naturally overlap through shared species. We can visualize this by
constructing a network where each state represents a node, and we
connect two nodes with an edge if they share at least one common
species (Fig.~\ref{fig:network}). The thickness of each edge indicates
the number of species shared between connected states. This network
reveals the intricate interconnectedness among states, with many
states sharing species across multiple connections. Rather than the
discrete blocks described in Appendix~\ref{app:blockCalc}, we find a
complex overlapping structure that reflects the natural complexity of
random disordered ecosystems.

One direction in which the analysis of this paper could be extended is
by taking further account of the detailed structure of individual
blocks and the connectivity between blocks.  This could likely lead to
a more refined statistical model.  As a first step in this direction,
and as a way of understanding the appearance of blocks, we 
\textcolor{black}{ outline here a simple statistical model based on $k$-cliques in random graphs that seems to capture the structure of the size and number of block/states for the random ecosystems considered here}

\textcolor{black}{ We begin by noting}
  that
even though in the random interaction matrix each interaction is
chosen independently according to a given mean and variance, the
combinations of species that persist in a stable block in general have
smaller inter-species interactions, as illustrated
 in Fig.~\ref{fig:fig2}.  Thus, to estimate for example the number
 of  stable blocks of size $k$ in a given system of size $S$, we can
make a rough model in which a block is persistent when all
inter-species interactions within the block lie below some specific
value $a$.  If we denote the probability that this occurs for a given
interaction under the normal distribution on $A_{i j}$ by $p$, then
the probability that any specific subset of $k$ species has all
inter-species interactions less than or equal to $a$ is $p^{k (k
  -1)/2}\sim p^{k^2/2}$, and the number of such subsets is $S!/((S -
k)!  k!)\sim S^k$, where the asymptotics are at large $k, S$, with $k
\ll S$.  This is essentially
the problem of finding $k$-cliques in a random graph of size $S$.  

In this simple model, the
expected number of blocks of size $k$ is given by
\textcolor{black}{
\begin{equation}
N (S, k) \cong\binom{S}{k} p^{\binom{k}{2}} (1 - p^k)^{S - k}\,,
\label{eq:}
\end{equation}
}
where the last factor comes from the condition that the block is uninvadable, hence is not a sub-block of a larger feasible stable block. \textcolor{black}{Each of the $(S-k)$ species outside the block can be added to it only if all $k$ of its interactions with the block lie below $a$, which occurs with probability $p^k$.} The largest clique/block in this context has size $k \sim 2 \ln S/\ln (1/p)$, which can be determined by finding the largest $k$ such that the expected number of blocks of size $k$ is ${\cal O}$(1);
\textcolor{black}{ this is a classic result in graph theory
\cite{bollobas1976cliques}.
Similarly, a saddle point analysis shows that the typical
maximal clique/block has size
$k^* \sim  \ln S/\ln (1/p)$, and the number of maximal cliques/blocks goes as $N (S) \sim S^{\frac{\ln S}{2 \ln (1/p)}}$.}
This argument explains why the stable states that we find in random
matrices of size, e.g., $S = 100$ are relatively small ($k \sim 6$).

\begin{figure}[ht!]
    \centering
    \includegraphics[width=1\textwidth]{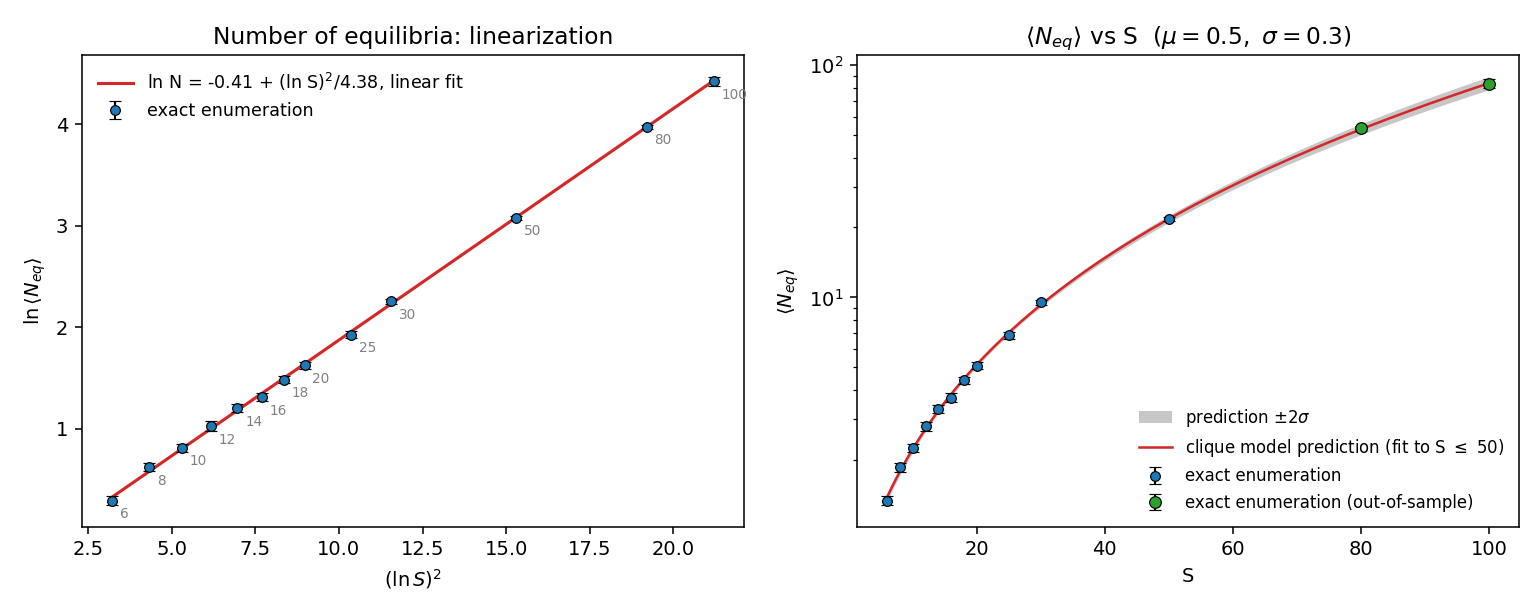} 
    \caption{\justifying \textcolor{black}{ The clique model for equilibrium state
  distribution suggests a number of equilibria going as $N_{\rm
    eq}\sim S^{\frac{\ln S}{2 c}}$, where $c$ is a constant depending
  on the matrix ensemble.  This matches explicit
  computation very closely.  Left: a linear fit between $\ln N_{\rm
    eq}$ and $(\ln S)^2$ matches  data for number
of equilibria with $S$ from 4 to 100, explicitly calculated exactly
for a sample of matrices using a pruned tree search; Right: a fit with numerical data
from $S \leq 50$ (80 matrices each at lower values, 100 matrices
at $S = 50$ gives a precise prediction of values at $S = 80, 100$
(estimated  from exact enumeration over 50, 10 matrices respectively).}}
    \label{fig:block-number-matching}
\end{figure}

\textcolor{black}{
This simple model for the number of maximal blocks matches very well
with explicit computations; see Figure~\ref{fig:block-number-matching}.
Because in the parameter ranges we are considering here, the number
and size of  stable equilibrium states is relatively small, a simple
pruned tree algorithm can enumerate all possible stable states for
e.g. $S = 100, \mu = 0.5, \sigma = 0.3,$ where across an ensemble of
10 matrices the number of distinct equilibrium states is $N_{\rm
  eq}\equiv  83 \pm 4$.
The pruning can be done efficiently using the fact that if a subset $S'$ has an
unstable $A$ matrix (negative eigenvalue), then there cannot be any
stable subset $S''$ where $S' \subset S''$.
Note that for this kind of system the standard 
ecological method of
predicting the number of states based on a limited
probabilistic sampling 
using, e.g., the Chao1 estimator \cite{chao1984nonparametric,chao2016species} based on the Good-Turing
discovery rate \cite{good1953population,chao2009sufficient,chao2012coverage} may not be trustworthy, particularly when the sampling
only touches  the large-probability tail of the distribution.
}

\textcolor{black}{
Using similar logic,
a} similar estimate can be given for, e.g., the number of pairs of
cliques of size $k, k'$ that overlap in $s$ species.
It would be interesting to develop this kind of model further to
understand in more detail the expected structure of
\textcolor{black}{interactions between}
 the stable states
in this kind of random system and use that structure to make more
precise predictions about domain of attraction sizes, but we leave
further analysis in that direction for future work.

\section{Method to reveal emergent block structure in a random interaction matrix}
\label{app:blockGrouping}

The same species can appear in multiple stable states for any given
random disordered interaction matrix, and as we show in
Appendix~\ref{app:overlappingStates}, the interaction matrix does not
form a perfect block structure as assumed in our theory
(Appendix~\ref{app:blockCalc}) due to overlapping states. In this
appendix, we show that nevertheless, there is an emergent block-like
organization when we reorder species in specific ways. Particularly,
here we detail a possible systematic approach---which we use---that
focuses on states that do not overlap.

Note that in this analysis, we use blocks that are identified as
stable solutions from simulations with random initial conditions.  An
alternative approach that would be worth exploring is to try to use
statistical properties of the matrix to identify stable blocks using
{\it a
priori} rubrics such as finding subsets of species where all
inter-species interactions within the subset are relatively weak.  As
discussed in Appendix~\ref{app:overlappingStates}, this is
analogous to the problem of identifying $k$-cliques in a random graph,
which is a computationally difficult, NP-hard problem.
The problem of identifying all stable states in a given random system is
likely at least this hard.
As discussed also in \cite{taylor2024structure}, the simulation approach,
which samples stable states with a given probability distribution
$p_s$ is much more effective at computing things like ensemble
averages over the set of states with that probability weighting than
over the full ensemble with equal weighting, which is computationally
much more difficult and may miss low-probability states without
exponential effort. The simulation approach that samples with distribution $p_s$ is also
more compatible with an experimental approach that would give states
with a similar distribution.

\begin{figure}[ht!]
    \centering
    \begin{subfigure}[b]{0.45\textwidth}
        \centering
        \includegraphics[width=\textwidth]{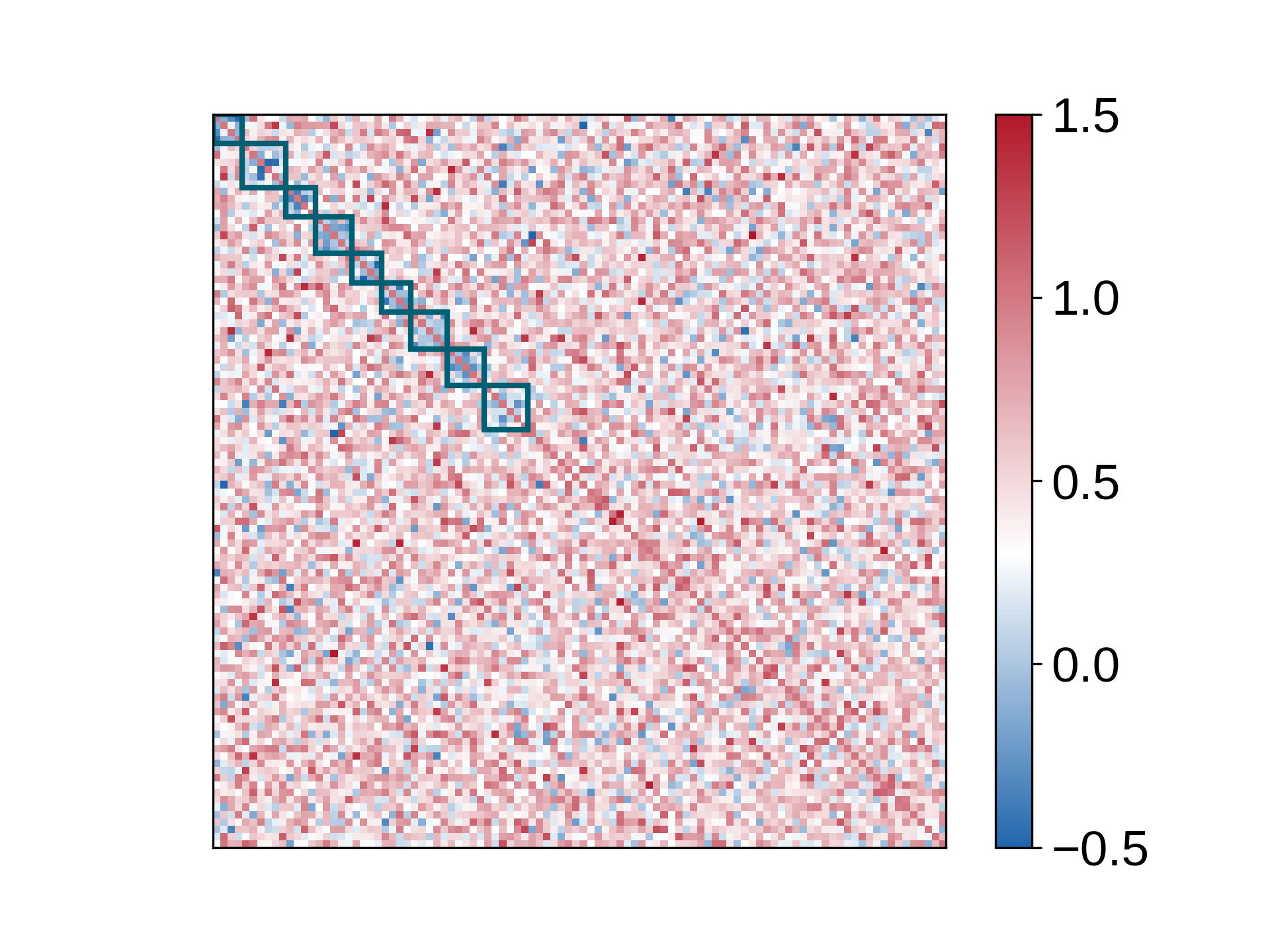}
        \caption{}
        \label{fig:block_like_full}
    \end{subfigure}
    \hfill
    \begin{subfigure}[b]{0.45\textwidth}
        \centering
        \includegraphics[width=\textwidth]{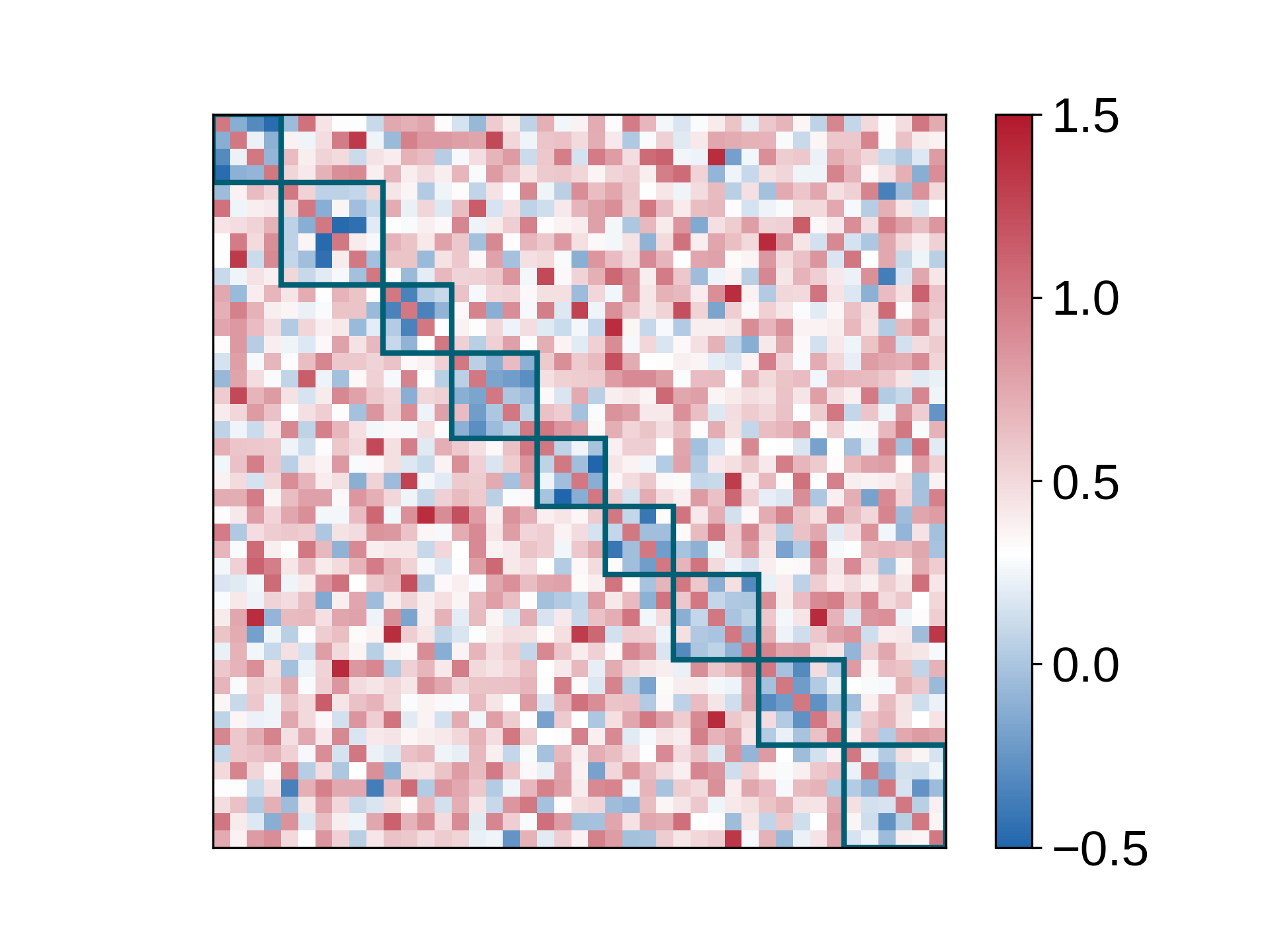}
        \caption{}
        \label{fig:block_like_zoom}
    \end{subfigure}
    \caption{\justifying Emergent block structure revealed by selecting non-overlapping states. (a) Full interaction matrix. The chosen non-overlapping states are placed first, followed by the remaining species. (b) Subset of the matrix showing only the chosen states.}
    \label{fig:block_like}
\end{figure}

\begin{figure}[ht!]
    \centering
    \includegraphics[width=0.6\textwidth]{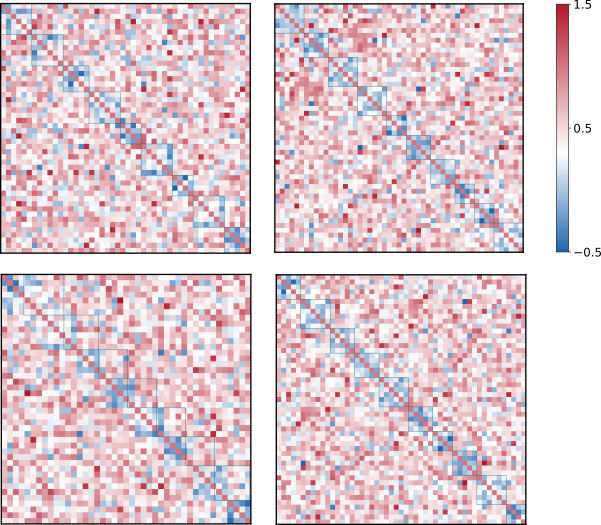}
    \vskip 10pt
    \caption{\justifying Robustness of block structure to reordering
      species in different ways. Shown are reordered subsets of the
      same interaction matrix. Each example shows a different
      reordered sub-matrix obtained by choosing different
      non-overlapping sets of states (see text in
      Appendix~\ref{app:blockGrouping}). 
      Although the species and states selected in each submatrix vary, we always observe that a clear block structure emerges when we cluster species from the same state together—at least in the submatrices containing only species found exclusively in one state.    }
    \label{fig:block}
\end{figure}

We begin by ordering all states in descending order of their likelihood. We then select all species from the most likely state and reorder the matrix to have rows and columns corresponding to them appear first. Then we choose species in the next most likely state. We check whether it shares any species with the previously selected state. If there is even one species that overlaps, we discard this state and move to the next one. If there is no overlap of species between states, we then move the rows and columns corresponding to species in this state right after those belonging to the first state. In this way we iteratively build a submatrix of the full interaction matrix until we have considered all states. This submatrix contains a set of states with no overlaps in species. This submatrix of the full interaction matrix reveals a clear block structure (Fig.~\ref{fig:block_like}a--b).

This method emphasizes high-likelihood states and reveals a clear emergent block structure in at least part of the full interaction matrix. However, this emergent block structure appears not just for highly likely states but also those with low likelihood. We demonstrate this by selecting states uniformly randomly rather than following descending order of likelihood as in the previous scheme. That is, we begin by picking a random state regardless of its likelihood and proceed iteratively. This procedure yields different sets of non-overlapping states each time, yet a clear block structure emerges in all cases (Fig.~\ref{fig:block}). Importantly, we do not alter any species interactions---we only reorganize their ordering within the matrix to reveal the latent structure.

These results show that block-like organization represents a fundamental property of our random ecological matrices, not merely an artifact of selecting high-likelihood states. The consistent emergence of block patterns across different selection schemes validates our theoretical approach of treating states as effective blocks in competition with each other. 

\FloatBarrier

{\color{black}

\section{Accounting for overlaps in the block model}
\label{app:blockOverlaps}

In the block models analyzed in Appendix~\ref{app:blockCalc}, each
species belongs to exactly one block. In random disordered matrices,
however, states overlap through shared species
(Appendix~\ref{app:overlappingStates}). 
Here, we study
our block model to
systematically vary the extent of overlap between blocks while keeping
the number of states fixed. We then use this model to show that when
the extent of overlap is sufficiently high, the true block size is a
worse predictor of block likelihoods than the mean block size. We also
find that the random matrices we studied in the main text indeed have
sufficiently large overlap such that using the mean block size for all
blocks would serve as a better predictor. This justifies the
mean-field approximation of the block model to predict likelihoods in
random disordered matrices, which we use in the main text
(Fig.~\ref{fig:fig3}) and in
Appendix~\ref{app:likelihoodCalcRandMatrix}.

To build intuition, consider building an interaction matrix for an ecosystem with 20 blocks or states. Say we want $10$ blocks of size $6$ and $10$ blocks of size $4$. The total number of species slots required to fill these blocks is $10 \times 6 + 10 \times 4 = 100$. In a perfect block matrix, all $100$ slots are filled by $100$ distinct species, each species appearing in exactly 1 state. On the other hand, if there are only $50$ species in the pool to form the same $20$ blocks, each species must participate in $2$ states on average. We use this idea to define the \emph{extent of overlap} $m$ as the average number of stable states in which a species participates. Note that an extent of overlap equal to $m=1$ corresponds to the perfect block structure with no overlaps we studied before. Larger values $m>1$ introduce progressively more overlap between states.

\textit{Constructing block models with controlled overlap.}---To verify that our results are robust to the specific way in which we introduce overlaps, we used two different methods to construct overlapping block matrices with a given mean extent of overlap $m$, for a fixed set of blocks (number of stable states) with fixed sizes and biomass values. We systematically increased $m$ and studied the prediction error of two block models (as described in Appendix~\ref{app:blockCalc}): one where we used the actual block size of each respective block, and the other where we used the average block size for all blocks.

Both methods rely on the same key idea to introduce overlaps. Namely, if the number of species in the pool is smaller than the total number of species slots $\sum_\alpha L_\alpha$, where $L_\alpha$ is the size of block $\alpha$, some species must appear in more than one block, since there are not enough distinct species to fill all slots without repetition. The fewer the species in the pool, the more frequently each species must be reused across blocks, and the greater the overlap between states. We thus used a reduced species pool of size $\sum_\alpha L_\alpha / m$, so that on average each species fills $m$ slots, giving a mean extent of overlap equal to $m$. In both methods, we assigned within-block interactions in the same way. For species belonging to block $\alpha$, we set the within-block interaction strength to $(\frac{L_\alpha}{B_\alpha^*}-1)/ ({L_\alpha - 1})$, which is weaker than $D$, ensuring that species within the block could coexist with required biomass $B_\alpha^*$ (Eq.~\eqref{eq:block_biomass}). We set all diagonal elements to $1$ and all off-diagonal elements outside any block to a constant $D$. If a pair of species appeared together in more than one block, we set their interaction to the average of the values assigned by those blocks. In the first method, we ensured that all blocks had exactly the same extent of overlap (``constant overlap''), while in the second, we allowed the extent of overlap to vary, while controlling the mean extent of overlap (``variable overlap'').

\textbf{Method 1 (Constant overlap):} We started with a pool of $\lfloor \sum_\alpha L_\alpha / m \rfloor$ species. We went through each block sequentially, and for each block $\alpha$, we randomly picked $L_\alpha$ species from the pool. Once a species was picked $m$ times, we removed it from the pool. If $\sum_\alpha L_\alpha$ is not divisible by $m$, we could not fill the remaining $r = \sum_\alpha L_\alpha \mod m$ slots from the pool, so we drew species again from the full set of already-used species. In such cases, some species may be used more or fewer than $m$ times, but the deviation was negligible and the mean extent of overlap remained effectively equal to $m$.

\textbf{Method 2 (variable overlap):} We filled each block by randomly drawing $L_\alpha$ species from the reduced pool, without any constraint on how many times a species could be selected. Since the pool was smaller than the total number of slots, we necessarily selected species multiple times, causing them to participate in multiple states. This resulted in different species having different extents of overlap, with a mean extent of overlap equal to $m$.

For random matrices, we observe that the extent of overlap of a species does not depend on the diversity of the states it is present in. Therefore, in both methods, we constructed our overlapping block matrices such that the extent of overlap of a species is independent of the block size it participates in.

\begin{figure}[h!]
    \centering
    \begin{subfigure}[b]{0.45\textwidth}
        \centering
        \includegraphics[width=\textwidth]{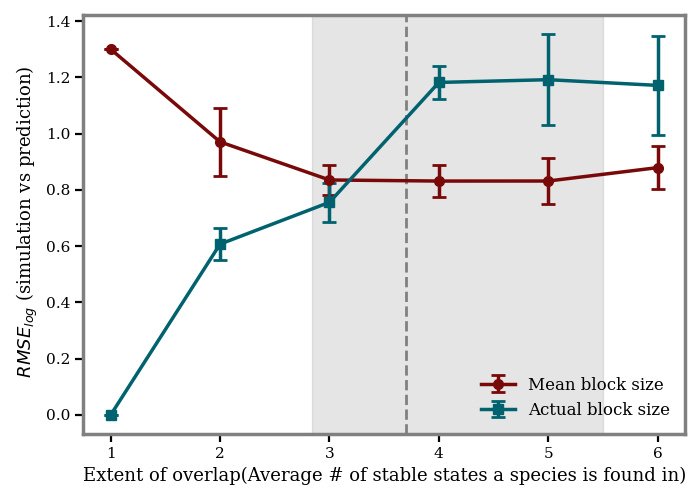}
        \caption{}
    \end{subfigure}
    \hfill
    \begin{subfigure}[b]{0.45\textwidth}
        \centering
        \includegraphics[width=\textwidth]{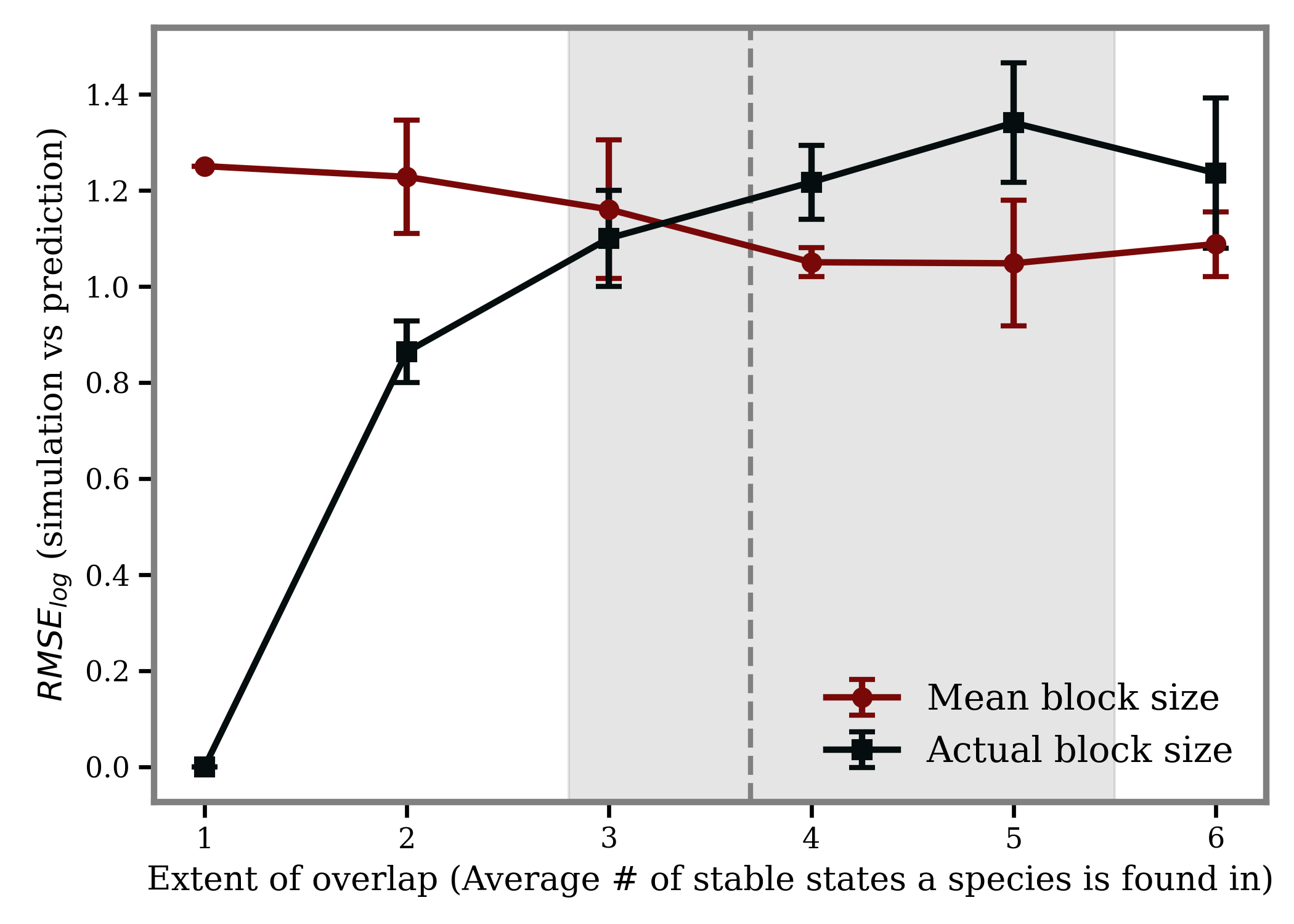}
        \caption{}
    \end{subfigure}
    \caption{\justifying\textcolor{black}{ Using mean block size for
        likelihood prediction gives better results in the presence of
        overlapping states. As overlap increases, the prediction error
        using actual block sizes grows, while the error using the mean
        block size decreases. Beyond an extent of overlap of
        approximately 3, the mean block size gives better
        predictions. (a) Fixed overlap: each species participates in a fixed number of blocks. (b) Variable overlap: Different species participates in different number of blocks.  The gray shaded region indicates the range of
        overlap corresponding to the random interaction matrix studied
        in the main text; the dashed line marks the measured extent of
        overlap for that matrix. Error bars indicate SEM of
        multiple realizations of the interaction matrix at fixed
        overlap structure.}}
        \label{fig:overlap_crossover}
\end{figure}

We compare two prediction schemes across varying extents of overlap: using each
state's actual diversity $L_\alpha$, and using the mean diversity $L =
\langle L_\alpha \rangle$ for all states. At zero overlap ($m= 1$), we find that using the actual block size gives better predictions, as
expected from the exact block theory. However, as the extent of
overlap increases, the error for the actual-block-size prediction
grows, while the error for the mean-block-size prediction
decreases. Beyond an extent of overlap of roughly $m=3$, the mean
block size consistently outperforms the actual block size in predictability
(Fig.~\ref{fig:overlap_crossover}). We also measured the extent of overlap metric for the random interaction
matrices we studied in the main text and found that they fall within this high-overlap regime (grey regions in Fig.~\ref{fig:overlap_crossover}).

\textit{A single overlap between blocks increases both their likelihoods.}---To understand why overlaps change predictions from the block model, here we study how introducing overlaps changes block likelihoods. For this, we
first consider the simplest case: a perfect block-structured matrix with no overlaps, into
which we introduce a single overlap between exactly two blocks by having them
share a single species. We find that the blocks participating in the
overlap now have higher likelihood compared to when these blocks did not overlap
(Fig.~\ref{fig:overlap_mutualism}). Consequently, the likelihoods of all other non-overlapping blocks decreases a bit to compensate. This is simply because the total likelihood is
conserved ($\sum_\alpha p_\alpha = 1$). This effect shows that by sharing overlapping species, blocks can effectively boost their likelihood.

\begin{figure}[h!]
    \centering
    \includegraphics[width=0.45\textwidth]{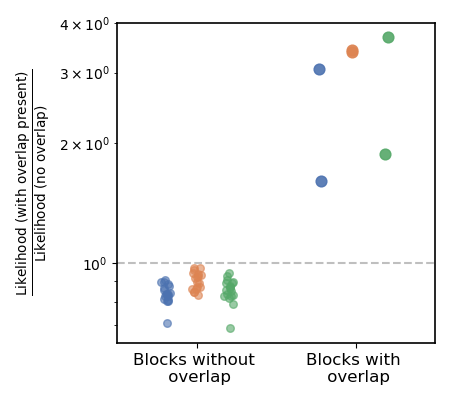}
    \caption{\justifying\textcolor{black}{Overlapping blocks experience
        a mutualistic increase in likelihood. The $y$-axis shows the
        ratio of likelihood with overlap present to that without
        overlap. This ratio exceeds 1 for blocks participating in the
        overlap and is less than 1 for all other blocks. Different
        colors represent different matrices where two blocks are
        randomly chosen to overlap.}}
    \label{fig:overlap_mutualism}
\end{figure}

To gain further insight into this mathematically, we examine the block-level dynamics in the
presence of overlap. When blocks $\alpha$ and $\beta$ share species,
the effective dynamics of block $\alpha$ acquires an additional term
compared to Eq.~\eqref{eq:neutral_block_2}:
\begin{equation}
    \frac{\dot{B}_\alpha}{B_\alpha}
    \approx 1 + (D - A_{\alpha\alpha}^{\mathrm{eff}}) B_\alpha
    - D \sum_\beta B_\beta
    + \frac{1}{B_\alpha}
    \sum_{\beta \in \Gamma(\alpha)}
    \sum_{j \in \alpha \cap \beta}
    (D - A_{\beta\beta}) N_j (B_\beta - N_j),
    \label{eq:overlap_dynamics}
\end{equation}
where $\Gamma(\alpha)$ denotes the set of blocks that overlap with
block $\alpha$, and the inner sum runs over species $j$ shared between
blocks $\alpha$ and $\beta$. The last term is the additional
contribution arising from overlap. Absorbing it into the effective
self-inhibition yields a modified value
\begin{equation}
    \tilde{A}_\alpha^{\mathrm{eff}}
    = A_\alpha^{\mathrm{eff}}
    - \frac{1}{B_\alpha^2}
    \sum_{\beta \in \Gamma(\alpha)}
    \sum_{j \in \alpha \cap \beta}
    (D - A_{\beta\beta}) N_j (B_\beta - N_j).
    \label{eq:overlap_selfinhib}
\end{equation}
Since $D > A_{\beta\beta}$ and $B_\beta > N_j$ for each shared
species, the correction term is positive, making
$\tilde{A}_\alpha^{\mathrm{eff}} < A_\alpha^{\mathrm{eff}}$. Blocks
participating in overlap thus have a reduced effective self-inhibition,
which increases their likelihood.

\textit{When overlaps are randomly distributed and widespread, smaller blocks typically benefit from overlap at the expense of bigger blocks.}---We next consider
the case where every block participates in overlap by randomly pairing
blocks and having each pair share one species. We find that smaller
blocks show a net increase in likelihood, while larger blocks show a
net decrease (Fig.~\ref{fig:overlap_size}(a)). This asymmetry can be
understood from Eq.~\eqref{eq:overlap_selfinhib}. Consider two blocks
$\alpha$ and $\beta$ with $B_\beta > B_\alpha$ that share a single species with
abundance $N_k$. The reductions in effective self-inhibition are
\begin{equation}
    \Delta A_{\alpha\alpha} = \frac{1}{B_\alpha^2}(D - A_{\beta\beta}) N_k (B_\beta - N_k),
    \qquad
    \Delta A_{\beta\beta} = \frac{1}{B_\beta^2}(D - A_{\alpha\alpha}) N_k (B_\alpha - N_k).
\end{equation}
Since $B_\alpha < B_\beta$, the smaller block has a smaller denominator
$B_\alpha^2$ and a larger numerator factor $(B_\beta - N_k)$, giving $\Delta
A_{\alpha\alpha} > \Delta A_{\beta\beta}$: the smaller block experiences a larger
reduction in self-inhibition and thus gains more benefit from the
overlap. For equal-sized blocks sharing a small overlap ($N_k < B$),
the benefit scales as $\Delta A \sim 1/B$, confirming that smaller
blocks gain disproportionately. Combining these results, the benefit
from overlap follows the ordering
\begin{equation}
    \text{small}\to\text{large} > \text{small}\to\text{small} > \text{large}\to\text{large} >
    \text{large}\to\text{small},
\end{equation}
where ``small$\to$large'' denotes the benefit received by a small block
from overlapping with a large block. 

The above explanation is for the case where each block overlaps with one other block irrespective of block size. As discussed earlier, in a random ecosystem, the extent of overlap of a species is independent of block size. Furthermore, we observe that in our random disordered matrices, larger blocks overlap with more blocks compared to smaller blocks, presumably because they contain more species. After including this effect, we still find that smaller blocks become more likely, while larger blocks become less likely (Fig.~\ref{fig:overlap_size}(b)).

\begin{figure}[h!]
    \centering
    \begin{subfigure}[b]{0.45\textwidth}
        \centering
        \includegraphics[width=\textwidth]{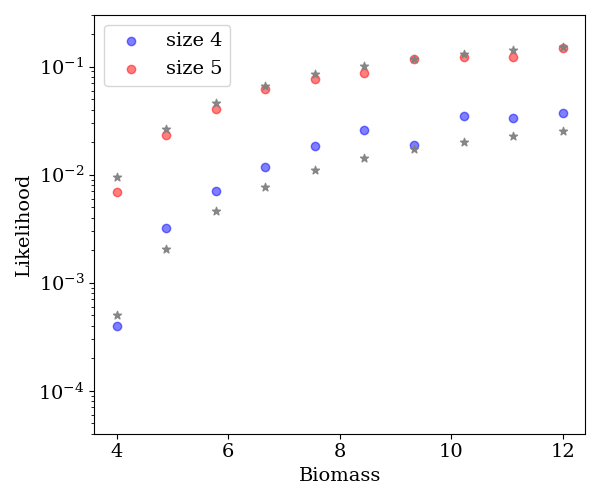}
        \caption{}
    \end{subfigure}
    \hfill
    \begin{subfigure}[b]{0.45\textwidth}
        \centering
        \includegraphics[width=\textwidth]{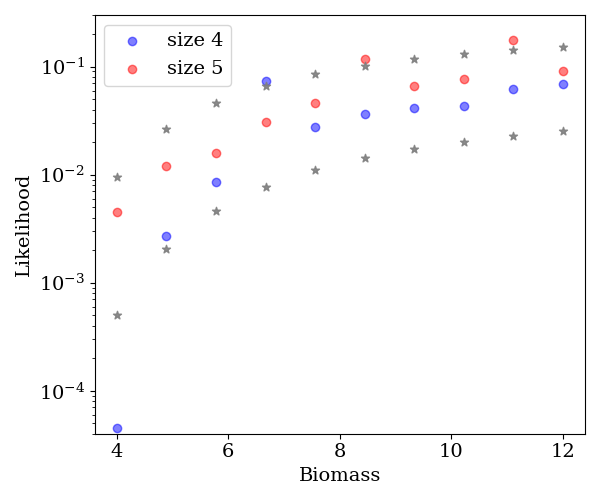}
        \caption{}
    \end{subfigure}
    \caption{\justifying\textcolor{black}{Overlap generally increases the likelihood of smaller blocks and decreases the likelihood of larger blocks. (a) We randomly pair each block with one other block and let them share a single species. Stars and circles represent likelihoods without overlap and with overlap, respectively. For each biomass value, two stars correspond to the two block sizes with the same biomass. The upper star represents the larger block, while the lower star represents the smaller block. For most states, smaller blocks (blue) show an increase in likelihood, while larger blocks (red) show a decrease in likelihood. (b) We pair each block of size 5 with two other blocks and let it share one species with each block, resulting in two distinct shared species, while blocks of size 4 share species with only one block. Even though the larger block shares more species, for most states smaller blocks (blue) show an increase in likelihood, while larger blocks (red) show a decrease in likelihood. In both (a) and (b), we use the same set of blocks, consisting of 10 blocks of size 4 and 10 blocks of size 5, with the same biomass values.}}
        \label{fig:overlap_size}
\end{figure}


\textit{Connection to the mean-field approximation.}---The
disproportionate benefit to smaller blocks explains why the mean block
size produces better predictions than the actual block size in the
presence of widespread overlaps, at least for the ways of introducing overlaps that we tried. Such widespread and randomly distributed overlaps (which are equally likely for states roughly independent of their size/diversity) effectively change the likelihoods of small and large blocks closer to each other. In fact such an effect tends to make block likelihoods come closer to the likelihood they would have
if they all had the mean size: small blocks are boosted in likelihoods while large blocks
are suppressed. Using the actual $L_\alpha$ for each state in the block model therefore misses this
redistribution. The mean-field approximation $L_\alpha
\to L$ is a simple and implicit way to capture this effect. This is why, we believe, it produces
better predictions despite being a cruder description of the
individual states. 

} 

\FloatBarrier

\section{Calculation of likelihood in case of random interaction matrices}
\label{app:likelihoodCalcRandMatrix}

We saw in Appendix~\ref{app:blockGrouping} that random disordered
interspecies interaction matrices exhibit partial block-like
organization. In perfectly block-structured matrices, we can calculate
likelihood using only the steady-state biomass and block size, as
developed in Appendix~\ref{app:blockCalc}.
In this appendix, we apply this framework to random matrices,
elaborating on the methodology summarized in the main text, as well as
considering alternative approaches.

\begin{figure}[ht!]
    \centering
    \begin{subfigure}[b]{0.45\textwidth}
        \centering
        \includegraphics[width=\textwidth]{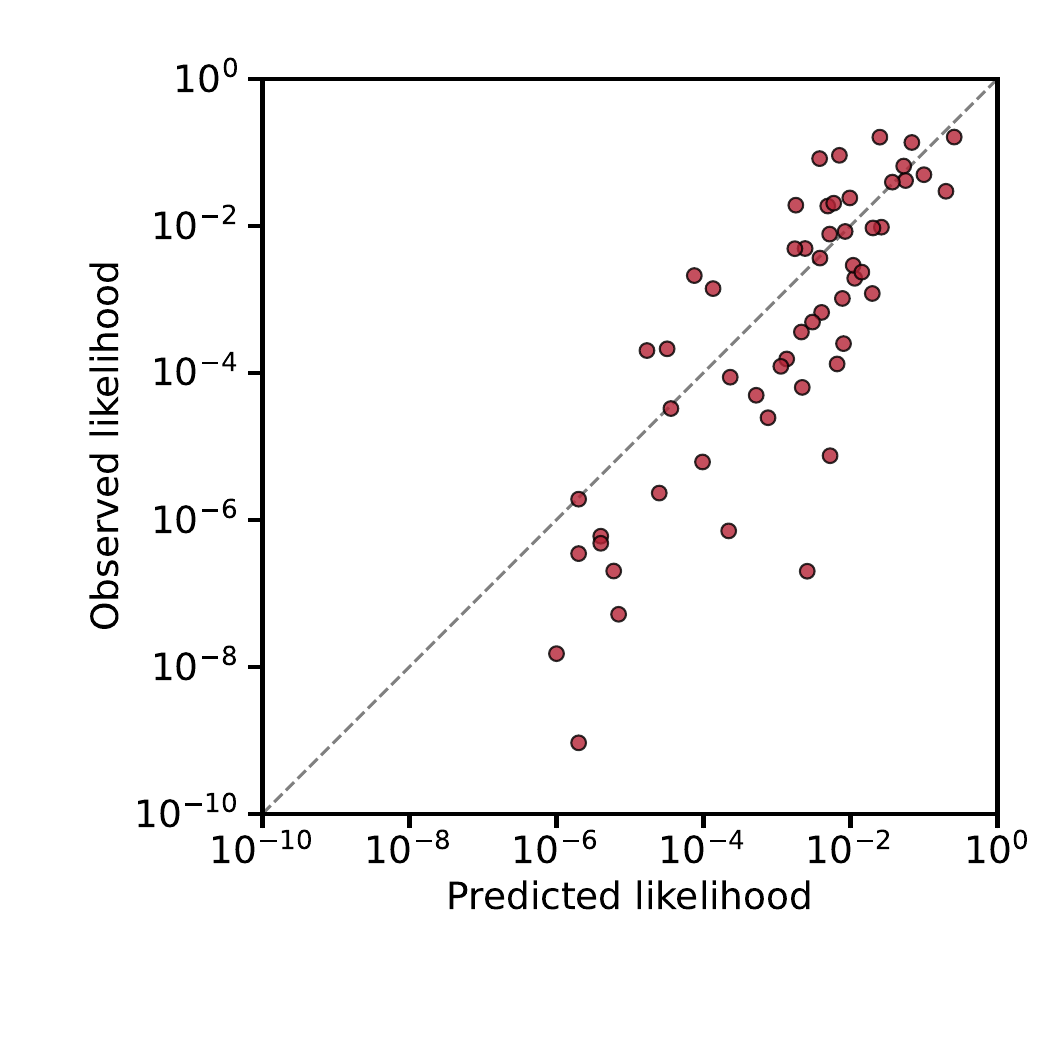}
        \caption{}
        \label{fig:pred_1}
    \end{subfigure}
    \hfill
    \begin{subfigure}[b]{0.45\textwidth}
        \centering
        \includegraphics[width=\textwidth]{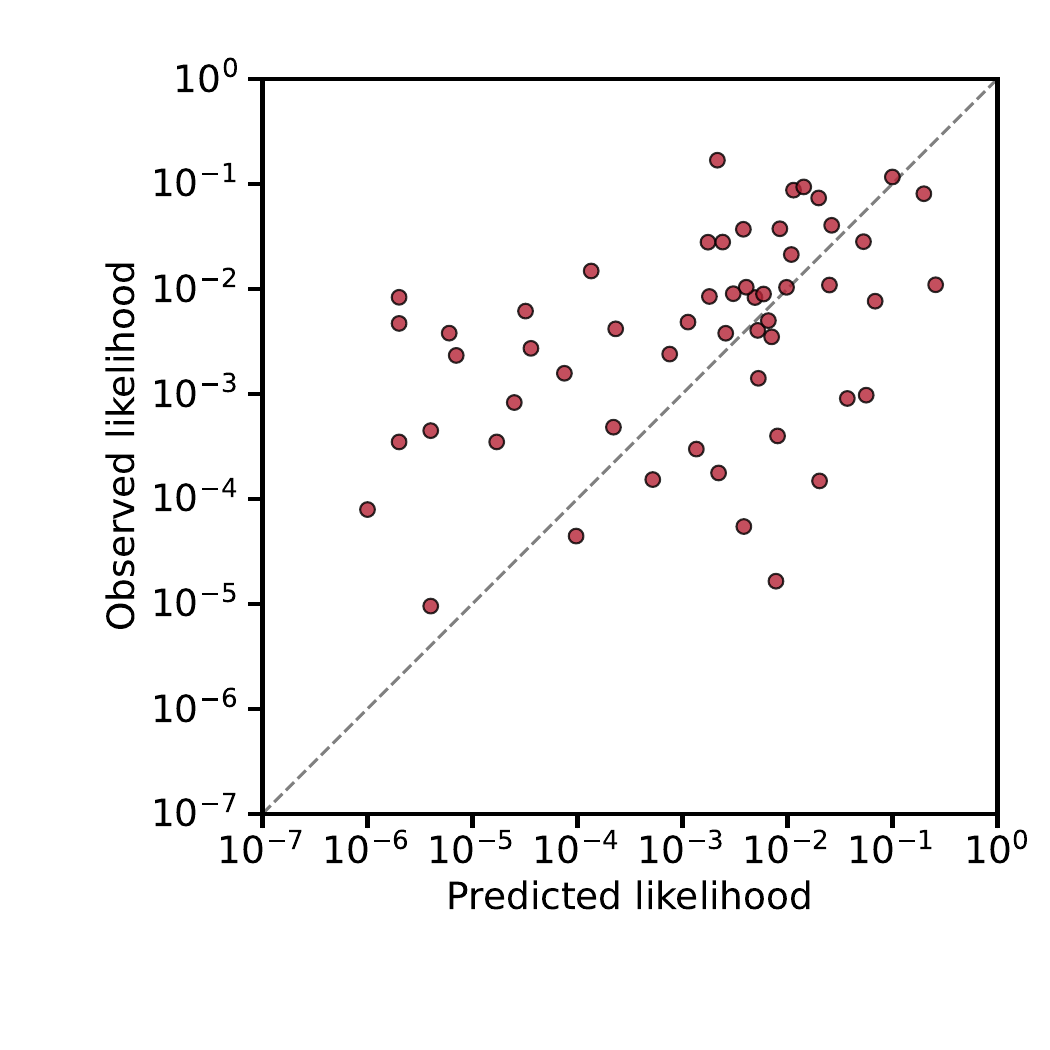}
        \caption{}
        \label{fig:pred_2}
    \end{subfigure}
    \justifying
    \caption{\justifying Observed vs. predicted likelihoods for the
      random matrix used in the main text Figs.\ 1c and 3, except here
      the predictions are made using different (less accurate) schemes. In the main
      text, we predicted likelihoods using the biomass of each state
      and the average species diversity as a common block size. In (a)
      we predict likelihoods using only the biomass of each state,
      setting the diversity of each state to 1, thus
        effectively using predictions
      from the monodominant model (Eq.~(4) in the main text). In (b)
      we use the biomass of each state and the diversity of each state
      (rather than a common diversity equal to the mean). Both schemes
      produce worse predictions than in Fig. 3b. In panel (a) we
      systematically overpredict likelihoods across all biomass
      ranges. Surprisingly we also produce worse predictions for (b),
      where we used the actual diversity of each state rather than a
      mean-field approximation. Comparing goodness of fit using $\chi^2$ values, we notice that the method in the main text Fig. 3b produces a $\chi^2\approx12$, whereas here we get significantly worse $\chi^2$ of (a) 29.15 and (b) 32.50, respectively.
    }
    \label{fig:pred}
\end{figure}

One option is to treat each state as an effective species --- like a
monodominant model --- and use only equilibrium biomass information
for predictions, while ignoring block size (diversity). So,
we then assume
the initial condition of each block to be a uniform distribution,
similar to the monodominant model.  We find that this approach tends
to overpredict strongly for low-likelihood
biomass. (Fig.~\ref{fig:pred_1}), indicating that biomass alone
provides insufficient information. A second option is to incorporate
both block size (diversity) and equilibrium biomass information ---
like the full-block model --- using the distinct species diversity
$L_\alpha$ of each state. Surprisingly, this approach also produces
poor predictions (Fig.~\ref{fig:pred_2}). The overlapping nature of
blocks in our random matrix (Appendix~\ref{app:overlappingStates})
suggests that effective size might differ from the simple species
count within each state. Moreover, the cross-state inhibitions are not
constant, but have disordered fluctuations.

\begin{figure}[ht!]
    \centering
    \begin{subfigure}{0.45\textwidth}
        \centering
        \includegraphics[width=\linewidth]{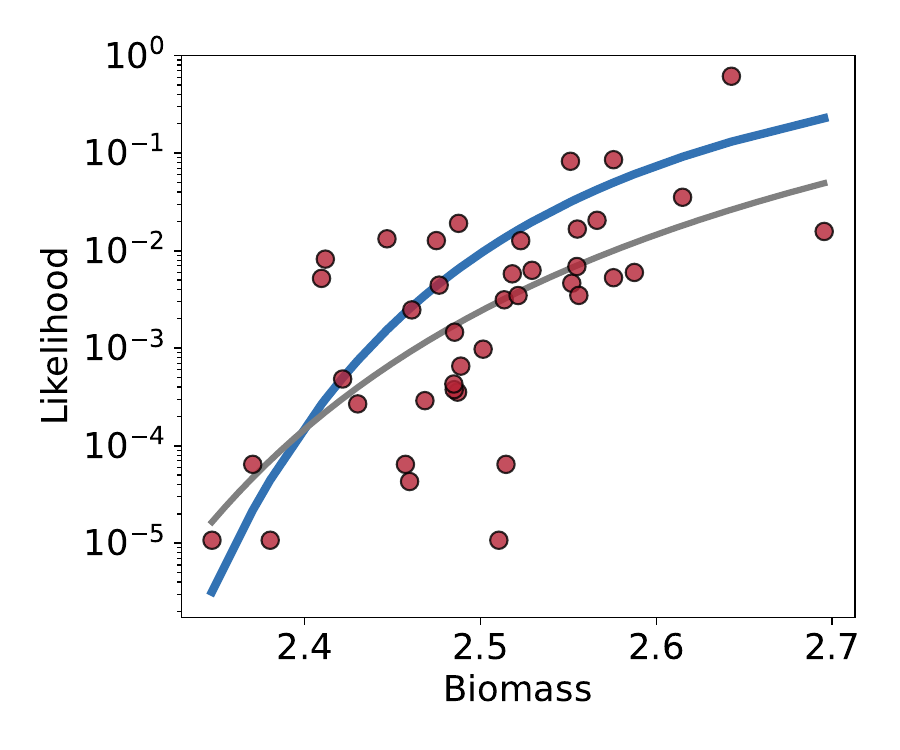}
        \caption{}
        \label{fig:}
    \end{subfigure}
    \hfill
    \begin{subfigure}{0.45\textwidth}
        \centering
        \includegraphics[width=\linewidth]{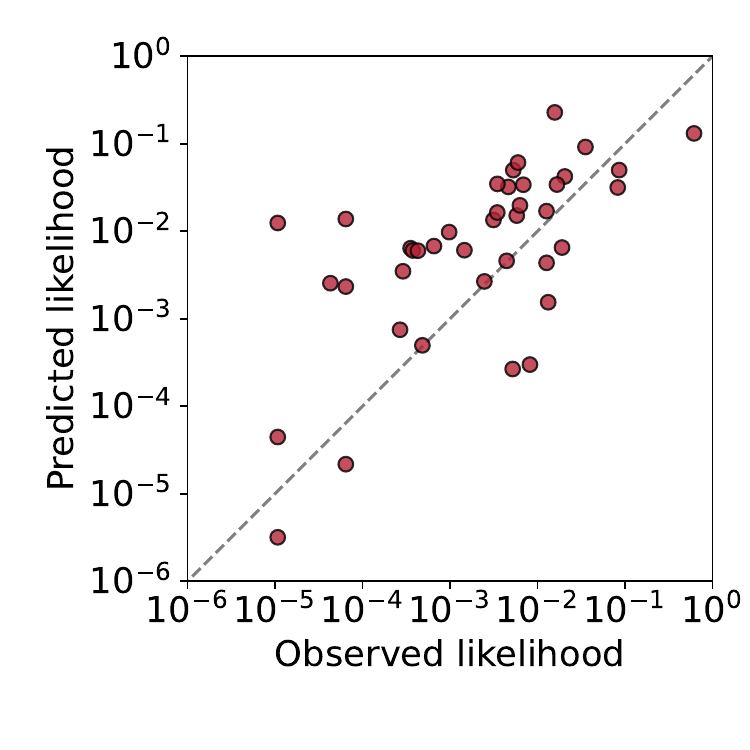}
        \caption{}
        \label{fig:}
    \end{subfigure}

    \begin{subfigure}{0.4\textwidth}
        \centering
        \includegraphics[width=\linewidth]{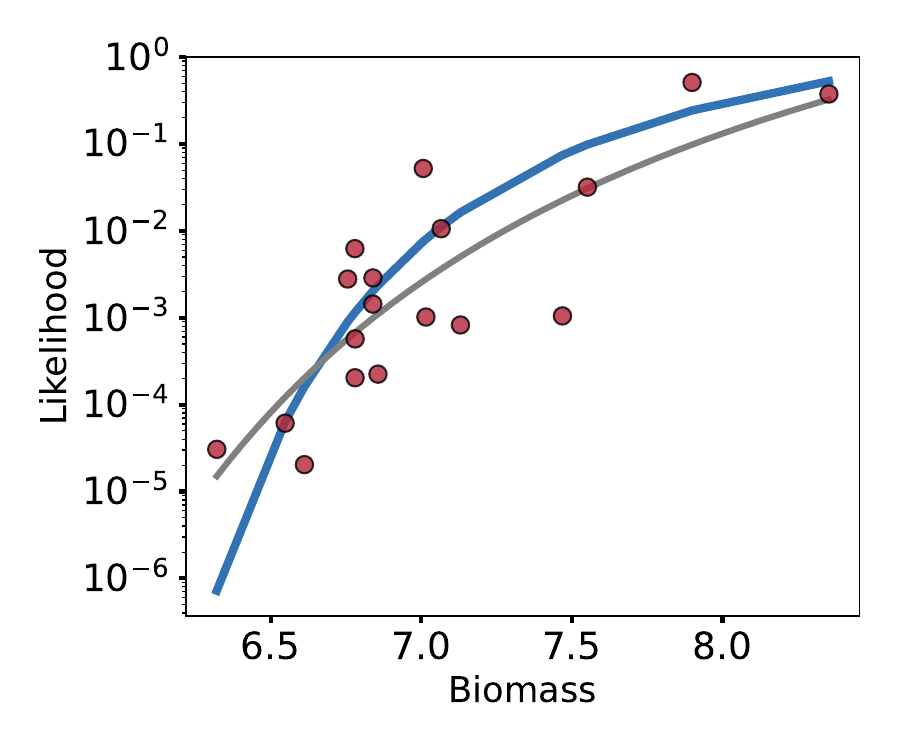}
        \caption{}
        \label{fig:3R}
    \end{subfigure}
    \hfill
    \begin{subfigure}{0.4\textwidth}
        \centering
        \includegraphics[width=\linewidth]{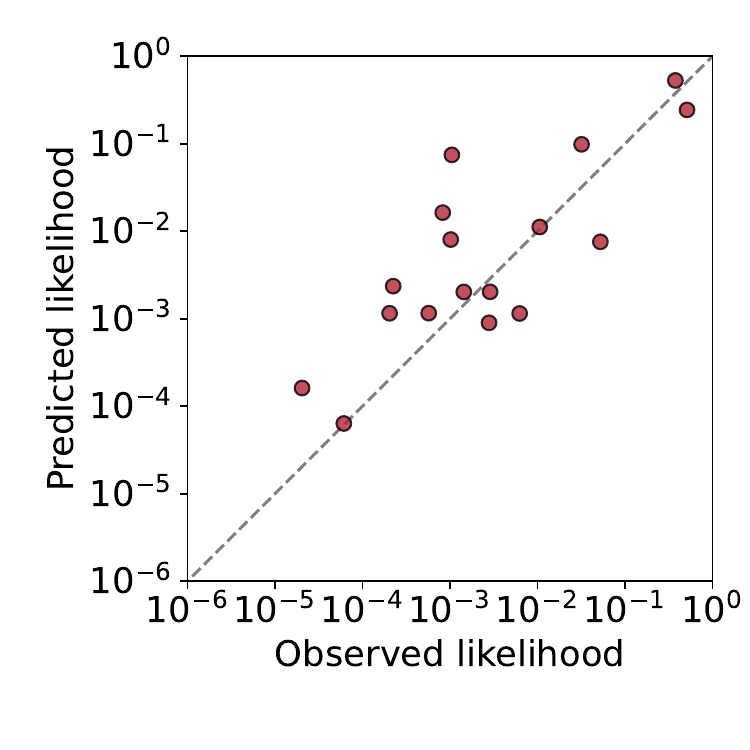}
        \caption{}
        \label{fig:}
    \end{subfigure}

    \caption{\justifying Biomass--likelihood relationship across
      different parameter regimes. The log-hyperbolic relationship
      emerges for different values of $\mu$ and $\sigma$: panels (a,b)
      show results for $\mu = 0.5$ and $\sigma = 0.2$, while panels
      (c,d) show results for $\mu = 0.2$ and $\sigma = 0.15$. We
      predict likelihoods using biomass and average diversity across
      all states. In panels (a) and (c), the blue curve shows
      predictions using Eq.~(5) from the main text, while the gray
      curve shows the hyperbolic fit $\log(p) \propto
      (B_c-B)^{-1}$. Panels (b) and (d) compare predicted versus
      observed likelihoods, demonstrating strong agreement across both
      parameter regimes. In both cases, we note that the predictions are slightly better than the fits in terms of their $\chi^2$ values. In (a)--(b), the fit has $\chi^2\approx19.05$, while the prediction has $\chi^2 \approx15.2$. Similarly, in (c)--(d), the fit has $\chi^2\approx5.2$, while the prediction has $\chi^2 \approx4.6$}

    \label{fig:general}
\end{figure}

\begin{figure}[ht!]
    \centering
    \begin{subfigure}{0.45\textwidth}
        \centering
        \includegraphics[width=\linewidth]{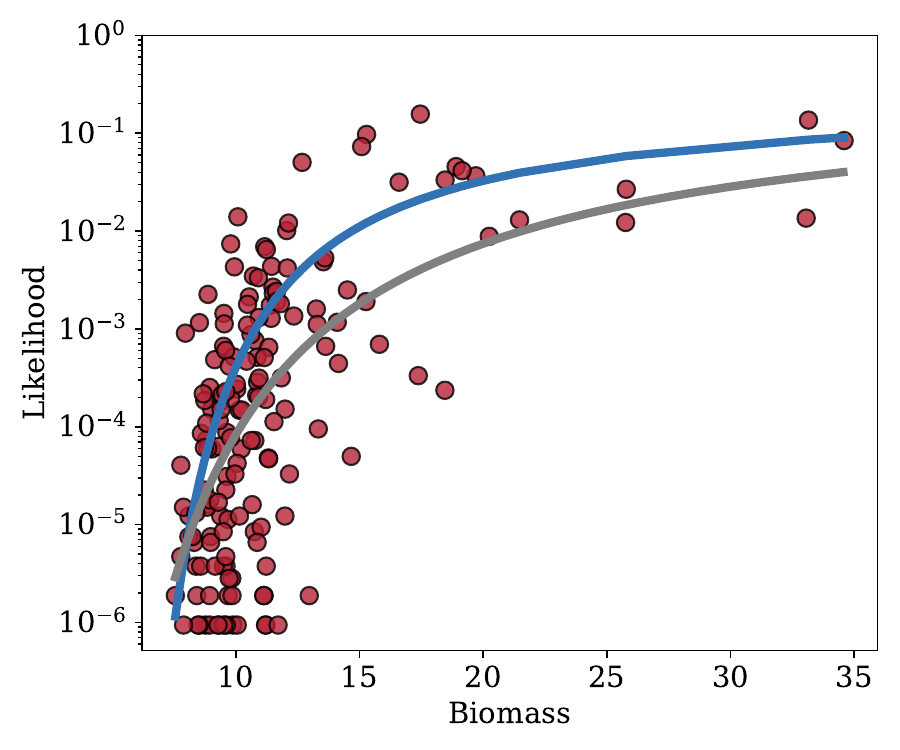}
        \caption{}
        \label{fig:}
    \end{subfigure}
    \hfill
    \begin{subfigure}{0.4\textwidth}
        \centering
        \includegraphics[width=\linewidth]{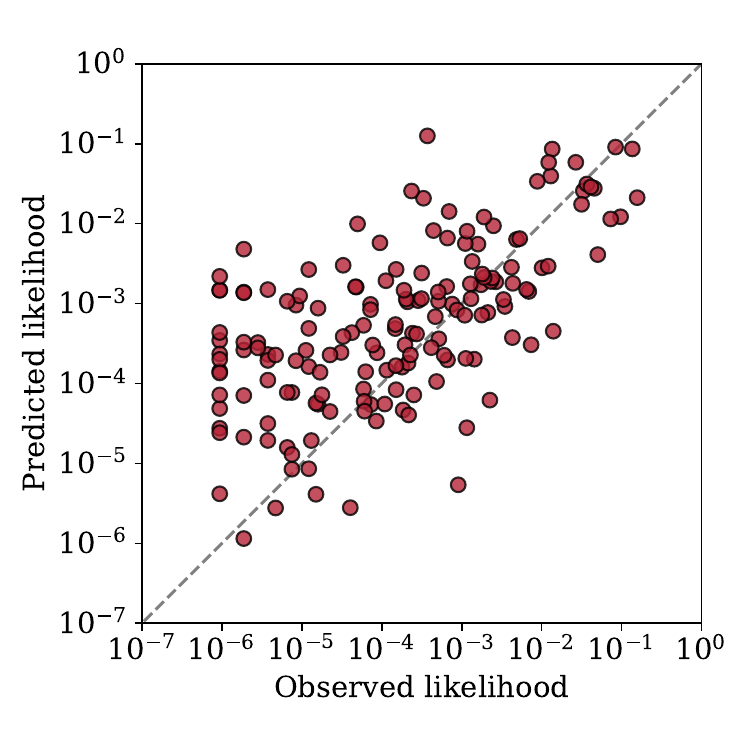}
        \caption{}
        \label{fig:}
    \end{subfigure}

    \begin{subfigure}{0.4\textwidth}
        \centering
        \includegraphics[width=\linewidth]{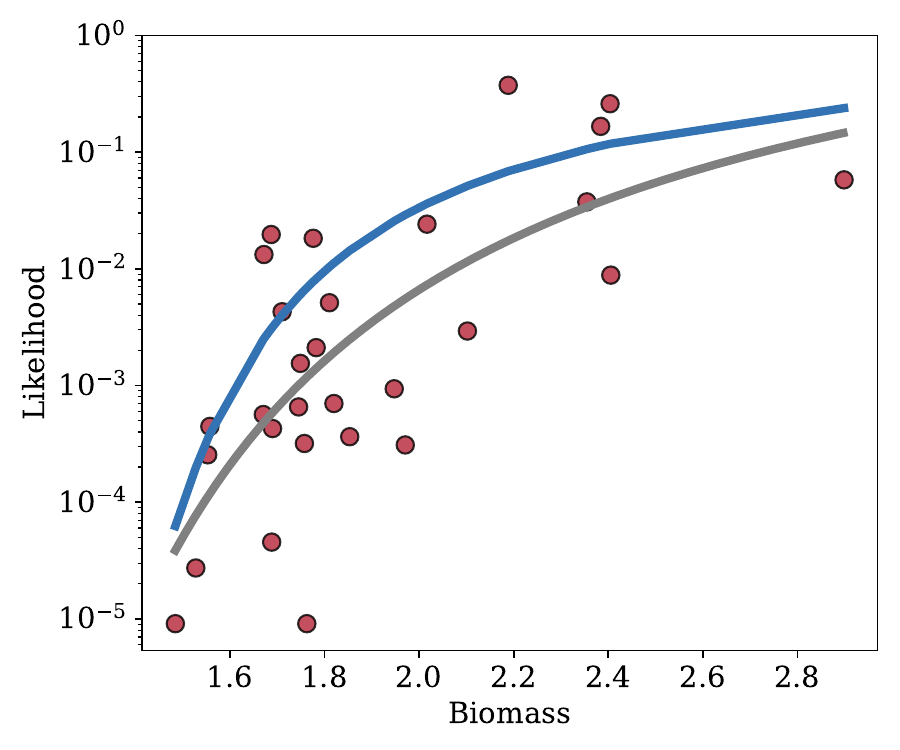}
        \caption{}
        \label{fig:3R}
    \end{subfigure}
    \hfill
    \begin{subfigure}{0.4\textwidth}
        \centering
        \includegraphics[width=\linewidth]{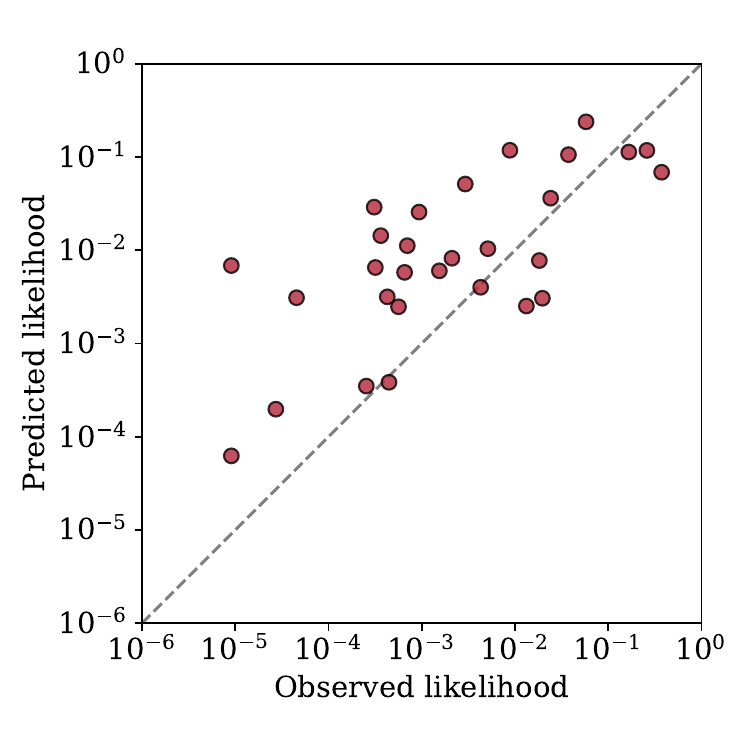}
        \caption{}
        \label{fig:}
    \end{subfigure}

    \caption{\justifying\textcolor{black}{Biomass--likelihood relationship across 
    different number of species. The log-hyperbolic relationship 
    emerges for different values of $S$: panels (a,b) 
    show results for $S=200$ ($\mu = 0.25$ and $\sigma = 0.21$), while panels 
    (c,d) show results for $S= 50$ $\mu = 1$ and $\sigma = 0.4$. We 
    predict likelihoods using biomass and average diversity across 
    all states. In panels (a) and (c), the blue curve shows 
    predictions using Eq.~(5) from the main text, while the gray 
    curve shows the hyperbolic fit $\log(p) \propto 
    (B_c-B)^{-1}$. Panels (b) and (d) compare predicted versus 
    observed likelihoods, demonstrating strong agreement across both 
    parameter regimes. In both cases, we note that the predictions are slightly better than the fits in terms of their $\chi^2$ values. In (a)--(b), the fit has $\chi^2\approx58.2$, while the prediction has $\chi^2 \approx40.1$. Similarly, in (c)--(d), the fit has $\chi^2\approx15.4$, while the prediction has $\chi^2 \approx12.2$}}
    \label{fig:diff_s}
\end{figure}

\begin{figure}[ht!]
\centering
\includegraphics[width=0.5\linewidth]{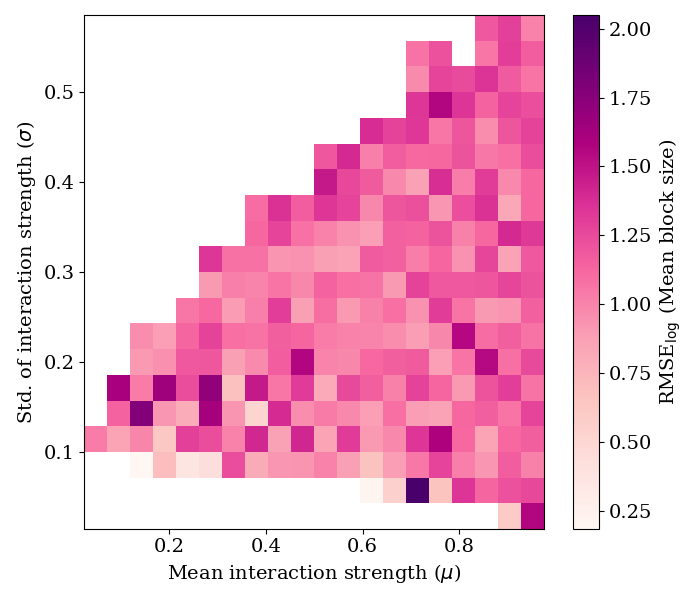}
\caption{\justifying \textcolor{black}{ The prediction error of using mean block size in Eq.~\eqref{eq:block_likelihood} to estimate likelihood of state remains consistent across the multistable regime. The error is similarly distributed across the  parameter space, suggesting that our prediction works equally well throughout the multistability regime. }}
\label{fig:rmse_1}
\end{figure}

\begin{figure}[ht!]
\centering
\includegraphics[width=0.5\linewidth]{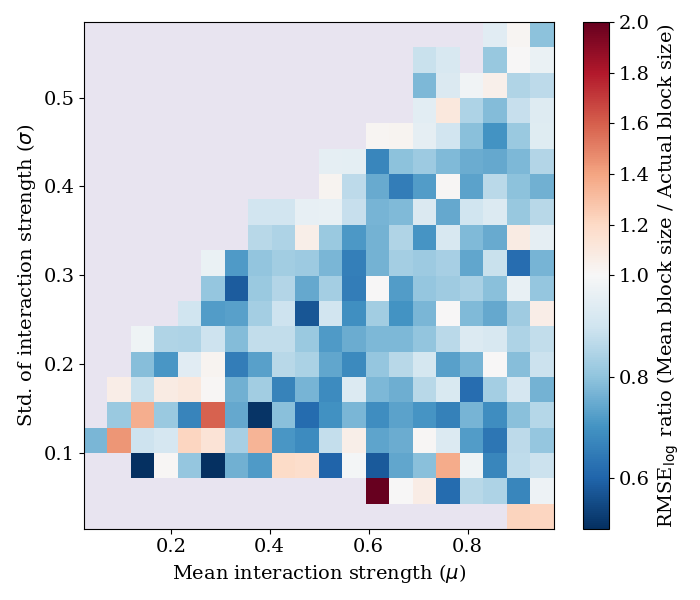}
\caption{\justifying \textcolor{black}{ Using the mean block size gives better predictions for likelihood across the multistable regime. The colorbar shows the ratio $\mathrm{RMSE}_{\log}(\text{mean block size})/ \mathrm{RMSE}_{\log}(\text{actual block size})$. Most values are below 1, indicating that using the mean block size leads to lower prediction error.}}
\label{fig:rmse_ratio}
\end{figure}

We find that we achieve most accurate predictions by using a hybrid
approach: that of using the true
equilibrium
biomass of each state along with the
average block size across all states (rather than using state-specific
diversity values). \textcolor{black}{As shown in Appendix~\ref{app:blockOverlaps}, as overlap increases, the mean block size becomes a better predictor of the likelihood. The predicted and observed likelihoods match reasonably well over a wide range (Figs.~\ref{fig:fig3}b , \ref{fig:general} and \ref{fig:diff_s}). Moreover, we find that this predictability persists throughout the entire multistable region of the model (Fig.~\ref{fig:rmse_1}, \ref{fig:rmse_ratio}).}  For a random matrix, this coarse-grained information proves
sufficient to estimate state likelihoods effectively.

We observe some discrepancies at very low likelihoods for two
reasons. First, our number of simulations, while extensive are
limited, and do not sample these rare states sufficiently to provide
reliable likelihood estimates. Second, we do not explicitly model the
detailed effects of overlaps and the exact structure of each state in
our mean-field approach. Nevertheless, across the multistable regime
for different values of $\mu$ and $\sigma$, we successfully predict
both individual state likelihoods and the overall biomass--likelihood
relationship using only biomass and average species richness
(Fig.~\ref{fig:general}). These results demonstrate that our
theoretical framework may capture the essential
dynamics
of state-state competition even in the presence of structural complexity.


\FloatBarrier

\section{Derivation of the multistability boundary}
\label{app:multistabilityboundary}
In the main text, we show that the GLV model with strong interactions exhibits a transition from a single stable state to multiple stable states. This transition occurs at a critical boundary that we derive here analytically and display as the black line in Fig.~\ref{fig:fig1}a.

We will follow the approach of previous work \cite{Bunin2017_EcoComLv}, which derived this boundary under weak interaction scaling where $ \text{mean}(A_{ij}) = \tilde{\mu}/S$ and $\text{std}(A_{ij}) =\tilde{\sigma}/\sqrt{S}$. The transition from single to multiple stable states for symmetric interaction matrices occurs at:
\begin{equation}
    \tilde{\sigma} = \frac{1-\frac{\tilde{\mu}}{S}}{\sqrt{2}}
\end{equation}

In this manuscript, we consider strong interactions that do not scale with the number of species $S$. To derive the multistability boundary in this regime, we remove the original scaling by substituting $\tilde{\mu} \to \mu S$ and $\tilde{\sigma} \to \sigma \sqrt{S}$. This transformation yields:
\begin{equation}
    \sigma = \frac{1-\mu}{\sqrt{2S}},
\end{equation}
which corresponds to Eq.~\eqref{eq:multistabBoundary} in the main text. This expression reveals important features of the strong interaction regime. For $\mu > 1$, no value of $\sigma$ produces a single stable state---the system always exhibits multistability. In the feasible parameter region, multistability dominates most of the phase space. Our numerical simulations confirm excellent agreement with this analytical prediction (Fig.~\ref{fig:fig1}a), validating our approach despite the approximations involved in extending the weak interaction theory to the strong interaction case (in particular, linear response which is central to the original approach is no longer valid, leading one to expect higher-order corrections to be important).

The boundary demonstrates that increasing the mean interaction strength $\mu$ expands the multistable regime by reducing the critical $\sigma$ required for the transition. This contrasts sharply with weak interaction models where the critical $\sigma$ remains constant \cite{Bunin2017_EcoComLv}, highlighting fundamental differences between these two interaction regimes in determining stability.

\FloatBarrier
\FloatBarrier

\section{The shape of the biomass--likelihood relationship}
\label{app:hyperboliclog}
\begin{figure}[h!]
    \centering
    \begin{subfigure}{0.45\textwidth}
        \centering
        \includegraphics[width=\linewidth]{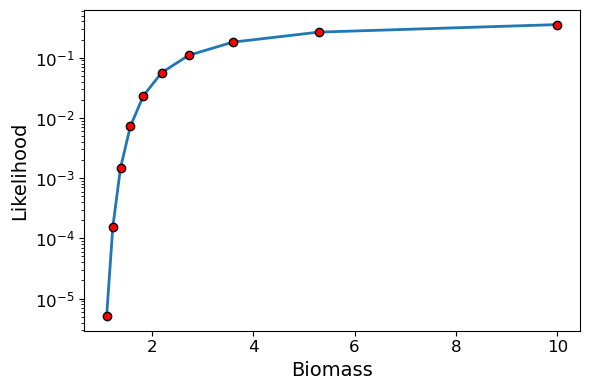}
        \caption{}
        \label{fig:5R}
    \end{subfigure}
    \hfill
    \begin{subfigure}{0.45\textwidth}
        \centering
        \includegraphics[width=\linewidth]{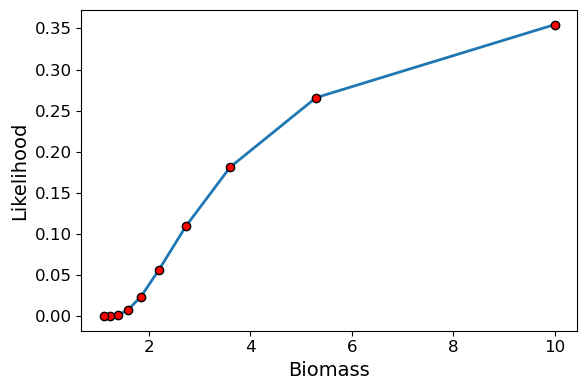}
        \caption{}
        \label{fig:6R}
    \end{subfigure}

    \begin{subfigure}{0.45\textwidth}
        \centering
        \includegraphics[width=\linewidth]{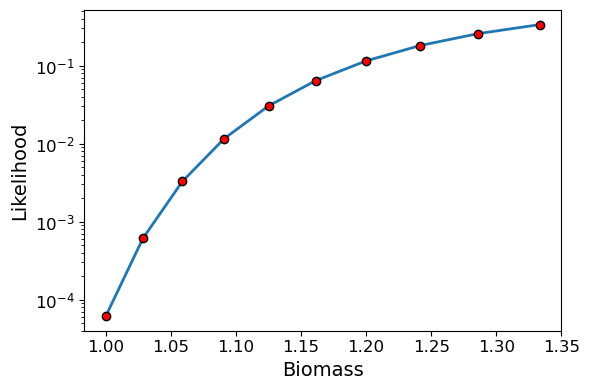}
        \caption{}
        \label{fig:3R}
    \end{subfigure}
    \hfill
    \begin{subfigure}{0.45\textwidth}
        \centering
        \includegraphics[width=\linewidth]{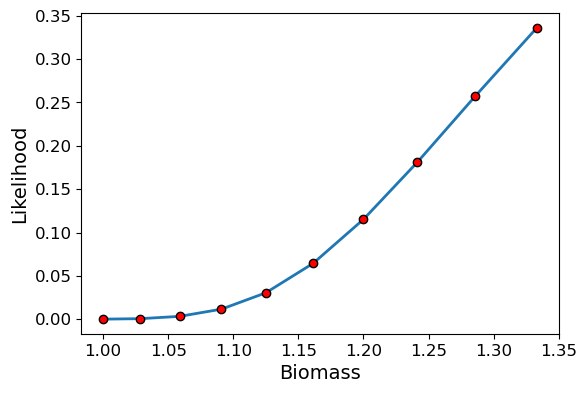}
        \caption{}
        \label{fig:4R}
    \end{subfigure}

    \begin{subfigure}{0.45\textwidth}
        \centering
        \includegraphics[width=\linewidth]{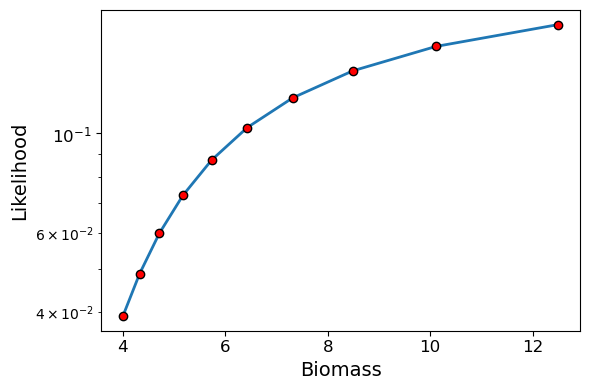}
        \caption{}
        \label{fig:1R}
    \end{subfigure}
    \hfill
    \begin{subfigure}{0.45\textwidth}
        \centering
        \includegraphics[width=\linewidth]{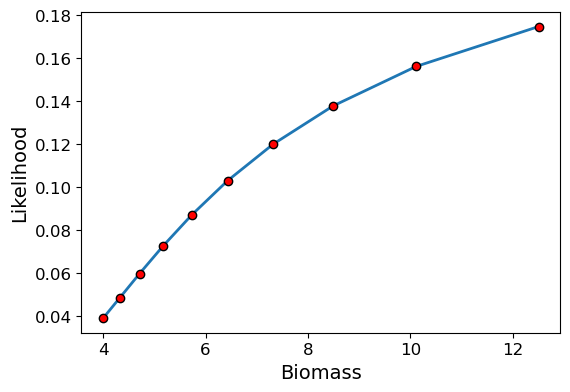}
        \caption{}
        \label{fig:2R}
    \end{subfigure}

    \caption{\justifying Likelihood versus biomass across different self-inhibition regimes. Note that the pairs of plots on the left and the right in each row represent the same data points, but the left has the $y$-axis on log scale while the right has it on a linear scale. Panels (a,b) show the case where self-inhibition terms span the full range $(0,D)$, panels (c,d) show the case where $A_{ii} \approx 0$, and panels (e,f) show the case where self-inhibition values $A_{ii} \approx D$. Each pair compares linear y-scale (a,c,e) with logarithmic y-scale (b,d,f). High self-inhibition $A_{ii} \approx D$ produces a concave-downward biomass--likelihood relationship (f), while low self-inhibition $A_{ii} \approx 0$ generates a concave-upward relationship (d). When self-inhibition values cover the complete range $(0,D)$, both curvature types appear across different biomass ranges (f), though the logarithmic y-scale masks this transition (b).}

    \label{fig:Rs}
\end{figure}

The relationship between biomass and likelihood exhibits different curvatures depending on the distribution of self-inhibition values across species. In our monodominant framework, diagonal self-inhibition terms $A_{ii}$ range between $0$ and the cross-inhibition strength $D$. Since biomass scales inversely with self-inhibition, large $A_{ii}$ values produce low-biomass while small $A_{ii}$ values generate high-biomass.

When self-inhibition values span the full interval $(0,D)$, the
biomass--likelihood relationship displays two distinct curvature
regimes in coordinates where we do not take the logarithm of
  the likelihood (Fig.~\ref{fig:6R}). At low-biomass, the curve exhibits concave-upward curvature, while at high-biomass, it becomes concave-downward. Plotting likelihood on a logarithmic scale often masks this transition between curvature regimes.

We can isolate each curvature type by constraining the distribution of
self-inhibition values. When all $A_{ii}$ values cluster near
$D$—producing uniformly low-biomass—we observe only concave-upward
curvature (Fig.~\ref{fig:4R}). Conversely, when $A_{ii}$ values
concentrate near zero—generating uniformly high-biomass—the
relationship shows concave-downward curvature
(Fig.~\ref{fig:2R})
.When self-inhibition values cover the complete range from zero to $D$, both curvature types emerge across different biomass ranges, creating the full biphasic relationship we observe in our simulations (Fig.~\ref{fig:6R}). In Fig.~\ref{fig:5R}, due to the log scale used for likelihood, the biphasic behavior is not clearly visible. 

{\color{black}
The two regimes of the biomass--likelihood relationship can be broken down as a
  roughly exponential part at low likelihoods ($p \lesssim 1/S$) and
  an
  approach to a plateau where
$p$ necessarily approaches a constant less than 1 as the biomass gets
  very large.
  The hyperbolic form $\log\left(p\right) \propto (B -
B_c)^{-1}$ is in some sense the simplest functional form that approaches a probability of 1 at large biomass and approaches a constant biomass at zero probability. To capture both of these limits, in the main text we thus used the log-hyperbolic relation as a simplified, rough description of the behavior:
\begin{equation}
    \log\left(\frac{p}{p^*}\right) = \frac{\kappa}{B_c - B}.
\end{equation}

Here, $B_c$ represents the lower biomass threshold. As $B \to B_c$, the denominator goes to zero and $p \to 0$,
which captures the sharp drop in likelihood at low-biomass. The critical biomass, $B_c$ decreases with increasing mean interaction strength $\mu$ and decreasing standard deviation of interaction strength $\sigma$ (Fig.~\ref{fig:13}).When $\mu$ decreases, the overall biomass increases, which leads to an increase in $B_c$. Similarly, when $\sigma$ increases, the variation in biomass becomes larger, producing more high-biomass states. This again increases the least likely biomass $B_c$.

\begin{figure}[t]
\centering
\includegraphics[width=0.5\linewidth]{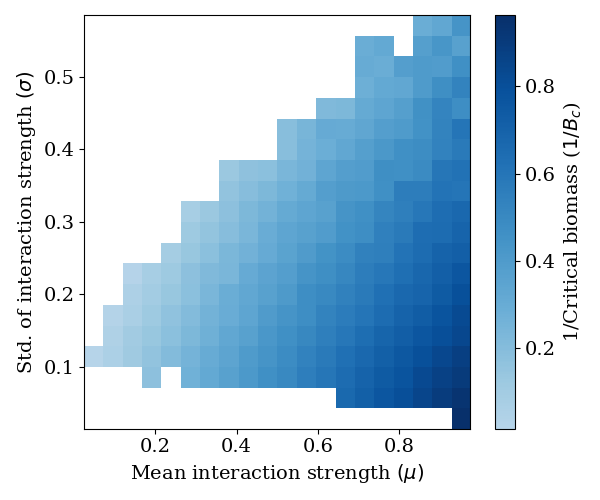}
\caption{\justifying  \textcolor{black}{inverse of the critical biomass ($1/B_c$) decreases as mean interaction strtength $\mu$ decreases and standard deviation of interaction strength $\sigma$ increases across the multistability regime. For each $(\mu, \sigma)$ point, we generated a random interaction matrix and fitted log-hyperbolic relation $\log\left(p\right) \propto (B_c - B)^{-1}$ to obtain $B_c$. When $\mu$ decreases, the overall biomass increases, which raises $B_c$ and hence lowers $1/B_c$. Similarly, increasing $\sigma$ increases the variation in biomass, producing more high-biomass states, which again increases $B_c$ and therefore decreases $1/B_c$.}}
\label{fig:13}
\end{figure}


\begin{figure}[t]
\centering
\includegraphics[width=0.45\linewidth]{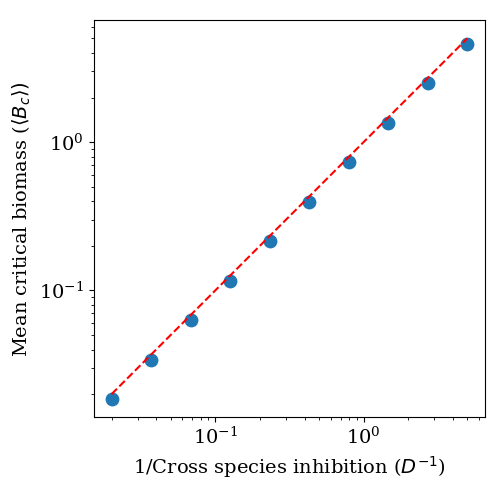}
\caption{\justifying \textcolor{black}{The critical biomass $B_c$ scales as inverse of cross-species inhibition $1/D$ for monodominant model. We generated multiple monodominant interaction matrices with self-inhibition of species uniformly sampled between $0$ and $D$, and the cross-species inhibition set to $D$. For each matrix, we computed the likelihoods of the states and averaged over $100$ such matrices for each value of $D$. The blue dots show the mean $B_c$ obtained by fitting log-hyperbolic relation $\log\left(p\right) \propto (B_c - B)^{-1}$. The mean $B_c$ shows a linear relation with $1/D$.}}
\label{fig:9}
\end{figure}

In  niche-structured models, it was shown in
Ref.~\cite{taylor2024structure}
that in a limit of large $S$ with exponentially many stable states
available, there is a precise sense in which probability scales
exponentially with biomass.
While these results share the common theme that states with greater biomass are
associated with larger basins of attraction and higher probability, it
is not completely clear exactly how these different functional
relations are related.
In the models of Ref.~\cite{taylor2024structure}, however, the exponential/log-linear biomass-probability relationship emerges most clearly in a large $S$ limit where the central limit theorem is relevant, corresponding to a limit in which the probabilities of almost all individual states become very low.
This is not related in any obvious way to any clear regime of the random matrix systems, but
 further inquiries in this
direction may lead to fruitful further insights.}

\FloatBarrier

\FloatBarrier

{\color{black}
\section{Details of simulation methods for main text figures, including parameters used}
\label{app:simulationdetails}
We used the simulation method described in Appendix~\ref{app:methods} to estimate the likelihoods for the biomass--likelihood relationship for all results. The code used for the simulations and for generating the figures is available at: \url{https://github.com/eltanin4/emergent-self-inhibition-biomass-likelihood}.

\textbf{Fig.~\ref{fig:fig1}a:}  
We considered a $35 \times 35$ grid in the phase space of mean interaction strength $\mu$ (ranging from $0$ to $1$) and standard deviation of interaction strength $\sigma$ (ranging from $0$ to $0.6$) . For each point, we generated five symmetric interaction matrices from a normal distribution. For every matrix, we ran simulations with $10^4$ random initial conditions and got the stable states (Appendix~\ref{app:methods}).

If all five matrices led to only one stable state, we colored the point white (single stable state). If, in any of the five matrices, the abundance diverged, we colored the point gray, indicating no stable states. If all five matrices converged and at least one of them showed more than one stable state, we defined the point multistable. We colored these points based on the average number of stable states obtained across the five matrices. We obtained the multistability boundary using the method described in Appendix~\ref{app:multistabilityboundary}.

\textbf{Fig.~\ref{fig:fig1}c:}  
The inset shows a specific interaction matrix drawn from the ensemble with mean interaction strength $\mu = 0.5$ and standard deviation of interaction strength $\sigma = 0.3$ from a normal distribution. For this matrix, we computed the biomass and likelihood of stable states using $10^6$ random initial conditions (Appendix~\ref{app:methods}), and plotted them as red points. We then repeated this process for 110 different matrices sampled with the same $\mu$ and $\sigma$ from a normal distribution. 

To obtain the average biomass-likelihood relationship from the resulting data points, we did the following. Starting from the lowest biomass state that we observed, we binned states with similar biomass. Because we observed many more low-biomass states than high-biomass states, we used geometric binning, starting with a bin size of $0.1$ and subsequently increasing bin width by a factor of $1.3$. Within each bin, we computed the mean log likelihood over all states in that bin $\log_{10}(p)$. This resulted in the average biomass-likelihood relationship across all 110 matrices we sampled, shown as the black curve.

We also fitted biomass--likelihood data for each of the 110 sampled matrices to a log-hyperbolic relation of the form
\begin{equation}
  \log(p) = \frac{\kappa}{B_c - B},  
\end{equation}
where $B_c$ and $\kappa$ were two fit parameters, which we found to be on average $2$ and $16$, respectively. We then averaged these fitted curves using the same geometric binning procedure described above, which resulted in the blue curve.

\textbf{Fig.~\ref{fig:fig2}a:}  
To check the difference between the interaction of the species which coexist with the ones which don't, we used the interaction matrix shown in the inset of Fig.~1c and its corresponding stable states. For each species, we divided its $N^2 - N$ cross-species interactions into two groups. In the first group (same state), we included interactions with species that co-occur with it in at least one stable state. In the second group (different state), we included interactions with species that do not co-occur with it in any stable state. We then plotted the distribution of interaction strengths for these two groups: same state (blue) and different state (red).

\textbf{Fig.~\ref{fig:fig2}b:}  
This panel shows the same interaction matrix as in the inset of Fig.~1c.

\textbf{Fig.~\ref{fig:fig2}c:}  
We constructed an effective coarse-grained model at the level of states for the matrix shown in Fig.~2b. For $55$ states, we built a $55 \times 55$ interaction matrix. The diagonal elements are given by $1/B_\alpha$, where $B_\alpha$ is the biomass of state $\alpha$. We set all off-diagonal elements to a constant value $\mu=0.5$.

\textbf{Fig.~\ref{fig:fig2}d:}  
The inset shows an example of a monodominant interaction matrix. We constructed a $10 \times 10$ matrix where the diagonal elements are linearly spaced from $0.1$ to $0.3$, and all off-diagonal elements are equal to $1.01$. In the main plot, we show the biomass and likelihood of all stable states obtained from simulations (red points) using $10^5$ random initial conditions (Appendix~\ref{app:methods}). We also computed the likelihood using $A_{ii}$, where $A_{ii}$ is the self-inhibition of species $i$, in Eq.~\eqref{app:monodominantCalc} and plotted it as the theory (blue curve).

\textbf{Fig.~\ref{fig:fig2}e:}  
To introduce the effect of species diversity on likelihood of the state, we constructed a block-diagonal interaction matrix with block size $4$. We set the diagonal elements to $1.01$. Within each block, species interacted weakly with a constant interaction strength. For species in different blocks, we set the interaction strength to $1.01$. We chose the inter-block self-inhibition values to vary across blocks and linearly spaced them from $0.2$ to $0.8$.

\textbf{Fig.~\ref{fig:fig2}f:}  
To check the prediction accuracy of likelihood for Eq.~\eqref{eq:block_likelihood}, we constructed a similar matrix for block size $6$ with the same in-block and off-block interaction structure (not shown). We computed the likelihood of stable states from simulations for both matrices using $10^5$ random initial conditions (Appendix~\ref{app:methods}). We also estimated the likelihood using biomass and block size in Eq.~\eqref{eq:block_likelihood}. The plot compares the predicted and observed likelihoods for both the matrices of different block sizes.

\textbf{Fig.~\ref{fig:fig3}a:} To obtain the average biomass-likelihood relationship for the same $110$ matrices used in Fig.~\ref{fig:fig1}c, we did the following. We divided the biomass into $15$ uniformly spaced bins between $3$ and $9$. Within each bin, we computed the mean likelihood over all states in that bin. This resulted in the average biomass-likelihood relationship across all 110 matrices we sampled, shown as the red curve. For the same set of states, we also estimated the likelihood using Eq.~\eqref{eq:block_likelihood} with the mean block size and biomass. We then averaged the estimated likelihoods using the same binning procedure described above, which resulted in the black curve.

We considered $7$ values of the number of sampled matrices, uniformly spaced between 10 and 110. For each case, we sampled that number of matrices and computed the mean observed and predicted likelihoods by binning the biomass into the same bins as mentioned above, which gave $15$ paired points $(p_\alpha^{\mathrm{obs}}, p_\alpha^{\mathrm{theory}})$ for each case. We then calculated $\mathrm{RMSE}_{\log}$ (Eq~\ref{eq:rmse_log}) as a measure of prediction error by taking the root mean square of the difference between their logarithms. In the inset, we showed how this error changed with the number of sampled matrices. The error decreased as the number of matrices increased, following a power law. The black line represents a fitted power law with exponent $-0.4$.

\textbf{Fig.~\ref{fig:fig3}b:}  
For the states obtained from the inset matrix shown in Fig.~\ref{fig:fig1}c, we used the biomass of the states together with the mean block size in Eq.~\eqref{eq:block_likelihood} to estimate the likelihood of the states. We then plotted the observed likelihoods against the predicted likelihood of the states to compare the two.
}

\end{document}